\documentclass[letterpaper, numberedappendix]{emulateapj} 

\usepackage{graphicx}
\usepackage{epsfig}
\usepackage{amsmath}
\usepackage{subfigure} 
\usepackage{hyperref}
\usepackage{amssymb}
\usepackage{lscape} 
\usepackage{apjfonts}
\usepackage{latexsym}

\newcommand{\vx}[0]{\mathitbf{x}}
\newcommand{\figref}[1]{Figure \,\ref{#1}}
\newcommand{\tabref}[1]{Table \,\ref{#1}}
\newcommand{\secref}[1]{Section \,\ref{#1}}
\newcommand{\appref}[1]{Appendix \,\ref{#1}}
\newcommand{\MW}[1]{$\text{MW}{#1}$}
\DeclareMathAlphabet{\mathitbf}{OML}{cmm}{b}{it}

\shorttitle{The  formation  history of  stellar  haloes through  model emulators} 

\shortauthors{F.~A. G\'omez \& C.E. Coleman-Smith et al.}

\begin{document}\title{Characterizing the formation history of Milky Way-like
  stellar haloes with model emulators}

\author{Facundo A. G\'omez}\affil{Department of Physics and Astronomy,
  Michigan State University, East Lansing, MI 48824, USA}\affil{Institute
  for Cyber-Enabled Research, Michigan State University, East Lansing, MI
  48824, USA}  
\author{Christopher E. Coleman-Smith}\affil{Department of Physics, Duke
  University, Durham, NC, 27708, USA}  
\author{Brian W. O'Shea}\affil{Department of Physics and Astronomy,
  Michigan State University, East Lansing, MI 48824,
  USA}\affil{Institute for Cyber-Enabled Research, Michigan State
  University, East Lansing, MI 48824, USA}\affil{Lyman Briggs College,
  Michigan State University, East Lansing, MI 48825, USA}\affil{Joint
  Institute for Nuclear Astrophysics}
\author{Jason Tumlinson}\affil{Space Telescope Science Institute,
  Baltimore, MD, USA} 
\author{Robert. L. Wolpert}
\affil{Department of Statistical Science, Duke University, Durham, NC
  27708-0251}. 

\label{firstpage}

\begin{abstract}
  We use the semi-analytic model ChemTreeN, coupled to cosmological
  $N$-body simulations, to explore how different galaxy formation histories
  can affect observational properties of Milky Way-like galaxies' stellar
  haloes and their satellite populations.  Gaussian processes are used to
  generate model emulators that allow one to statistically estimate a
  desired set of model outputs at any location of a $p$-dimensional input
  parameter space.  This enables one to explore the full input parameter
  space orders of magnitude faster than could be done otherwise.  Using
  mock observational data sets generated by ChemTreeN itself, we show that
  it is possible to successfully recover the input parameter vectors used
  to generate the mock observables if the merger history of the host halo
  is known.  However, our results indicate that for a given observational
  data set the determination of ``best fit'' parameters is highly
  susceptible to the particular merger history of the host.  Very different
  halo merger histories can reproduce the same observational dataset, if
  the "best fit" parameters are allowed to vary from history to history.
  Thus, attempts to characterize the formation history of the Milky Way
  using these kind of techniques must be performed statistically, analyzing
  large samples of high resolution $N$-body simulations.

\end{abstract}

\keywords{galaxies: formation -- Galaxy:  formation -- Galaxy: halo --
  methods: statistical --  methods: analytical -- methods: $N$-body
  simulations}

\section{Introduction}
\label{sec:intro}

Understanding the formation and evolution of galaxies is a central and
long-standing problem in astrophysics.  Over the past century, and
particularly in the past decade, a tremendous amount of information has
been gleaned about populations of galaxies and their temporal evolution,
and data have been collected on galaxies spanning more than six orders of
magnitude in stellar mass and over thirteen billion years in the age of the
Universe.  These observations show that the galaxies that we can see have
undergone radical changes in size, appearance, and content over the last
thirteen billion years \citep{2004ApJS..152..163R,
  2004MNRAS.351.1151B,2005ApJ...631..101P,2011ARA&A..49..525S}.
Complementary observations have provided a rich data-set on the kinematics
and elemental abundances of stars in our own Milky Way, including large
numbers of metal-poor stars in the halo of our own galaxy and in local
dwarf galaxies.  In principle, this `galactic fossil record' can probe the
entire merger and star formation history of the Milky Way and its
satellites, and complement direct observations at higher redshifts.

The quantity and quality of observational data on galaxy formation, which
is already staggering, is going to increase exponentially over the next
decade.  Surveys such as LAMOST \citep{2009AAS...21341614N}, SkyMapper
\citep{2007PASA...24....1K}, Gaia \citep{2001A&A...369..339P}, and,
ultimately, the Large Synoptic Survey Telescope \citep{2009arXiv0912.0201L}
will produce petabytes of data on billions of individual objects, both
galactic and extra galactic, that will strongly inform our understanding of
galaxy behavior.

Despite this wealth of observational information, we currently lack the
detailed and self-consistent theoretical models necessary to adequately
interpret such observational data sets.  Purely analytic (\emph{i.e.},
``pencil-and-paper'') theoretical models are insufficient to address the
questions that are currently being asked about galaxy formation, due in no
small part to the range of physical components that must be simultaneously
modeled (e.g., gravity, dark matter, gas dynamics, radiative cooling, star
formation and feedback), and the complex and non-linear coupling of these
components.  As a result of these complications, two separate theoretical
methods are commonly used to study galaxy formation: multiphysics
hydrodynamical simulations and semi-analytic models.

Multi-physics numerical simulations are typically used to model galaxy
formation by implementing relevant  physical processes in as realistic
a  manner  as  is  technically and  computationally  feasible.   These
calculations  are typically  based  on $N$-body  dark matter  dynamics
simulations  of  cosmological  structure  formation, and  include  gas
dynamics, the radiative cooling and heating of gas, subgrid models for
star formation and feedback, and possibly more complex physics such as
magneto-hydrodynamics, radiation transport,  and the formation of, and
feedback from, super massive black holes.  Commonly-used codes of this
type   include  Enzo  \citep{2004astro.ph..3044O,2007arXiv0705.1556N},
Gadget              \citep{2005MNRAS.364.1105S},              Gasoline
\citep{2004NewA....9..137W},  RAMSES  \citep{2002A&A...385..337T}, and
more recently AREPO  \citep{2010MNRAS.401..791S}.  These codes produce
broadly similar results, although some important differences remain to
be     resolved    \citep{2005ApJS..160....1O,    2007MNRAS.380..963A,
  2008MNRAS.390.1267T,  2011arXiv1109.3468S}.  The  main  advantage of
such  calculations is that  they attempt  to faithfully  reproduce the
relevant physical  processes in as  accurate of a manner  as possible,
and by virtue of their construction automatically include any complex,
non-linear interaction between important physical processes.  For
  example, this  approach is able  to naturally handle  the non-linear
  hydrodynamics of gas ejecta (once initialized) due to bursts of star
  formation, as does the return  of such gas into later generations of
  structure  formation.   The  main  disadvantage  of  this  sort  of
simulation  lies in  its  cost: current-generation  calculations of  a
single Milky Way-like galaxy performed at high ($\sim 100$ pc) spatial
resolution  \citep[e.g.][]{2011MNRAS.410.1391A}   can  easily  consume
hundreds of  thousands of  CPU hours and  months of time  to complete,
making  it challenging to  model statistically-significant  numbers of
galaxies  or to  perform  a  meaningful study  of  variations in  free
parameters within the models.

A  second  approach  is  often  referred to  as  ``semi-analytic''  or
``phenomenological'' modeling of galaxy formation.  This type of model
typically is based upon  either the extended Press-Schechter formalism
or $N$-body  cosmological simulations, which  provide the evolutionary
histories  for  a  population  of galaxies.   Prescriptions  are  then
applied  on  top  of  these  evolutionary histories  to  describe  the
behavior  of the  gas and  stellar populations  contained  within, and
surrounding,  the  dark matter  halos  that  drive  dynamics on  large
scales,  as well  as  the observational  properties  of the  resulting
galaxies.  These models are then  calibrated by comparison to some set
of observations.  Some examples of  this sort of model include GALFORM
\citep{2003ApJ...599...38B, 2006MNRAS.370..645B, 2010MNRAS.405.1573B},
Galacticus       \citep{2012NewA...17..175B},       and      ChemTreeN
\citep[][]{tumlinson1,tumlinson2}.   Two important  strengths  of this
type  of model  are flexibility  and speed:  one can  easily implement
variations on  a model  (gas ejection from  galaxies as a  function of
halo mass  and redshift versus a  constant value) and  then see within
minutes  how   this  affects  the  modeled   population  of  galaxies.
Alternately, one  can rapidly sweep through large  swaths of parameter
space,  finding the  best-fit model  to a  given set  of observational
constraints.    The  disadvantages  of  this  modeling  technique
  include the  extent to which the observable  properties of simulated
  galaxies depend  on the models of specific  physical phenomena, such
  as  the behavior  of galaxies  during mergers,  as well  as  the large
  number of  free parameters.   Note that the  latter also  pertains to
  hydrodynamical simulations, although to a lesser extent.  Even with
these substantial downsides,  however, semi-analytic models are useful
for exploring  the consequences of  various physical phenomena  on the
observable properties of galaxies.

One   consequence  of  the   large  number   of  free   parameters  in
semi-analytic models is that there is rarely a single ``best fit'' set
of parameters  to a  given set of  observations.  Rather,  given N
  free parameters  that define  an $N$-dimensional space  of potential
  values, it is likely that  there is an $M$-dimensional surface, with
  $M  < N$,  of  approximately  equally good  statistical  fit to  the
  observations in question, or even possibly several discrete regions
within the  parameter space  that provide comparably  good statistical
fits  to the  chosen observations.   When multiple  observational data
sets  are  used,  or  observations  of different  populations,  it  is
possible  that a  given model  parameter may  only be  relevant  for a
subset of the observations  (and, thus, the other observations provide
no significant  constraints upon this parameter).   This challenge has
only recently begun to be  explored in the context of galaxy formation
by     \citet{2009MNRAS.396..535H}, \citet{2010MNRAS.407.2017B}     and
\citet{2011MNRAS.416.1949L,2012MNRAS.421.1779L}.

It is also desirable to attempt to model the formation of the Milky Way as
an exemplar of galaxy formation in general.  The primary reason for doing
so is that, by virtue of the Earth being embedded within the Milky Way, we
possess a massive amount of data on this particular object \citep[e.g.][and
many others]{2001ApJ...547..792D, 2003ApJ...588..824Y,2007Natur.450.1020C,
  2008ApJ...684..287I, 2008ApJ...673..864J, 2008ApJ...689.1044G,
  2010ApJ...716....1B}.  One important drawback of modeling a single
object, however, is that the most fundamental part of either type of galaxy
formation model described above -- the gravitationally-driven merger of
successive generations of dark matter halos -- is unique to a given halo,
and may profoundly affect many observational quantities.  As a result,
there may be critical degeneracies between the merger history of a single
galaxy and the model parameters.  This is not an issue when studying large
numbers of galaxies, as is done by \citet[][hereafter
B10]{2010MNRAS.407.2017B}, since the varieties of merger histories are
included in the overall population.  However, it presents substantial
complications when trying to duplicate observations of the Milky Way using
either hydrodynamic or semi-analytic models.

In this paper, we combine semi-analytic models of the formation of the
Milky Way (including several different $N$-body simulation-based merger
histories) with modern statistical techniques to explore how one might
meaningfully constrain the formation of the Milky Way's stellar halo and
population of satellite galaxies both from a theoretical standpoint and in
terms of guiding future observations.  This paper is arranged as follows:
in \secref{sec:methods}, we describe our methodology, including the
cosmological simulations, semi-analytic models, and in \secref{sec:mod_emu}
the statistical tools.  In \secref{sec:general_results}, we examine how a
small set of model parameters affects observational quantities of our own
Milky Way in a qualitative way, and do the same in a statistical sense in
\secref{sec:statistical_analysis}.  We discuss the results and limitations
of this work in \secref{sec:conclusions}.

\section{Numerical methods}
\label{sec:methods}

In this Section, we briefly describe the $N$-body simulations analyzed in
this work and provide a brief summary of the main characteristic of the
semi-analytical model, ChemTreeN. For a detail description of our numerical
methods, we refer the reader to \citet[][hereafter T10]{tumlinson2}.

\subsection{$N$-body simulations}

Four different simulations of the formation of Milky Way-like dark matter
(DM) haloes are analyzed in this work.  The simulations were run using
Gadget-2 \citep{springel2005} in a local computer cluster.  Milky Way-like
haloes were first identified in a cosmological simulation with a particle
resolution of $128^{3}$ within a periodic box of side 7.32 $h^{-1}$ Mpc.  A
WMAP3 cosmology \citep{wmap} was adopted, with matter density $\Omega_{m} =
0.238$, baryon density $\Omega_{b}= 0.0416$, vacuum energy density
$\Omega_{\Lambda}= 0.762$, power spectrum normalization $\sigma_{8} =
0.761$, power spectrum slope $n_{s} = 0.958$, and Hubble constant $H_{0} =
73.2$ km s$^{-1}$ Mpc$^{-1}$.  The candidates were selected to have
gravitationally-bound dark matter halos with virial masses of M$_{200}
\approx 1.5 \times 10^{12}$~M$_{\odot}$ at $z=0$ and no major mergers since
$z = 1.5$ - 2.  These Milky Way-like dark matter haloes were subsequently
re-simulated at a resolution of $512^{3}$ by applying multi-mass particle
'zoom-in' technique.  At this resolution, each DM particle has a mass of
$M_{p} = 2.64 \times 10^{5}$ M$_{\odot}$.  Snapshots were generated at
intervals of 20 Myr before $z=4$ and 75 Myr intervals form $z = 4$ to
$z=0$.  A six dimensional friends-of-friends \citep{fof} algorithm was
applied to identify dark matter halos in each snapshot.  The gravitational
softening length was 100 comoving pc in all simulations.  The main
properties of the resulting dark matter halos are listed in
\tabref{table:sims}.

\begin{deluxetable}{lcccc}
  \tabletypesize{\footnotesize} \tablecaption{Main properties at $z=0$ of
    the four Dark Matter halos analyzed in this work.\label{table:sims}}
  \tablewidth{200pt} \tablehead{\colhead{Name}&
    \colhead{$R_{200}$\tablenotemark{a}} &
    \colhead{$M_{200}$\tablenotemark{b}} & \colhead{$c$} & \colhead{$z_{\rm
        LMM}$}} \startdata
  \MW1 & 381 & 1.63 & 12.2 & 2.1 \\
  \MW2 & 378 & 1.59 & 9.2 & 3.5 \\
  \MW3 & 347 & 1.23 & 15.5 & 2.0 \\
  \MW4\tablenotemark{c} & 366 & 1.44 & 13.6 & 3.0
\enddata
\tablenotetext{a}{Distances are listed in kpc}
\tablenotetext{b}{Masses are listed in $10^{12}~M_{\odot}$}
\tablenotetext{c}{\MW4 corresponds to the simulation MW6 presented in T10}
\end{deluxetable}

\subsection{Galactic chemical evolution model}

Dark matter only simulations are a useful tool for self-consistently
characterizing the mass assembly history of dark matter halos in a CDM
cosmology.  Yet they don't provide any information about the stellar
populations of a galaxy such as our own Milky Way. For this purpose, we
have coupled to our $N$-body simulations the semi-analytical model,
ChemTreeN (T10).  A semi-analytical model consists of a set of analytic
prescriptions that describe the underlying physical mechanisms driving the
evolution of the baryons.  Processes such as star formation, stellar winds
or chemical enrichment are introduced in the model through differential
equations that are controlled via a set of adjustable input parameters.
These parameters are commonly set to simultaneously match different
observable quantities, such as the galaxy luminosity functions or a variety
of scaling relations.  The general approach used in our models is to assume
that each dark matter halo found in the simulations, and followed through
the merger tree, possesses gas that has been accreted from the
intergalactic medium (IGM), that this gas forms stars, that these stars
return metals and energy to the host halo and to the larger environment,
and that future generations of stars form with the now metal-enriched gas.
\tabref{table:param} summarizes the numerical values of the parameters used
for our fiducial models.  The set of parameters that are considered in the
model emulator analysis, as well as the range of values over which they are
allow to vary, are also listed in this table. In what follows we briefly
describe the implementation of the prescriptions that are most relevant to
this work.

\subsubsection{Parcels and Particle Tagging}

\begin{deluxetable*}{@{}ccccc}
  \tabletypesize{\footnotesize}
  \tablecaption{Model Parameters.\label{table:param}}
  \tablewidth{360pt}  
  \tablehead{\colhead{Parameter} &
  \colhead{Fiducial Value} &
  \colhead{Range} &
  \colhead{Description} &
  \colhead{Explored}}
\startdata
  $z_{\rm r}$ & 10 & 5 -- 19 & Epoch of re-ionization & Yes \\
  $f_{\rm bary}$ & 0.05 & 0 -- 0.2 & Baryonic mass fraction & Yes \\
  $f_{\rm esc}$ & 50 & 0 -- 110 & Escape factor of metals & Yes \\
  $\epsilon_{*}$ & $1 \times 10^{-10}$ & 0.2 -- 1.8 
     & Star formation efficiency  ($10^{-10}$ yr$^{-1}$) & Yes\\
  $m^{\rm II}_{\rm Fe}$ & 0.07 & 0.04 -- 0.2 &  SN II iron yield ($M_{\odot}$) & Yes\\
 $f_{Ia}$ & 0.015 & $\cdots$ & SN Ia probability & No \\
 $\epsilon_{\rm SN}$ & 0.0015 & $\cdots$ &SNe energy coupling & No \\
 $m^{\rm Ia}_{\rm Fe}$ & 0.5 & $\cdots$ & SN Ia iron yield ($M_{\odot}$) & No
\enddata
\end{deluxetable*} 

ChemTreeN follows  the evolution of stellar mass,  metal abundance and
gas  budgets of  galaxies  as a  function  of cosmic  time.  For  each
endpoint or  ``root halo''  in which the  star formation  and chemical
enrichment  history  will  be  determined,  the  code  identifies  its
highest-redshift  progenitor and starts  the calculation  there.  Note
that, due to tidal disruption, the endpoint of some halos can be found
at  $z  > 0$.   The  star formation  history  is  calculated using  10
timesteps between  each redshift snapshot,  so timesteps are  2-8 Myr.
At each timestep a star  formation ``parcel'' is created with a single
initial mass,  metallically, and IMF.  The metallicity  for the parcel
is  derived  from  the  present  gas metallicity.   Each  parcel  thus
represents a single-age stellar  population with a unique metallicity.
When  halos  merge, their  lists  of  parcels  are concatenated.   The
calculation concludes  when the ``root halo'' or  endpoint is reached.
Note that  this process is performed independently  for each satellite
galaxy or  building block that gave  rise to the  final Milky Way-like
halo.  To explore the  spatial, kinematic, and dynamical properties of
stellar populations in  the resulting halos it is  necessary to assign
``stars'' to  dark matter particles in the  $N$-body simulation.  This
is performed  as follows.  At each  snapshot output a  fraction of the
most bound  particles in each  halo are selected.  The  star formation
parcels that occurred between this snapshot and the previous snapshots
are then identified, and an equal fraction of star-forming parcels are
assigned to each  of the selected particles.  Note  that only the 10\%
most gravitationally  bound particles in each halo  are considered, in
order to approximate the effect of stars forming deep in the potential
well  of  the  galaxy,  where  dense  gas  would  be  most  likely  to
collect.  As shown by  \citet{cooper}, to reproduce  the observed
  relationship between  size and luminosity  of dwarf galaxies  in the
  Local Group, a  fraction of 1 - 3\% is required.  This choice is not
  feasible   here    due   to   the   limited    resolution   of   our
  $N$-simulations. However, gross properties of dwarf galaxies such as
  total  luminosity  are  not  strongly  affected  by  our  particular
  choice. Furthermore,  rather than comparing  with real observational
  data,  in  this  work  we  focus on  mock  observational  data  sets
  generated by the model itself.

\subsubsection{Baryon Assignment}
The number  of stars that a  galaxy has formed  throughout its history
strongly depends  on the amount of  gas it contained.  It is therefore
important to  define a prescription  to model baryonic  accretion into
dark matter halos.   Our models adopt a prescription  based on that of
\citet{bj05} that  heuristially takes into account the  influence of a
photoionizing  background from  the  aggregate star  formation in  all
galaxies.  This model assigns a fixed mass fraction of baryons,
$f_{\rm  bary}$, to all  DM halos  before re-ionization,  $z_{\rm r}$.
After $z_{\rm r}$, gas accretion and therefore star formation in small
halos are suppressed  below $v_{\rm c} = 30$  km s$^{-1}$.  Between 30
km s$^{-1}$  and 50 km  s$^{-1}$, the assigned baryon  fraction varies
linearly from 0 to $f_{\rm bary}$.  This baryon assignment is intended
to capture  the IGM ``filtering  mass'' \citep{gne} below  which halos
are too  small to retain baryons  that have been heated  to $T \gtrsim
10^{4}$ K by global re-ionization.  

\subsubsection{Star Formation Efficiency}

Stars are  formed in discrete ``parcels'' with  a constant efficiency,
$\epsilon_{*}$,  such  that  the  mass  formed  into  stars  $M_{*}  =
\epsilon_{*} M_{\rm gas}  \Delta t$ in time interval  $\Delta t$.  The
star formation efficiency is  equivalent to a timescale, $\epsilon_{*}
= 1/t_{*}$, on  which baryons are converted into  stars.  The fiducial
choice  for this parameter  is $t_{*}  = 10$  Gyr, or  $\epsilon_{*} =
10^{-10}$ yr$^{-1}$.

\subsubsection{Stellar Initial Mass Function}

An invariant stellar  initial mass function (IMF) at  all times and at
all metallicities  is assumed.  The  invariant IMF adopted is  that of
\citet{kroupa2001}, $dn/dM \propto (m/M_{\odot})^{\alpha}$, with slope
$\alpha =  -2.3$ from  $0.5$ -- $140$  $M_{\odot}$ and slope  $\alpha =
-1.3$ from $0.1$  -- $0.5$ $M_{\odot}$.  The impact  that variations of
the IMF may  have on the stellar populations  of the resulting stellar
halos will be studied in a follow-up work.

\subsubsection{Type Ia SNe}

Type  Ia  SNe  are  assumed  to arise  from  thermonuclear  explosions
triggered  by the collapse  of a  C/O white  dwarf precursor  that has
slowly accreted mass from a  binary companion until it exceeds the 1.4
$M_{\odot}$  Chandrasekhar limit.   For stars  that evolve  into white
dwarfs as  binaries, the SN occurs  after a time  delay from formation
that is roughly equal to  the lifetime of the least massive companion.
In our  models, stars  with initial mass  $M = 1.5-8$  $M_{\odot}$ are
considered eligible to  eventually yield a Type Ia  SN.  When stars in
this  mass range  are formed,  some  fraction of  them, $f_{Ia}$,  are
assigned status  as a Type Ia  and given a binary  companion with mass
obtained from a suitable probability distribution \citep{gregren}. The
chemical  evolution results  are sensitive  to the  SN  Ia probability
normalization, $f_{Ia}$. This parameter is fixed by normalizing to the
observed relative rates of Type II and Type Ia SNe for spiral galaxies
in the local universe \citep{1994ApJS...92..487T}.  This normalization
gives  a ratio  of SN  II to  Ia of  6 to  1. 
\subsubsection{Chemical Yields}

ChemTreeN tracks the time evolution of galaxies' bulk metallicities by
considering Fe  as the proxy reference  element. For Type  Ia SNe with
1.5 -- 8 $M_{\odot}$ the models adopt the W7 yields of \citet{nomo1997}
for Fe, with 0.5 $M_{\odot}$ of Fe  from each Type Ia SN.  Type II SNe
are assumed  to arise from  stars of 10  to 40 $M_{\odot}$,  with mass
yields  provided by  \citet{2009ApJ...690..526T}.  They  represent the
bulk yields  of core-collapse SNe  with uniform explosion energy  $E =
10^{51}$ erg.   These models have $M  = 0.07$ --  $0.15$ $M_{\odot}$ Fe
per event.

\subsubsection{Chemical and Kinematic Feedback}

One possible cause of  the observed luminosity-metallicity ($L$-$Z$)
relation for  Local Group dwarf  galaxies is SN-driven mass  loss from
small  DM halos \citep{2003MNRAS.344.1131D}.   To model  this physical
mechanism, ChemTreeN tracks mass loss  due to SN-driven winds in terms
of the number of SNe per timestep in a way that takes into account the
intrinsic time variability in the  star formation rate and rate of SNe
from a stochastically sampled IMF.  At each timestep, a mass of gas
\begin{equation}
M_{\rm lost} = \epsilon_{\rm SN} \sum_{i} \dfrac{N^{i}_{\rm SN}
E^{i}_{\rm SN}} {2v_{\rm circ}^{2}}
\end{equation}

becomes  unbound and is  removed permanently  from the  gas reservoir.
Here $v_{\rm circ}$ is the maximum circular velocity of the halo,
  $N_{\rm SN}$ is the number of  SNe occurring in a given timestep and
  $E_{\rm SN}$  is the energy released  by those SNe.   The only free
parameter,  $\epsilon_{\rm  SN}$, expresses  the  fraction  of the  SN
energy that is converted to kinetic  energy retained by the wind as it
escapes.  The sum  over index i sums over all  massive stars formed in
past timesteps  that are just  undergoing an explosion in  the current
timestep.  Note that this approach allows for variations in the number
and energy of SNe from timestep to timestep.

The selective  loss of metals that  should arise when  SNe drive their
own  ejecta out  of the  host galaxy  is captured  by the parameter
$f_{\rm  esc}$,  which  expresses the  increased  metallicity of  the
ejected winds with respect to the ambient interstellar medium. At each
timestep, a total mass in iron $M^{\rm Fe}_{\rm lost}$ is removed from
the gas reservoir of the halo:
\begin{equation}
  M^{\rm Fe}_{\rm lost} = f_{\rm  esc} M_{\rm lost} \dfrac{M^{\rm Fe}_{\rm ISM}}{M_{\rm gas}}
\end{equation}

where $M^{\rm Fe}_{\rm  ISM}$ is the total mass of  iron in the ambient
interstellar  medium, $M_{\rm  gas}  \times 10^{\rm  [Fe/H]}$ .   This
prescription ensures  that, on average, the ejected  winds are $f_{\rm
  esc}$  times  more  metal-enriched  than  the  ambient  interstellar
medium.  Alternatively, the fraction of  metal mass lost from the halo
is  $f_{\rm esc}$ times  higher than  the total  fraction of  gas mass
lost.

\subsubsection{Isochrones and Synthetic Stellar Populations}

To compare these model halos to  observational data on the real MW and
its dwarf  satellites, it is  necessary to calculate  the luminosities
and   colors  of  model   stellar  populations   using  pre-calculated
isochrones  and  population  synthesis  models.  Each  star  formation
parcel  possesses  a  metallicity,   age  and  a  total  initial  mass
distributed  according to  the  assumed IMF.   These three  quantities
together uniquely  specify an isochrone  and how it is  populated. The
models  adopt the isochrones  of \citet{girardi2002,  girardi2004} for
the UBVRIJHK and SDSS  {\it ugriz} systems, respectively, as published
on the Padova group website\footnote{\url{http://stev.oapd.inaf.it/}}.
The  lowest available  metallicity  in these  isochrones  is [Fe/H]  =
-2.3. Thus, this  value is used to represent  stellar populations with
lower  metallicities.

\section{Statistical Methods}
\label{sec:mod_emu}

As discussed in the introduction, semi-analytic models such as ChemTreeN
need to be calibrated by comparing their outputs to an observational data
set.  It is important to explore the input parameter space as thoroughly as
possible to identify regions where good fits to the data can be achieved.
The simplest method of exploring the dependence of interesting model
outputs on these input parameters would be to run the model over a grid of
points, densely sampling the parameter space.  Although ChemTreeN can
generate a new model within a few minutes, this technique quickly becomes
prohibitively expensive as the dimensionally of the input parameter space
increases.  As an alternative to this approach we can construct a Gaussian
Process emulator \citep{OHag:2006, Oakl:Ohag:2002, Oakl:Ohag:2004,
  Kenn:OHag:2000}, which acts as a statistical model of our computer model,
ChemTreeN. An emulator is constructed by conditioning (\emph{i.e.},
constraining fits as described below) a prior Gaussian Process on a finite
set of observations of model output, taken at points dispersed throughout
the parameter space.  Once the emulator is trained it can rapidly give
predictions for both model outputs and an attendant measure of uncertainty
about these outputs at any point in the parameter space.  The probability
distribution for the model output at all points in parameter space is a
very useful feature of Gaussian Process emulators -- \emph{simpler
  interpolation schemes, such as interpolating polynomials, produce an
  estimate of the model output at a given location in the parameter space
  with no indication as to how much this value should be trusted}.
Furthermore, numerical implementations of Gaussian Process emulators are
very computationally efficient (producing output in microseconds rather
than minutes), making it feasible to predict vast numbers of model outputs
in a short period of time.  This ability opens many new doors for the
analysis of computer codes which would otherwise require unacceptable
amounts of time \citep[][B10]{Higd:Gatt:etal:2008,
  Baya:Berg:Paul:etal:2007}.

\subsection{Gaussian Process model emulator}
\label{sec:gp}

\begin{figure}
  \centering
  \includegraphics[width=0.45\textwidth, clip=true]{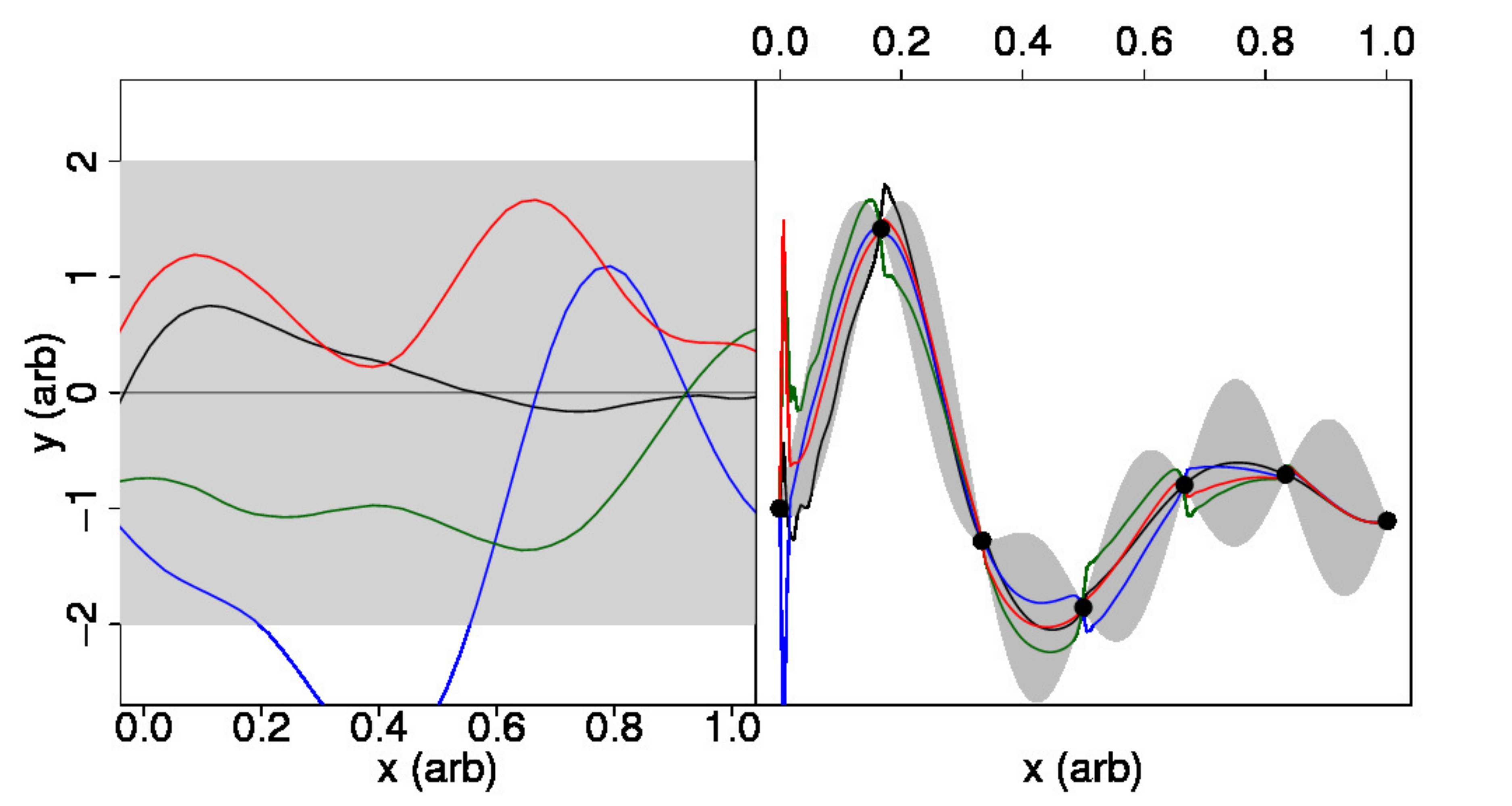}
  \caption{Left  panel: unconditioned  draws from  a  Gaussian Process
    $GP(0, 1)$ with a mean of  zero and constant unit variance. In our
    work, the  indexing variable  $x$ represents an input  parameter of
    the model,  whereas the variable  $y$ a desired  observable. Right
    panel:  draws  from  the  same  process after  conditioning  on  7
    training points (black circles).  The  gray band in both panels is
    a pointwise $95\%$ confidence  interval.  Note how the uncertainty
    in the right panel grows when away from the training points.}
\label{fig-gp-example} 
\end{figure}

We construct an emulator for a model by conditioning a Gaussian Process
prior (see \figref{fig-gp-example}) on the training data
\citep{Chil:Delf:1999, Cres:1993, Rasmussen05}.  A Gaussian Process is a
stochastic process with the property that any finite set of samples drawn
at different points of its domain will have a multivariate-normal (MVN)
distribution.  Samples drawn from a stochastic process will be functions
indexed by a continuous variable (such as a position, time or, in our case,
a parameter of the model) as opposed to a collection of values as generated
by, e.g., a normally-distributed random variable.  A Gaussian Process is
completely specified in terms of a mean and covariance, both of which can
be functions of the indexing variable.  The unconditioned draws (solid
lines) shown in the left panel of \figref{fig-gp-example} are smooth
functions over the domain space labeled $x$.  If enough samples are drawn
from the process the average of the resulting curves at each point would
converge to zero.  A posterior distribution function can be obtain by
conditioning this process on the training points obtained from the model.
This forces samples drawn from the process to always pass through the
training points, as shown in the right hand panel of
\figref{fig-gp-example}.  Repeated draws from the conditioned posterior
distribution would on average follow the underlying curve with some
variation, shown by the gray confidence regions.  These confidence bubbles
grow away from the training points, where the interpolation is least
certain, and contract to zero at the training points where the
interpolation is absolutely certain.  The posterior distribution can be
evaluated to give a mean and variance at any point in the parameter space.
We may interpret the mean of the emulator as the predicted value at a
point, the variance at this point gives an indication of how close the mean
is the true value of the model.  Again we emphasize that \emph{simpler
  interpolation methods, such as interpolating polynomials or splines,
  generally do not provide any measure of the accuracy of the method at a
  given point in parameter space}.

\begin{figure}
  \centering
  \includegraphics[width=0.45\textwidth,
  clip=true]{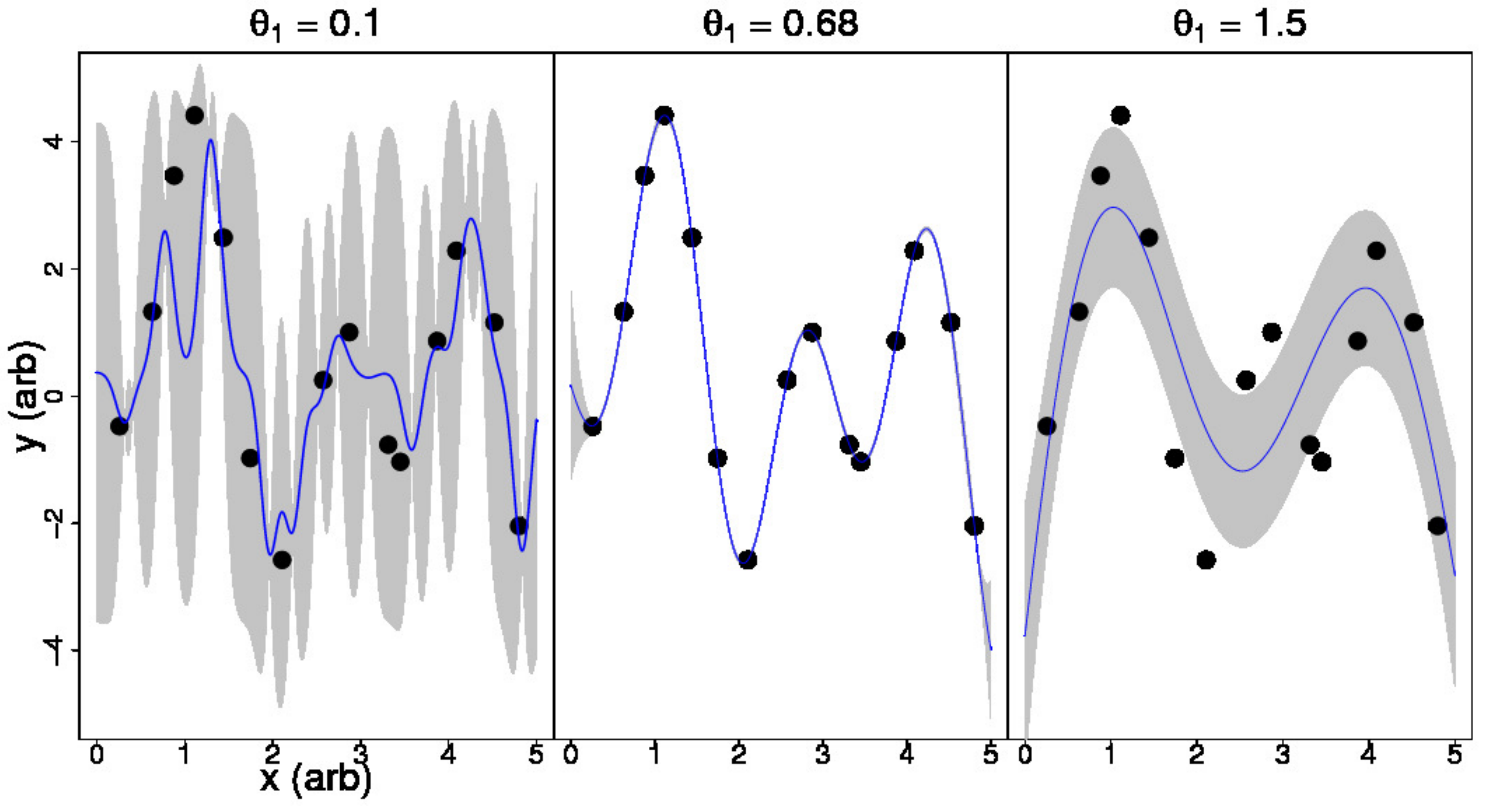}
  \caption{Demonstration of emulator behavior as a function of correlation
    length, $\theta_1$.  In all panels, the solid blue line shows the mean
    of the emulator and the solid gray region is a $95\%$ confidence
    interval around this region.  Left panel: fitting with a value of
    $\theta_1$ that is too small (under-smoothing).  Right panel: shows
    over-smoothing by using a value of $\theta_1$ that is too large.
    Central panel: smoothing with a value of $\theta_1=0.68$ that was
    obtained by a maximum likelihood estimation method.}
  \label{fig-gp-theta}
\end{figure}

To construct an emulator we need to fully specify our Gaussian Process
(GP)  by  choosing  a  prior  mean  and  a  form  for  the  covariance
function. The  model parameter space  is taken to  be $p$-dimensional.
We  model the  prior  mean by  linear  regression with  some basis  of
functions $\mathitbf{h}(x)$,  we use $\mathitbf{h}(x) = \{  1 \}$.  We
specify  a power  exponential form  for the  covariance  function with
power $\alpha \simeq 2$ which ensures smoothness of the GP draws,
 \begin{equation}
   \label{eqn-emu-cov}
   c(\mathitbf{x}_i, \mathitbf{x}_j) = \theta_0 \exp\left(-\frac{1}{2}
     \sum_{k=1}^{p} \left(\frac{x_i^{k} -
       x_j^{k}}{\theta^{k}}\right)^{\alpha}\right) + \delta_{ij} \theta_{N}. 
 \end{equation}
 Here $\theta_0$ is the overall variance, the $\theta^{k}$ set
 characteristic length scales in each dimension in the parameter space and
 $\theta_N$ is a small term, usually called a nugget, added to ensure
 numerical convergence or to model some measurement error in the code
 output. The shape of the covariance function sets how the correlations
 between pairs of outputs vary as the distance between them in the
 parameter space increases.  The scales in the covariance function
 $\theta^{k}$ are estimated from the data using maximum likelihood methods
 \citep{Rasmussen05}, in \figref{fig-gp-theta} we demonstrate their
 influence on an artificial data set. The linear regression model handles
 large scale trends of the model under study, and the Gaussian Process
 covariance structure captures the residual variations.

 Given a set of $n$ design points $\mathcal{D} = \{\vx_1, \ldots, \vx_n\}$
 in a $p$-dimensional parameter space, and the corresponding set of $n$
 training values representing the model output at the design locations
 $\mathitbf{Y} = \{y_1, \ldots, y_n\}$ , the posterior distribution
 defining our emulator is
\[ 
\mathcal{P}(\vx,  \mathitbf{\theta})  \sim \mbox{GP}\left(\hat{m}(\vx,
  \mathitbf{\theta}), \hat{\Sigma}(\vx, \mathitbf{\theta})\right),
\]
for conditional mean $\hat{m}$ and covariance $\hat{\Sigma}$.

\begin{align}
  \label{eqn-emu-mean-var}
  \hat{m}(\vx) &= \mathitbf{h}(\vx)^{T}\hat{\beta} +
  \mathitbf{k}^{T}(\vx) \mathbf{C}^{-1} ( \mathitbf{Y} - {\bf H} \hat{\beta}), \notag \\
  \hat{\Sigma}(\vx_i, \vx_j) &= c(\vx_i, \vx_j)  + \mathitbf{k}^{T}(\vx_i) \mathbf{C}^{-1} \mathitbf{k}(\vx_j) + \Gamma(x_i, x_j), \notag\\
  \mathbf{C}_{ij} &= c(\vx_i, \vx_j) \\
  \Gamma(x_i,   x_j)  &=  \left(   \mathitbf{h}(\mathitbf{x_i})^{T}  -
    \mathitbf{k}^{T}(\mathitbf{x_i})\mathbf{C}^{-1}               {\bf
      H}\right)^{T}
  \left({\bf H}^{T} \mathbf{C}^{-1} {\bf H}\right)^{-1} \notag \\
  &\left(\mathitbf{h}(\mathitbf{x_j})^{T} -
    \mathitbf{k}^{T}(\mathitbf{x_j})\mathbf{C}^{-1}{\bf H} \right), \notag \\
  \mathitbf{k}(\vx)^{T}  &= \left(  c(\vx_1, \vx)  ,  \ldots, c(\vx_n,
    \vx) \right).
\end{align}
Where $\hat{m}(\mathitbf{x})$ is the posterior mean at $\mathitbf{x}$,
$\hat{\Sigma}(\mathitbf{x}_i, \mathitbf{x}_j)$ is the posterior covariance
between points $\mathitbf{x}_i$ and $\mathitbf{x}_j$, $\mathbf{C}$ is the
$n \times n$ covariance matrix of the design $\mathcal{D}$, $\hat{\beta}$
are the maximum-likelihood estimated regression coefficients,
$\mathitbf{h}$ the basis of regression functions and ${\bf H}$ the matrix
of these functions evaluated at the training points.

The elements of the vector $\mathitbf{k}(\vx)$ are the covariance of an
output at $\vx$ and each element of the training set.
It is through this vector $\mathitbf{k}(\vx)$ that the emulator ``feels
out'' how correlated an output at $\vx$ is with the training set and thus
how similar the emulated mean should be to the training values at those
points.  Note that the quantities defined in \eqref{eqn-emu-mean-var}
depend implicitly upon the choice of correlation length scales
$\mathitbf{\theta} = \{ \theta_0, \theta^{k}, \theta_N\}$ which determine
the shape of the covariance function.

The  expression  for  the  mean  in  \eqref{eqn-emu-mean-var}  can  be
decomposed into  a contribution from the prior,  the linear regression
model  $\mathitbf{h}(\mathitbf{x})^{T}\hat{\beta}$  plus a  correction
applied  to  the  residuals  determined by  the  covariance  structure
$\mathitbf{k}^{T}(\mathitbf{x})\mathbf{C}^{-1}( \mathitbf{Y} - {\bf H}
\hat{\beta})$.   Similarly the  covariance  can be  decomposed into  a
contribution    from    the    prior,    the    covariance    function
$c(\mathitbf{x}_i, \mathitbf{x}_j)$ plus  corrections arising from the
prior  covariance structure  and the  covariance of  the  new location
$\vx$  through  $\mathitbf{k}(\vx)$.  These  terms weight  the  points
$\mathitbf{x}_i, \mathitbf{x}_j$  more highly  the closer they  are to
the training  points through $\mathitbf{k}$.  The  $\Gamma$ term gives
the corrections to the covariance arising from the regression model.

A  Latin Hyper  Cube (LHC)  design is  used to  generate  the training
locations  in the  parameter space.  This is  an efficient  design for
space-filling     in     high     dimensional    parameter     spaces.
\citep{Sack:Welc:Mitc:Wynn:1989,  Sant:Will:Notz:2003}.   LHC sampling
divides  the range  of  each dimension  of  the $p$-dimensional  input
parameter space into  $N$ equal length intervals. $N$  points are then
randomly distributed  in the resulting grid,  satisfying the condition
that each row and column are  occupied by only one point for each 2D
projection of the design.

\subsection{Model to data comparison}
\label{sec:mtod}

To compare model output (via the emulator) to experimental data, it is
convenient   to  introduce   the   notion  of   \emph{implausibility}.
Following  B10,  we  define  an  implausibility  measure  as  follows.
Consider a model  with a single output for which  we have generated an
emulator  with  posterior  mean $\hat{m}(\mathitbf{x})$  and  variance
$\hat{\Sigma}(\mathitbf{x},    \mathitbf{x})$.    The   implausibility
$I(\mathitbf{x})$ of the emulated model output at a point $\vx$ in the
parameter space is determined by
\begin{equation}
  \label{eqn-implaus-scalar}
  I^{2}(\mathitbf{x}) 
    = \frac{(\hat{m}(\mathitbf{x}) - E[Y_f])^2}
           {\hat{\Sigma}(\mathitbf{x}, \mathitbf{x}) + V[Y_f]},
         \end{equation}
          where $Y_f$ represents the distribution  of field data that we seek to
           compare our  model against, $E[Y_f]$  the expected value of  $Y_f$ and
           $V[Y_f]$ the observational uncertainties.   Large values of $I(\mathitbf{x_{t}})$
         indicate  that  the  input  parameter vector  $\mathitbf  {x_{t}}$  is
         unlikely to give a good fit to the observable data.  Here we have only
         accounted for  the variation  from the emulator  itself and  the field
         data.  In  the following  work we carry  out comparisons of  the model
         output with idealized field data  generated from the model itself.  As
         we will  show, this can be  a very useful exercise  in understanding a
         model and its dependence upon the underlying parameters.  It is common
         practice \citep [introduced by][] {Kenn:OHag:2001} to introduce a bias
         or  discrepancy   term  representing   the  deviation  of   the  model
         predictions from  observed reality;  this is beyond  the scope  of the
         present analysis,  in which we  compare the model output  at different
         locations in the parameter space to certain selected default values.

The  output   of  ChemTreeN  is  multivariate---   the  code  produces
predictions for many observables, such as the distributions of stellar
populations  in  stellar  halos  of  Milky Way-like  galaxies  or  its
satellite   galaxy  luminosity  function   and  metallicity-luminosity
relation.  It is possible to compare separately each observable with a
model  emulator   generated  from  the   corresponding  model  output.
However,  considerably  more  powerful  conclusions can  be  drawn  by
examining the  joint properties of the observables  and model outputs.
Consider a $t$-dimensional  vector of model outputs $\mathitbf{y(\vx)}
= \{y_1,  \ldots, y_t  \}$ with a  corresponding vector of  field data
$\mathitbf{Y_f}$.  We extend  our training set to be  the $t \times n$
matrix    $\mathitbf{Y}     =    \{    \mathitbf{y(\vx_1)},    \ldots,
\mathitbf{y(\vx_n)} \}$.

We  apply  a  Principal Component  Analysis  \citep[PCA,][]{jollife02}
decomposition  to our  training data  set $\mathitbf{Y}$  to  obtain a
reduced  set of  $r\ll  t$ approximately  independent and  numerically
orthogonal  basis vectors nearly  spanning the  $t$-dimensional output
space,  discarding terms in  the eigen-decomposition  which contribute
less  than $5\%$  of  the total  variation.   We construct  individual
independent  emulators from  the training  values projected  onto each
basis.  When we wish to evaluate the model output at a new location we
invert this transformation to obtain predictive distributions for each
of the $t$  model outputs at a given location  in the parameter space.
This procedure is detailed in \appref{sec:appen-pca-emu}.

The implausibility \eqref{eqn-implaus-scalar}  has a natural extension
to multivariate outputs  (B10, \S3.7).  From the emulator  we obtain a
$t$-dimensional vector of predictions  for the model output with means
$\mathitbf{\hat{m}}(\mathitbf{x})$.    The  emulated   $t   \times  t$
dimensional  covariance matrix $\mathbf{K}(\mathitbf{x})$  between the
model   outputs   at   the   point  $\mathitbf{x}$   in   the   design
space\footnote{Here      $\mathbf{K}_{ij}(\mathitbf{x})     =     {\rm
    Cov}[y_{i}(\mathitbf{x}),y_{j}(\mathitbf{x})]$.                 See
  equation~\ref{eqa7} for more details.}  can also be constructed from
the PCA decomposition.  With these two quantities, we define the joint
implausibility $J(\mathitbf{x})$ for observables $\mathitbf{Y_f}$ with
measurement    variance    $V[\mathitbf{Y_f}]$    and   mean    values
$E[\mathitbf{Y_f}]$:
\begin{multline}
  \label{eqn-implaus-joint}
  J^{2}(\mathitbf{x}) =
    \left(E[\mathitbf{Y_f}]- \mathitbf{\hat{m}}(\mathitbf{x}) \right)^{T} \\
    \left(\mathbf{K}(\mathitbf{x}) + I \cdot V[\mathitbf{Y_f}] \right)^{-1}
    \left(E[\mathitbf{Y_f}]- \mathitbf{\hat{m}}(\mathitbf{x}) \right),
\end{multline}
This covariance-weighted combination of the multiple observables gives a
reasonable indication of which input values $\mathitbf{x}$ are predicted by
the emulator to lead to model predictions close to the observed values
$\mathitbf{Y_f}$.  The squared implausibility score $J^2(\vx)$ can be
thought of as approximately the sum of $r$ squared independent standard
normal random variables, hence with approximately a $\chi^2_r$
distribution, leading in the usual way to confidence intervals for
$J(\vx)$ that induce confidence sets in the input space.  In this work we
consider $75\%$ confidence sets for $\vx$.  In the univariate case, for
example, locations $\vx$ in the parameter space with $J(\vx) < 3$ are
viewed as ``plausible'', or likely to give model outputs closely
reproducing the observational data, given the experimental and
interpolation uncertainties.

The model parameters $\vx$ and output $\mathitbf{Y}$ were scaled and centered
prior to emulator analysis, which improves numerical convergence in the
matrix inversions and maximum likelihood estimation process.

\begin{figure*}
\centering
\includegraphics[width=84mm,clip]{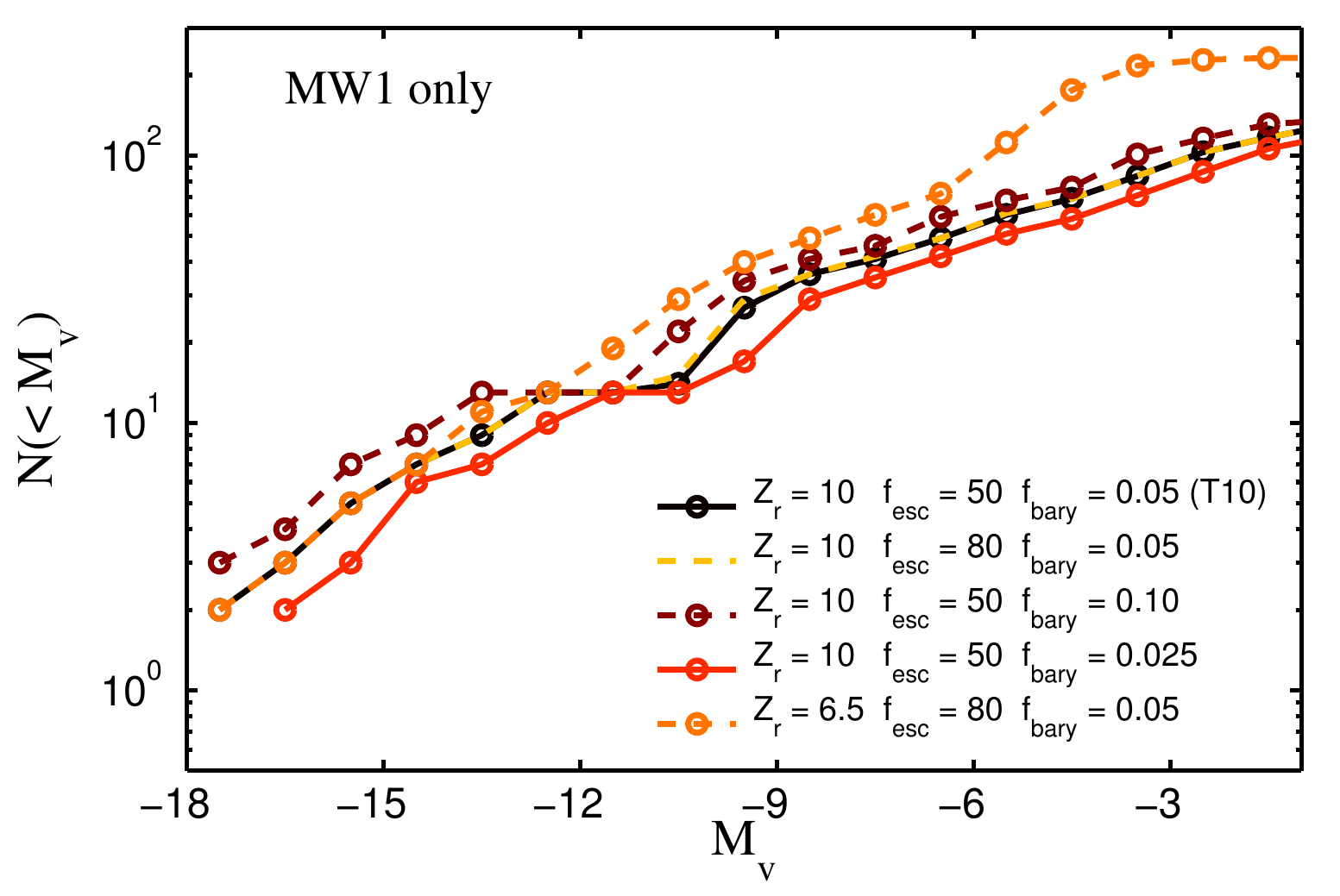}
\includegraphics[width=84mm,clip]{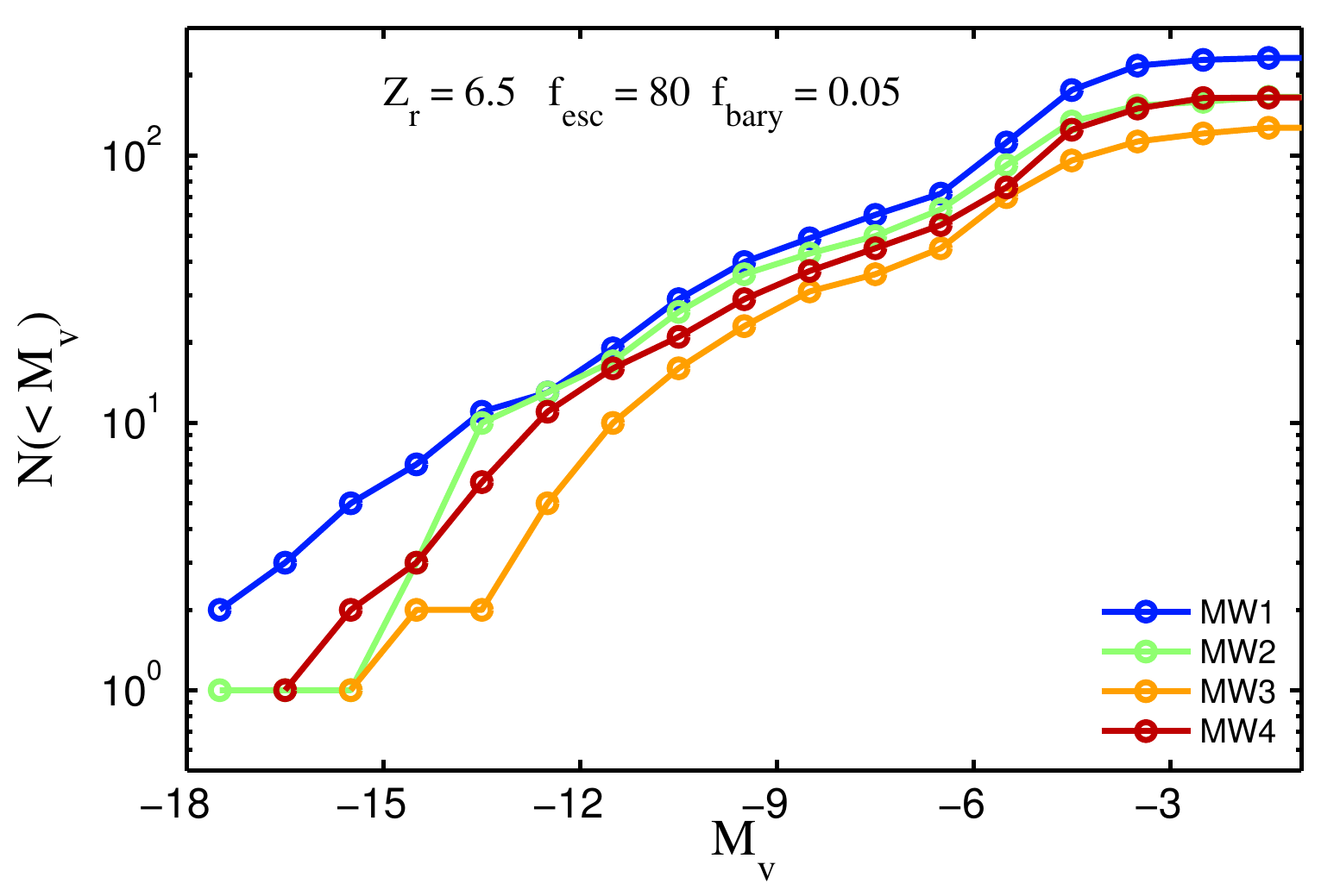}
\\
\includegraphics[width=84mm,clip]{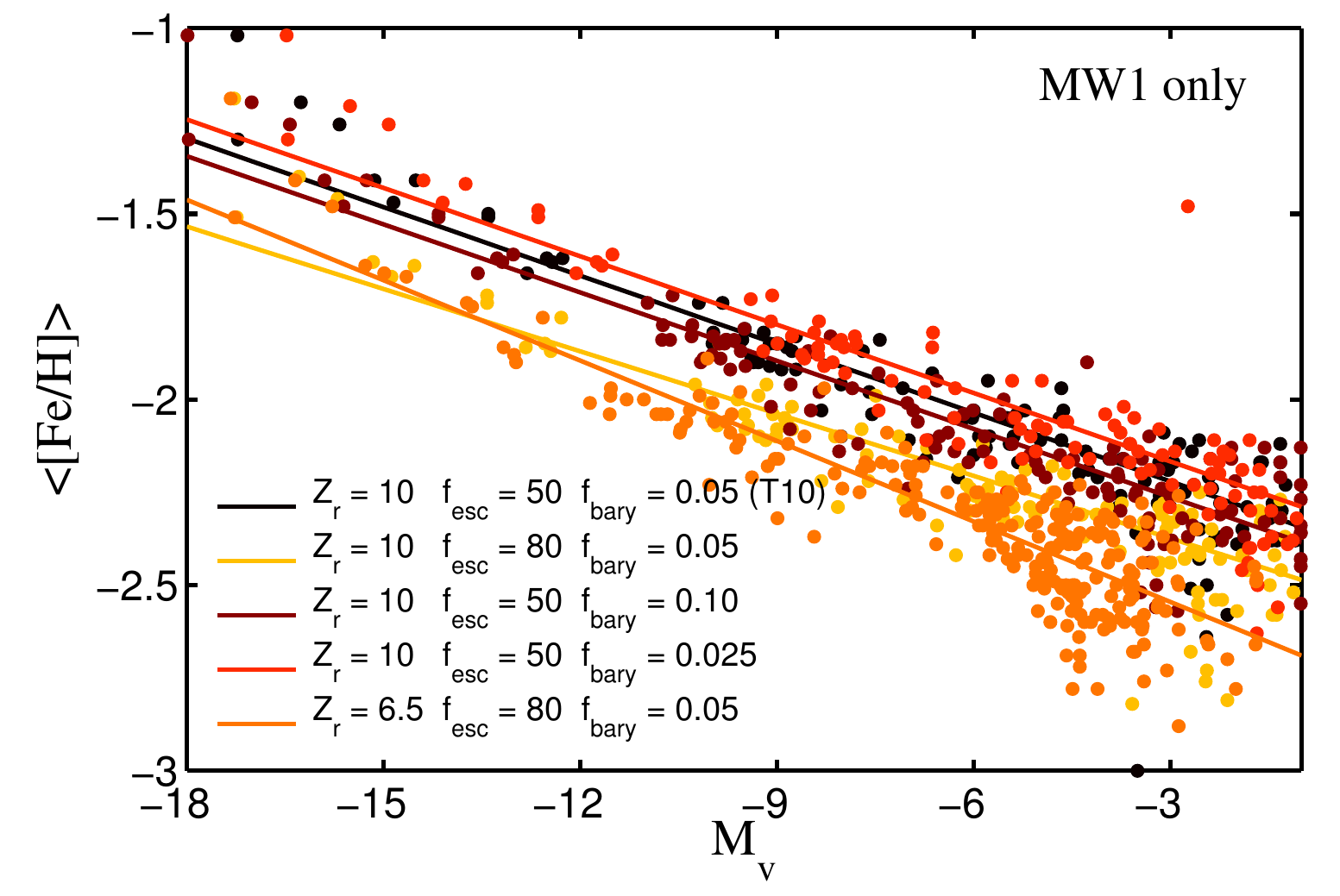}
\includegraphics[width=84mm,clip]{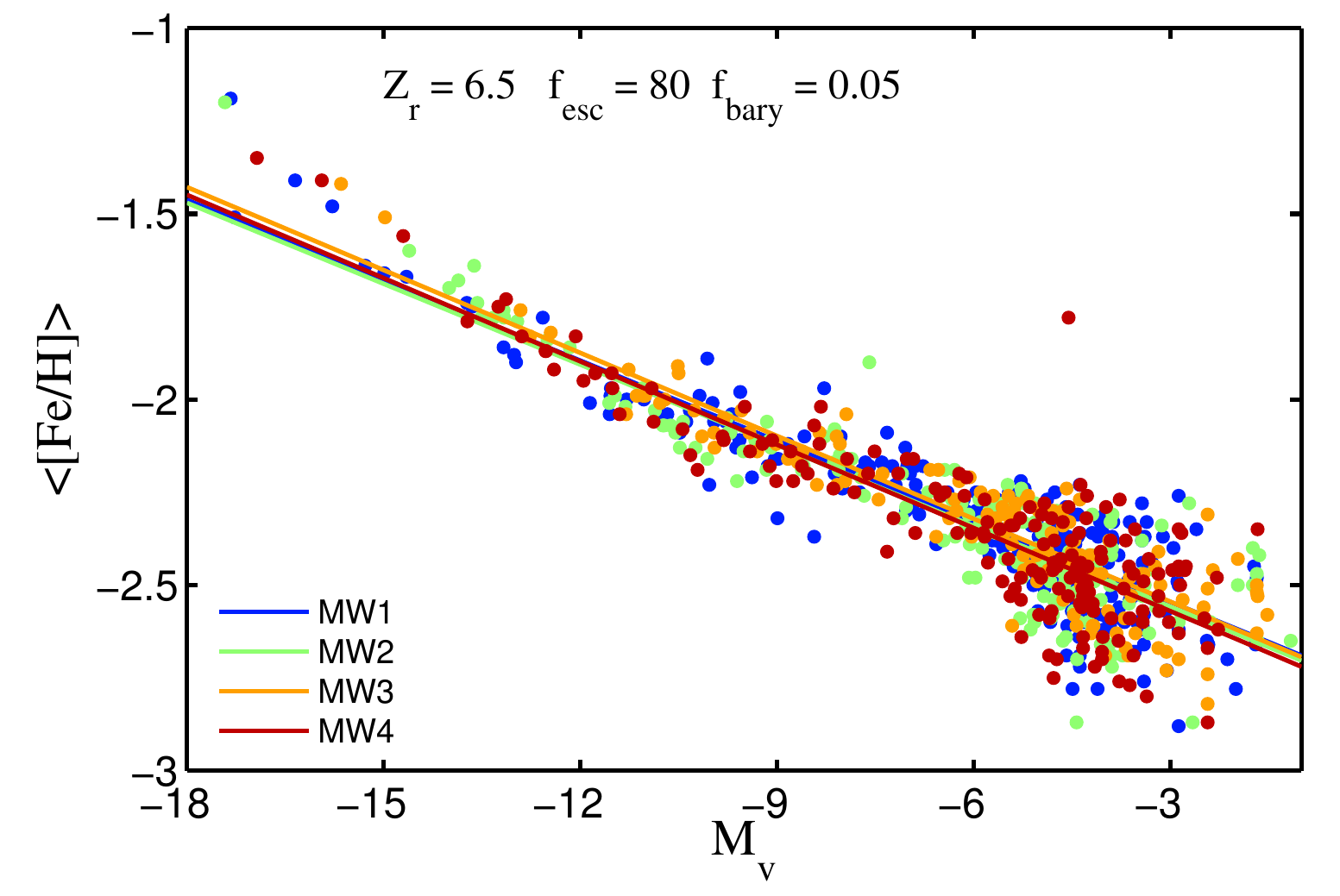}
\\
\includegraphics[width=84mm,clip]{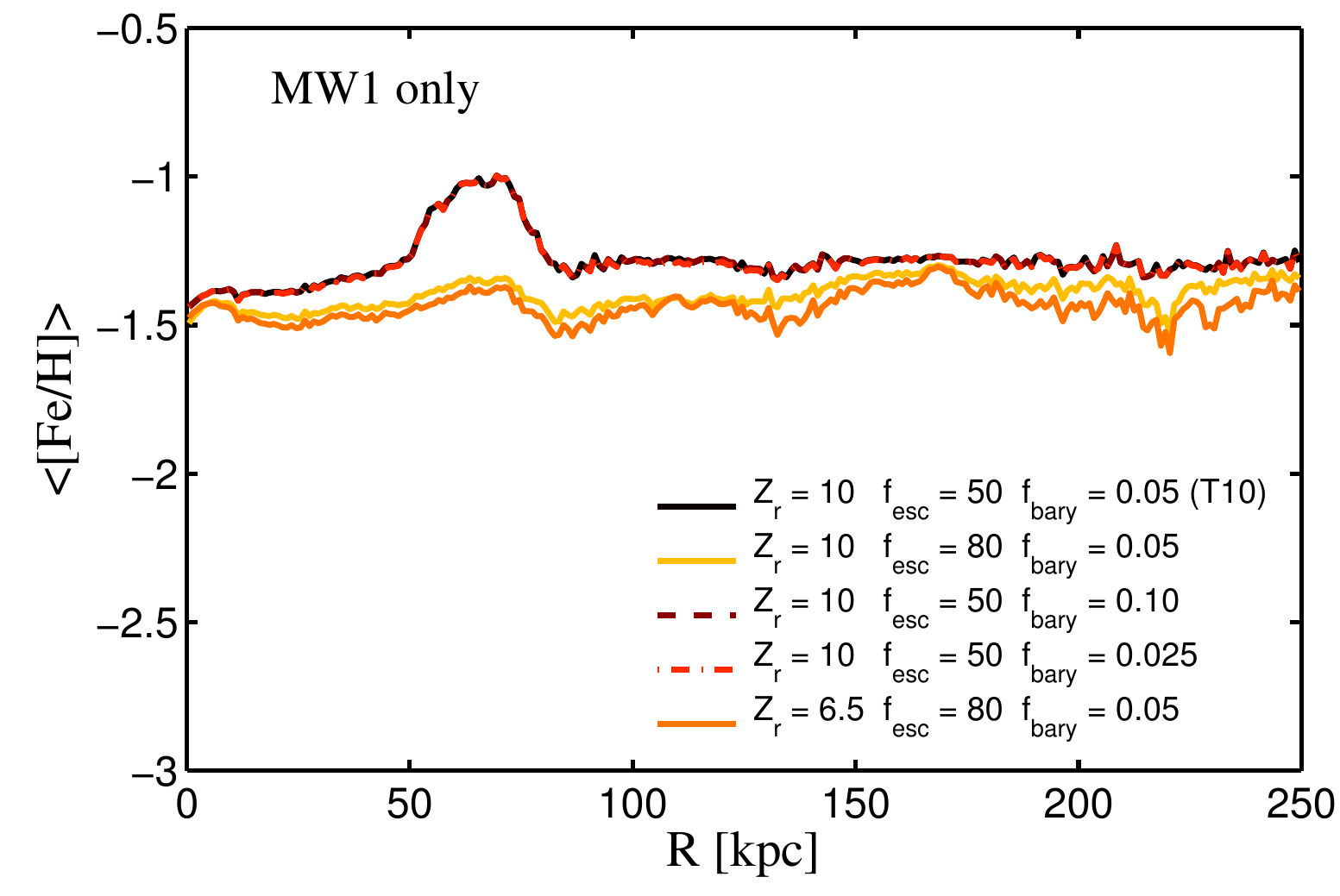}
\includegraphics[width=84mm,clip]{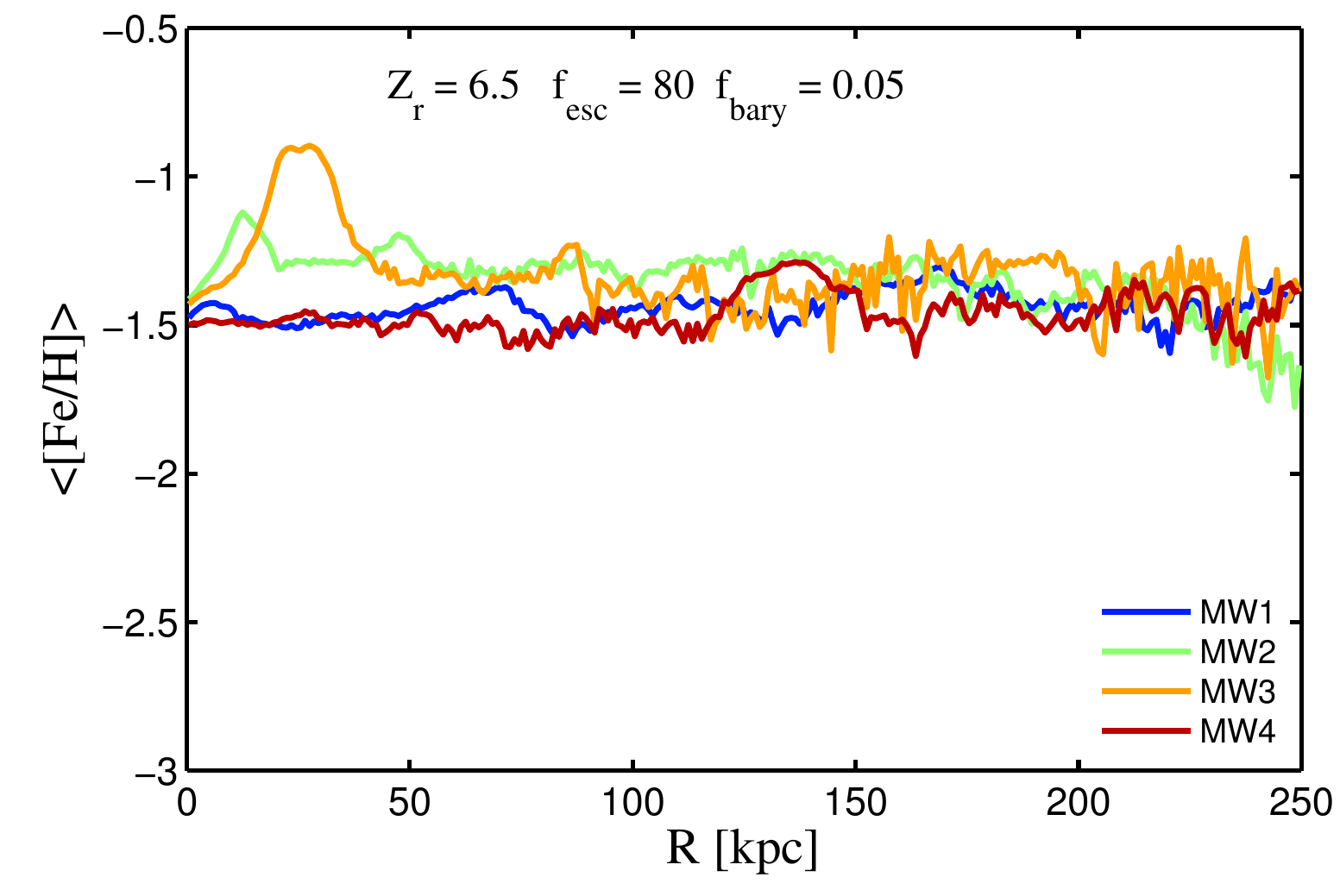}
\caption{Behavior of mock observables as a function of model parameters and
  dark matter halo merger history.  Left column: Dependence of different
  mock observables on the input parameters of the model.  From top to
  bottom panel, we show the satellite galaxy luminosity function, satellite
  luminosity-metallicity relation, and the mass-averaged metallicity
  profile of the primary galaxy's stellar halo for five different models.
  The models were obtained by coupling ChemTreeN with the $N$-body
  simulation \MW1.  The legend on each panel indicates the values of the
  parameters associated with each model.  Right column: Dependence of
  different mock observables on galaxy merger history.  From top to bottom
  panel, we show the satellite galaxy luminosity function, satellite
  luminosity-metallicity relation, and the mass-averaged metallicity
  profile of the primary galax's stellar halo for four different models.
  These models were obtained by coupling ChemTreeN with the four available
  $N$-body simulations.  To generate these models we fixed the values of
  all model parameters. The values chosen for this example are ($z_{\rm
    r},~f_{\rm bary},~f_{\rm esc}) = (6.5,~0.05,~80)$.}
\label{fig:diff_models}
\end{figure*}

\section{Effects of galaxy formation history on bulk
            stellar properties I: Generalities} 
\label{sec:general_results}

In this Section we examine how observable quantities of our final Galactic
models, such as properties of their dwarf galaxy population, depends on the
galaxy's merger history as well as on the model parameters that control
bulk properties of their star formation histories.  We start by
qualitatively comparing the output of models obtained by varying both the
merger histories and the model parameters, and then move on to a more
rigorous statistical analysis in the following Section.

In the left panels  of \figref{fig:diff_models} we explore the effects
that different parameter values have on Milky Way-like models obtained
after  coupling ChemTreeN  with the  merger tree  based  on simulation
\MW1.  The  top panel  shows how the  Luminosity Function (LF)  of the
surviving satellites at $z=0$ depends  on the redshift of the epoch of
re-ionization, $z_{\rm  r}$, the baryon fraction,  $f_{\rm bary}$, and
the  fraction of  metal lost  through  SNe winds,  $f_{\rm esc}$.   As
previously discussed  by T10, for a  given $f_{\rm bary}  = 0.05$, the
number of  bright satellite galaxies (\emph{i.e.}, $M_{\rm  v} < -10$)
does not  strongly depends on $z_{\rm  r}$.  However, we  do observe a
strong dependence  of the number of  satellites on $z_{\rm  r}$ at the
faint end of the LF. Note that the number of faint dwarfs increases by
a factor  of $\approx 2$  - 3 when  shifting $z_{\rm r}$ from  $10$ to
$6.5$.   This  is  not  surprising,  since  by  moving  the  epoch  of
re-ionization to lower redshifts we allow star formation to take place
in  small  halos (\emph{i.e.},  $v_{\rm  c}  <  30$ km  s$^{-1}$)  for
extended periods  of time.   On the other  hand, the number  of bright
satellites  does depend  on the  parameter controlling  the  amount of
available gas to form stars, $f_{\rm bary}$.  This parameter basically
acts  as  a  re-normalization   of  the  satellite  galaxy  luminosity
function, shifting it up or down for higher or lower values of $f_{\rm
  bary}$, respectively.   In our models,  the escape factor  of metals
does not have  a significant impact on the  luminosity function.  This
is because we have assumed a fixed star formation efficiency, and thus
complicated physical  processes that affect a  galaxy's star formation
history, such as metal cooling, are neglected \citep[e.g.][]{choi}.

\begin{figure}
\includegraphics[width=82mm,clip]{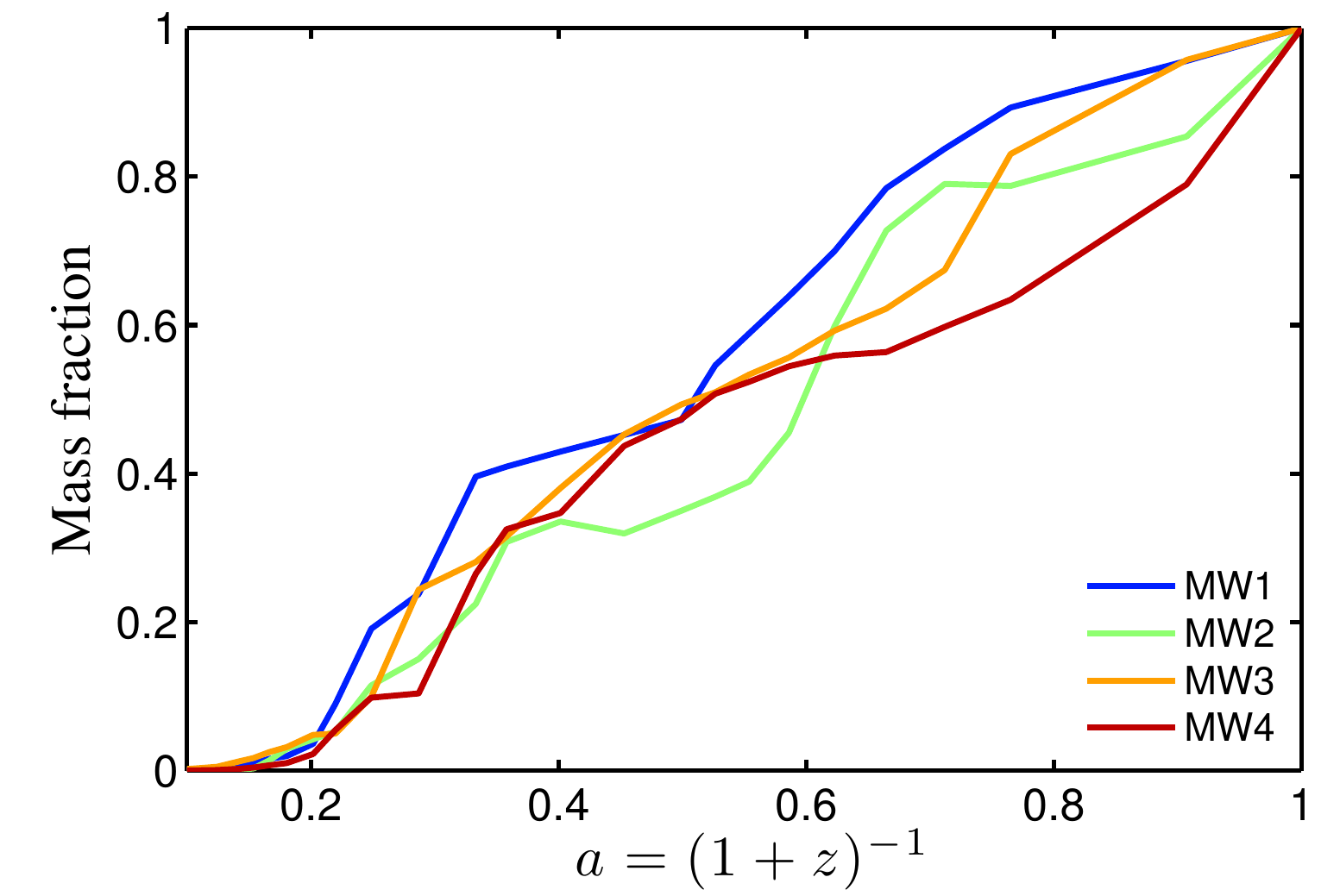}
\caption{Galaxy formation history as shown by the virial mass of the most
  massive progenitor of our Milky Way-like dark matter halos as a function
  of the expansion factor.  In all cases, the mass is normalized to the
  $z=0$ mass of the galaxy.  Notice the range of merger histories spanned
  by these halos.}
\label{fig:mhist}
\end{figure}

It is very interesting to compare these LFs with those obtained from
galaxies that have had different merging histories, after fixing the values
of all model parameters.  \figref{fig:mhist} shows the fraction of mass
within $R_{200}$ at any given redshift of our four Milky Way-like dark
matter halos. Notice the range of merger histories spanned by these halos.
The luminosity functions obtained after coupling each halo merger tree to
ChemTreeN are shown in the top right panel of \figref{fig:diff_models}.
The values chosen for the input parameters in this analysis are $z_{\rm r}
= 6.5$, $f_{\rm bary} = 0.05$ and $f_{\rm esc} = 80$.  The number of bright
dwarf galaxies is very sensitive to the merger history of the parent halo,
though there is a significant halo-to-halo variation (approximately factor
of two) in the overall number of dwarf galaxies between the four
simulations.  It is important to notice that the scatter observed in these
luminosity functions is similar to that previously observed when varying
model parameters.  This suggests a degeneracy between the effects that
different parameters and merger histories have on the observable properties
of our Milky Way-like galaxies.

The situation is different for the Luminosity-Metallicity ($L$-$Z$)
relation of the surviving satellite galaxies.  We derived each satellite's
metallicity by obtaining a model metallicity distribution function (MDF)
from the galaxy's particle-tagged stellar populations and taking its mean
value.  The right middle panel of \figref{fig:diff_models} shows that this
relation is not sensitive to the merger history of a galaxy.  For
comparison, we show as well the linear fits derived from each data set.
This result highlights the importance of scaling relations for this kind of
analysis, as they allow us to put constraints on our models independently
of the particular galaxy's merger history.  In the left middle panel we
explore dependencies of this relation with model parameters.  The parameter
with the strongest impact is the escape factor of metals.  Note that by
decreasing the value of $f_{\rm escp}$ from 80 to 50 we modify not only the
slope of the $L$-$Z$ relation but also its intercept, increasing the mean
metallicity of all satellites.  Instead, by varying the baryon fraction we
re-normalize the $L$-$Z$ relation, keeping its slope constant. Although
the mean metallically of a given satellite does not depends on the value
chosen for $f_{\rm bary}$, its absolute magnitude $M_{\rm v}$ does.  For a
smaller (larger) value of $f_{\rm bary}$ the whole distribution of
satellites is shifted towards fainter (brighter) absolute magnitudes, thus
effectively decreasing (increasing) the average satellite's metallicity a
given at $M_{\rm v}$.  We finally notice that variations of $z_{\rm r}$
mainly induce changes in the slope of this relation.  A later epoch of
re-ionization allows star formation to take place on the smallest galaxies
for longer periods of time.  As a consequence, faint satellites that were
fed by the smallest building blocks have a larger absolute magnitude at
$z=0$.  Note, however, that this mechanism does not strongly affect the
brightest satellites ($M_{\rm v} \lesssim -12$), since accretion of these
smallest building blocks does not significantly modify their overall
luminosity.  This is because they are big enough to accrete gas and keep
forming stars after re-ionization.

\begin{figure*}
\centering
\includegraphics[width=87mm,clip]{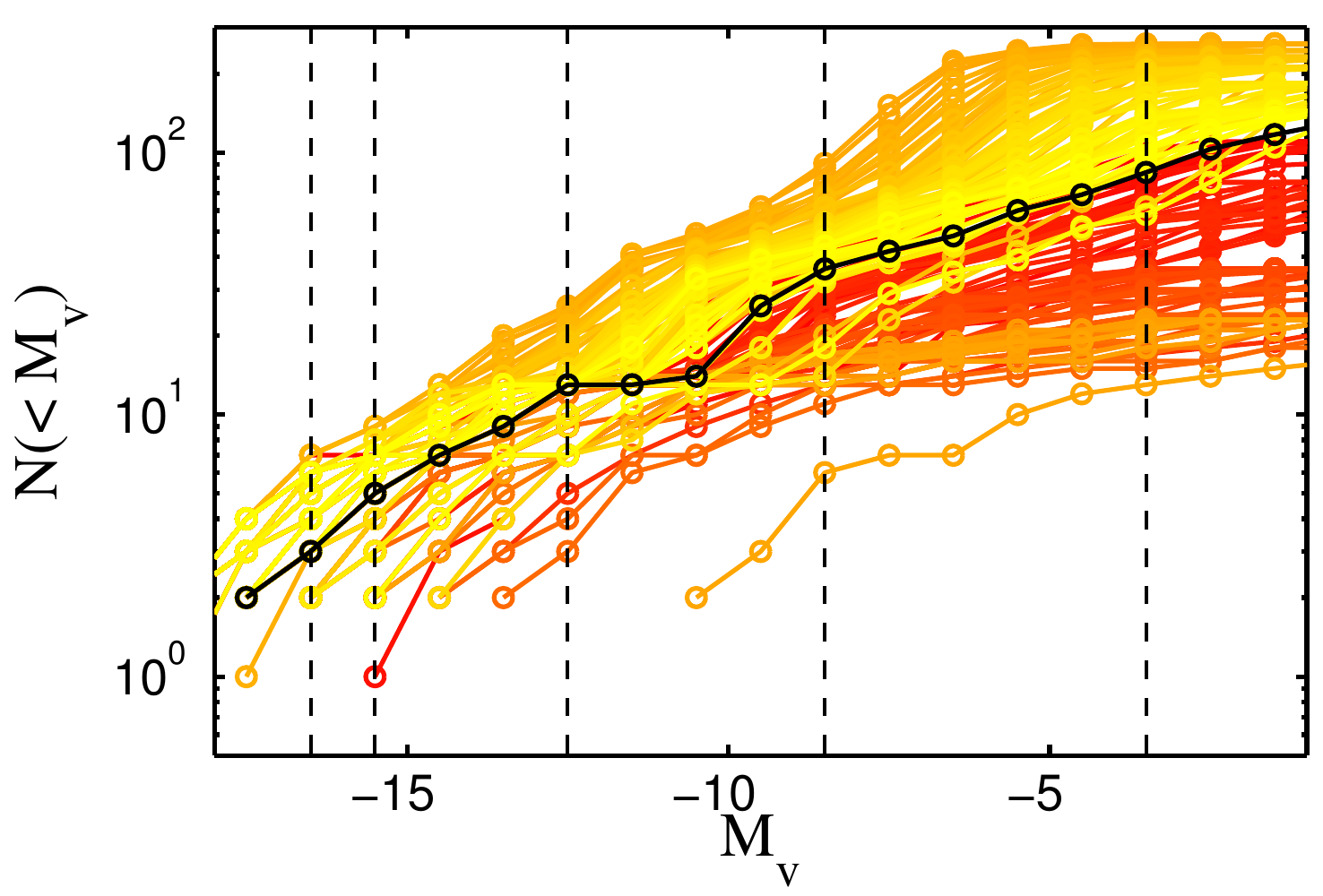}
\includegraphics[width=87mm,clip]{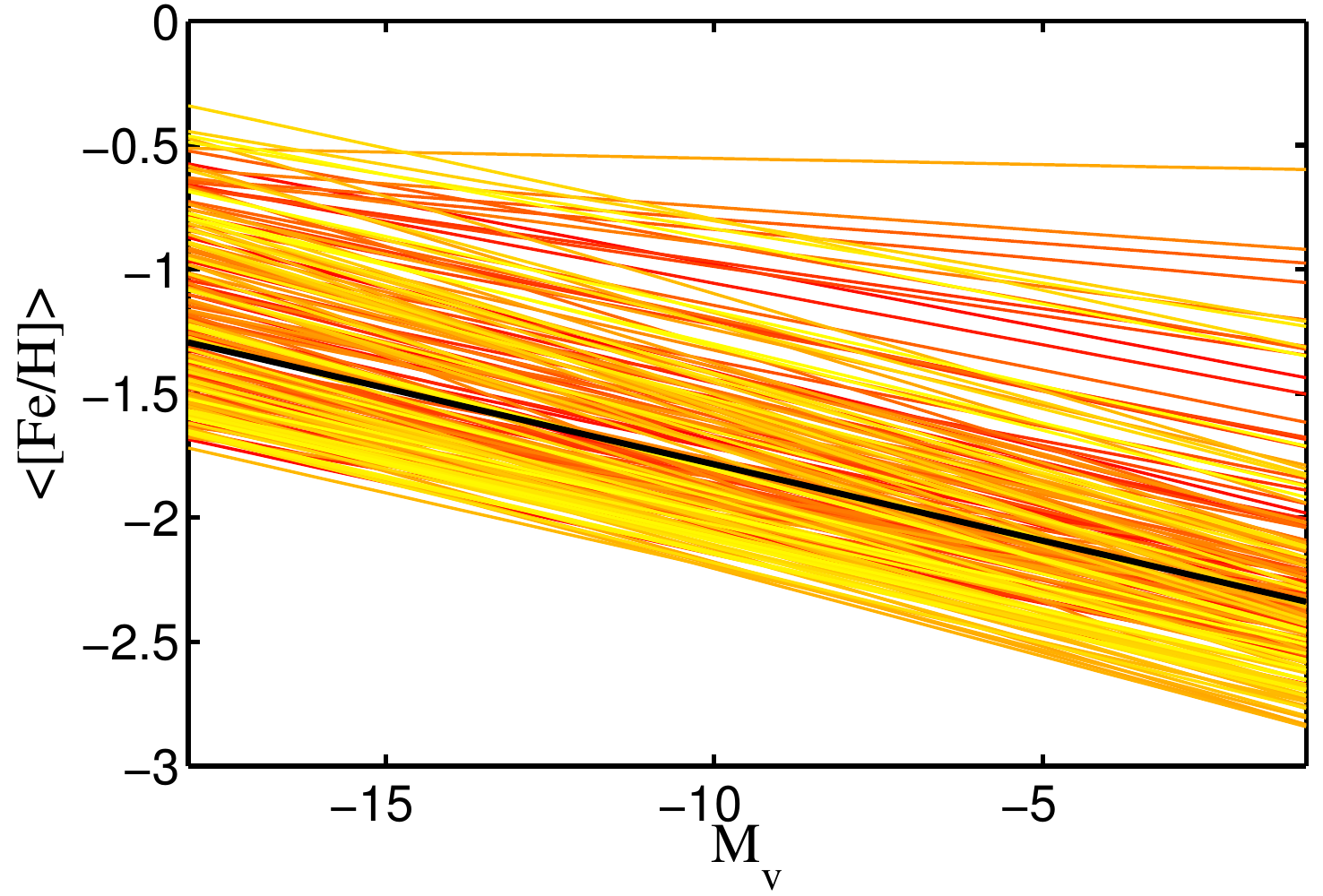}
\caption{The yellow/red solid lines show the satellite galaxy luminosity
  functions (left panel) and satellite galaxy luminosity-metallicity
  relations (right panel) extracted from a set of 200 models used to train
  our suite of model emulators.  For clarity, only the result of a linear
  fit to each luminosity-metallicity relation is shown. Note the great
  diversity of outputs that can be obtained by varying the different input
  parameters. The models were obtained after coupling ChemTreeN with the
  $N$-body simulation \MW1.  The black solid line on both panels indicate
  the model considered to be the galaxy's ``true'' observational
  quantities, obtained after running ChemTreeN with the input parameter
  vector $\mathitbf{x}_{\rm obs}$. The vertical dashed lines on the left
  panel indicate the five values of $M_{\rm v}$ chosen to sample the LFs.
  We sample each $L$-$Z$ relation by extracting slopes and intercepts from
  the corresponding linear fits.  A different model emulator was
  constructed for each of these seven model outputs.}
\label{fig:design}
\end{figure*}

It  is also  interesting  to compare  the  stellar halos'  metallicity
profiles  as  function of  galactocentric  radius  obtained from  both
simulations  with  different merger  histories  and model  parameters.
Notice  that after  the  last major  merger  event, most  of the  star
formation  in  the  host  galaxy  is  expected  to  take  place  in  a
dissipationally  collapsed, baryon-dominated  disc.  The  formation of
this galactic component  cannot be properly accounted for  in our dark
matter-only $N$-body simulation. Therefore, for this analysis, we have
decided to neglect  all star formation episodes that  have taken place
in  the  host  halo  after  the  redshift of  the  last  major  merger
(LMM). The right bottom  panel of \figref{fig:diff_models} shows that,
independently  of  the  merger  history,  our  four  halos  present  a
noticeably flat  spherically-averaged metallicity profile  up to $\sim
150$ kpc, with mean metallicities  between $-1.3$ dex and $-1.6$. dex.
This  is in  agreement  with  previous studies  that  have found  flat
metallicity profiles on halos formed purely through accretion of large
number   of   satellites  \citep[see   e.g.][Monachesi   et  al.,   in
prep.]{lh,cooper}.  Furthermore,  recent observations on  the halos of
M31 \citep[e.g.][]{rich,kali},  M81 (Monachesi et al.,  in prep.)  and
the  Milky  Way  \citep[][in  prep.]{2012AAS...21925214M}  have  shown
indications of  flat or  relatively mild metallicity  profiles outside
$\sim  20-30$  kpc;  radius  at  which accretion  from  satellites  is
expected  to become the  dominant contribution  as opposed  to in-situ
star                                                          formation
\citep{2009ApJ...702.1058Z,2010ApJ...721..738Z,2011MNRAS.416.2802F}.
Note that our  results are not in contradiction  with the dual stellar
halo proposed by  \citet{2007Natur.450.1020C,c10} since, as previously
discussed, in  situ star  formation is not  properly accounted  for in
these simulations.   In addition, \citet{cooper}  showed that stronger
metallicity gradients  can be  found in halos  with a small  number of
relatively massive progenitors.

On the left bottom panel of \figref{fig:diff_models} we show that this
result is robust to variations of the model's parameters.  The parameter
that most strongly affect this profile is $f_{\rm esc}$ since, as discussed
above, this controls the mean metallicity of the host halo's building
blocks.  Nevertheless, variations of this parameter result on final stellar
halos with approximately flat metallicity profiles.  Note that the scatter
induced by varying the model parameters is similar to that obtained by
sampling different merger histories.

\section{Effects of  galaxy formation history on bulk 
            stellar properties II: Statistical analysis}
\label{sec:statistical_analysis}

In the previous Section we have shown qualitatively that certain observable
quantities, commonly used to constrain the input parameter space of our
semi-analytic model, depend strongly on the particular merger history of a
galaxy.  Furthermore, as shown previously by B10 in a different context,
simultaneous variations of certain groups of parameters can yield similar
observables, revealing important connections among the underlying physical
mechanisms.  The present discussion highlights the need for robust
statistical tools capable of exploring the input parameter space densely,
as well as the need for quantitative characterization of the effects that
different merger histories have on the present-day distribution of
observable quantities.  In what follows we will show how model emulators
are well-suited to address this kind of problem.

\subsection{Parameter space exploration}
\label{sec:param_explor}

\begin{figure*}
\includegraphics[width=180mm,clip]{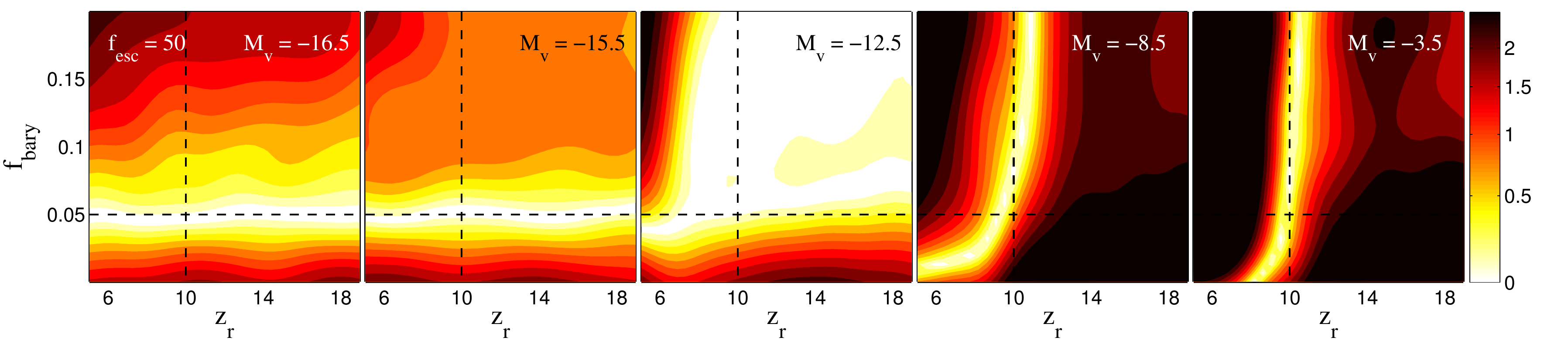}
\\
\includegraphics[width=180mm,clip]{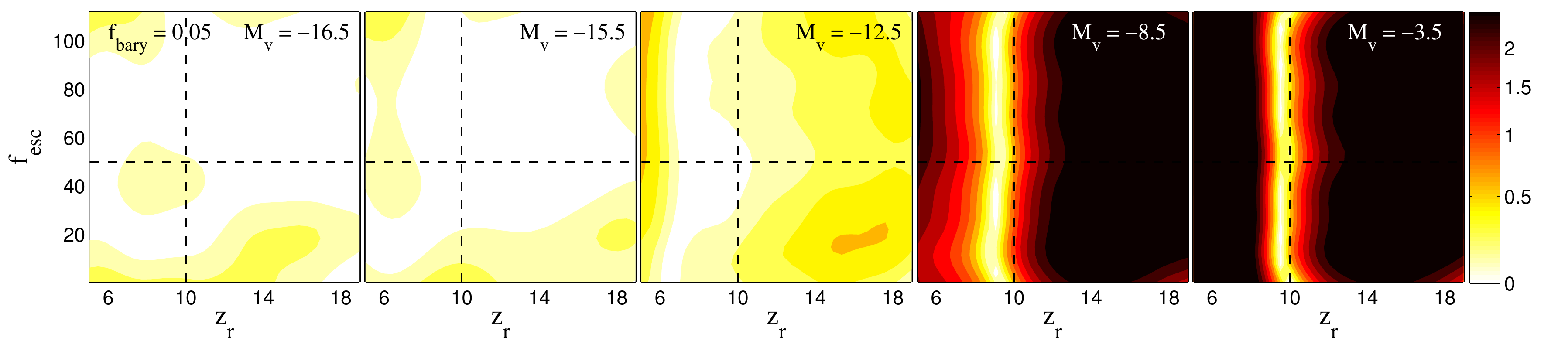}
\\
\includegraphics[width=180mm,clip]{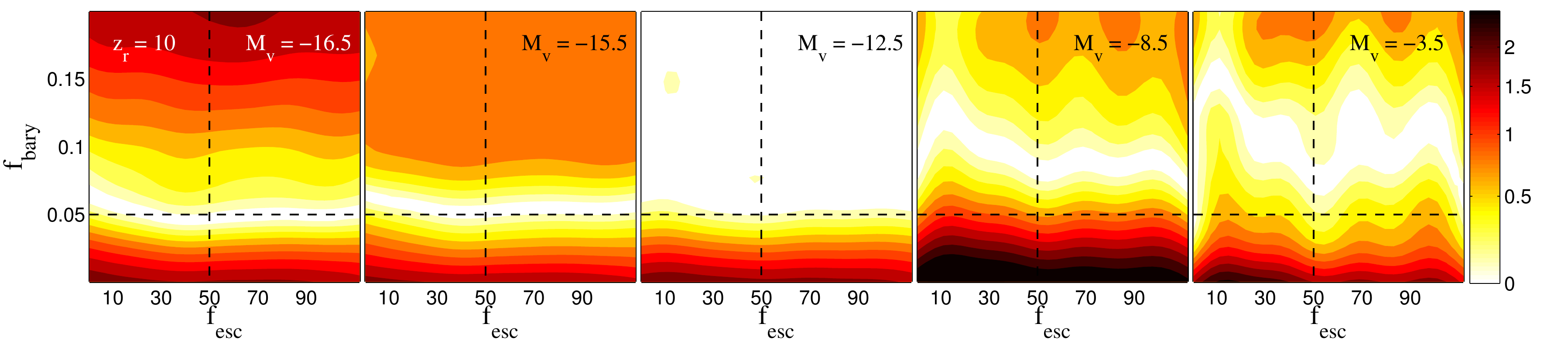}
\caption{Three different sections of each Implausibility surfaces,
  $I(\mathitbf{x})$, obtained from the five model emulators constructed for
  the LF's outputs. The output being emulated is indicated on the top right
  corner of each panel.  Note that columns correspond to different
  observables.  The $3D$ implausibility surfaces are sliced with three
  orthogonal planes as defined by the components of $\mathitbf{x}_{\rm
    obs}$.  The different colors show different values of $I(\mathitbf{x})$
  in logarithmic scale. Low implausibility values are indicated in light
  colours.  Note that, given an input parameter vector $\mathitbf{x_{t}}$,
  the larger the value of the $I(\mathitbf{x_{t}})$, the less likely a good
  fit to the observable data can be obtained.  The top, middle and bottom
  row panels show the $f_{\rm esc} = 50$, $f_{\rm bary} = 0.05$ and $z_{\rm
    r} = 10$ sections of the $I(\mathitbf{x})$ surfaces, respectively.  The
  black dashed lines indicate the values of the remaining two components of
  $\mathitbf{x}_{\rm obs}$.  Model emulators are compared to the mock
  observable data obtained after running ChemTreeN with the input parameter
  vector $\mathitbf{x}_{\rm obs}$. Both mock observables and training data
  set are obtained by coupling ChemTreeN with the $N$-body simulation \MW1.
  From these model emulators it is possible to constrain strongly the
  parameters $f_{\rm bary}$ and $z_{\rm r}$, but not $f_{\rm esc}$.}
\label{fig:implaus}
\end{figure*}

As discussed in \secref{sec:mod_emu}, the first step in constructing a
model emulator is to train a suite of Gaussian process priors using a
finite set of model outputs.  These outputs are obtained by running
ChemTreeN using different sparsely sampled sets of input parameters drawn
from an experimental design $\mathcal{D} = \{\vx_1, \ldots, \vx_n\}$.  For
simplicity, we will first take ${\vx_i}$ to be a three-component vector,
$\vx_i = (z_{\rm r}^{i},~f_{\rm esc}^{i},~f_{\rm bary}^{i})$.  Within this
framework it is trivial to increase the dimensionality of $\vx_i$, but
interpreting and visualizing the final implausibility surfaces become
progressively more challenging.  (For examples of this kind of analysis
with larger dimensionality see \appref{sec:appen-5} or B10.)

The training set consists of a number $n=200$ of design points $\mathcal{D}
= \{\vx_1, \ldots, \vx_n\}\subset\mathbb{R}^3$, drawn using LHC sampling.
This gives an acceptable balance between adequate coverage of the input
space and acceptable run time.  The input parameters are allowed to vary
within the ranges specified in \tabref{table:param}.  The full model
ChemTreeN is then run at each of these $200$ training points.  The next
step is to select a set of outputs, $\mathitbf{Y} = \{y_1, \ldots, y_t\}$,
for which individual emulators will be constructed.  Note that the output
selection is determined purely by the set of $t$ observables or field data,
$\mathitbf{Y_f} = \{y_{f,1}, \ldots , y_{f,t} \}$, chosen.  Motivated by
our discussion in \secref{sec:general_results}, we chose to emulate five
values of the satellite galaxy luminosity function, each at a different
absolute magnitude, in addition to the slope and the intercept of the
satellite galaxy luminosity-metallicity ($L$-$Z$) relation.  This gives us
a total of $t=7$ outputs to be extracted from the models.  As we show
below, each of these outputs most strongly constrains different model
parameters.  In \figref{fig:design} we show the luminosity functions and
$L$-$Z$ relations extracted from the training set models, obtained after
running ChemTreeN on the design points $\mathcal{D}$.  For clarity, only
the result of a linear fit to each $L$-$Z$ relation is shown.  Note the
great diversity of outputs that can be obtained by varying the input
parameter of the model for a fixed merger history.  The black dashed lines
on the left panel indicate the five values of $M_{\rm v}$ chosen to sample
the LFs.

After training seven model emulators (one for each observable), we compare
the model (via the emulators) to the observable data by calculating
individual surfaces of implausibility, level sets of $I(\mathitbf{x})$ for
each observable.  As described in Section~\ref{sec:mtod}, these
three-dimensional surfaces bound the set of input parameter vectors
$\mathitbf{x}$ that are more likely to reproduce the observed data set
$\mathitbf{Y_{f}}$.  In principal, the observable data should be obtained
from the luminosity function and $L$-$Z$ relation of the Milky Way
satellite galaxies.  However, to test the constraining power of this
approach when the ``truth'' is known, a particular run of the ChemTreeN
model will be used as a mock observable data set.  This type of controlled
experiment can be very helpful in model performance assessment, since we
know exactly what values of the input parameters were used to obtain the
artificial ``field data;'' it also obviates the need for modeling
discrepancy.  The black solid lines in \figref{fig:design} show the
luminosity function and $L$-$Z$ relation of the model used as the mock
observations.  The values of the input parameters used are
$\mathitbf{x}_{\rm obs} = (z_{\rm r},~f_{\rm esc},~f_{\rm bary}) =
(10,~50,~0.05)$.  As shown by T10, this set of parameters provides a
reasonable fit to the actual field data.  \emph{It is important to note
  that this input parameter vector is not included among our design points}
$\mathcal{D}$.

\begin{figure}
\hspace{-0.1cm}
\includegraphics[width=84mm,clip]{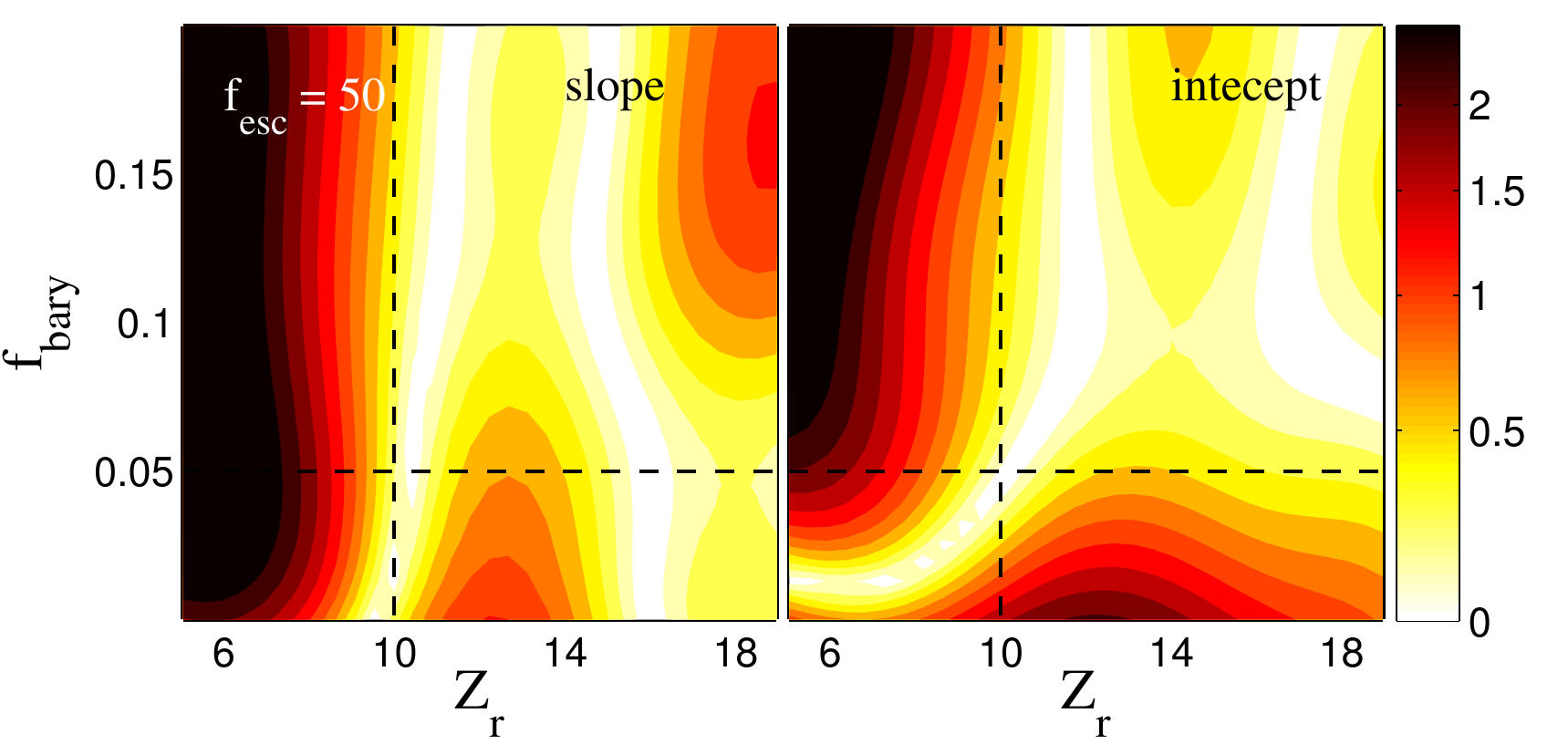}
\\
\includegraphics[width=84mm,clip]{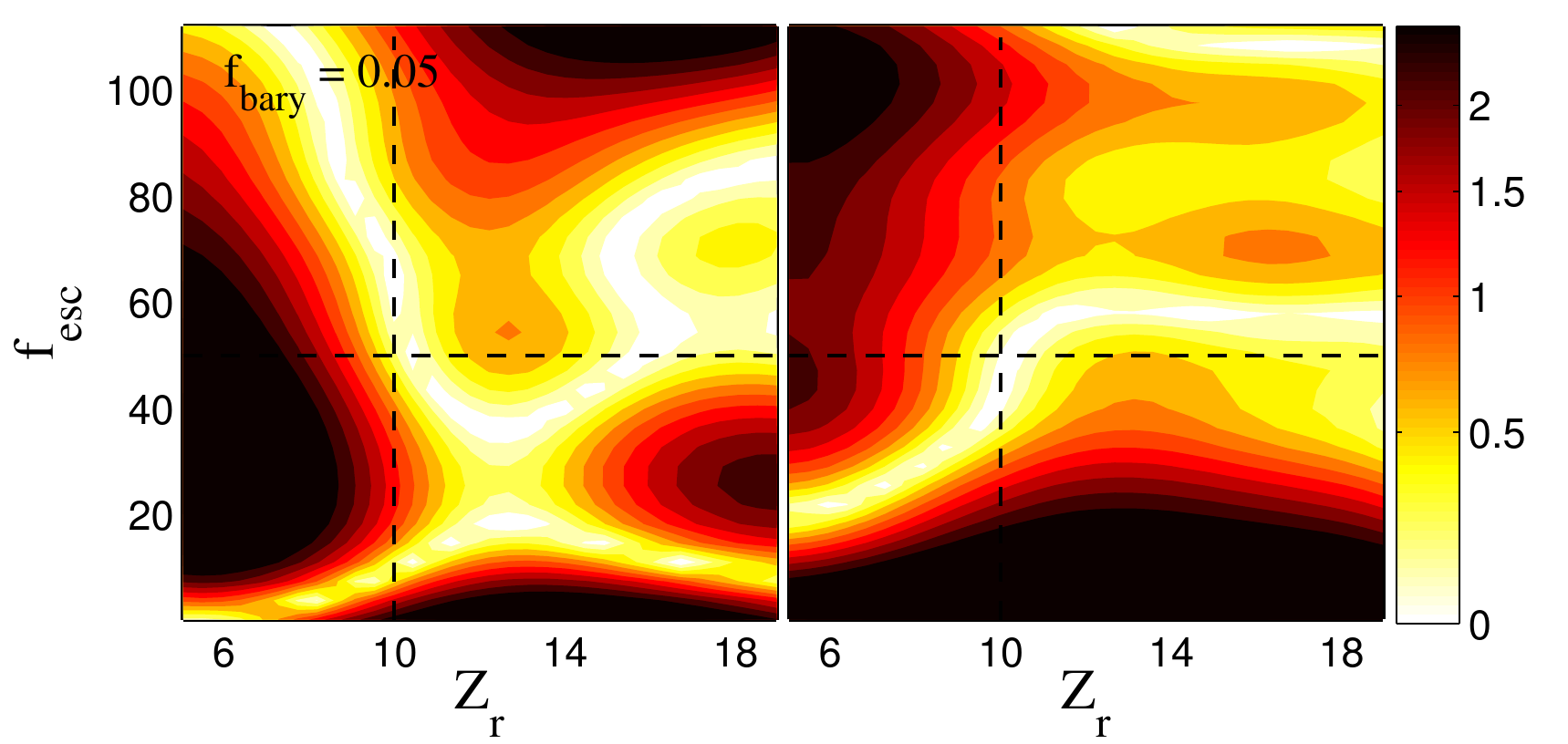}
\\
\includegraphics[width=84mm,clip]{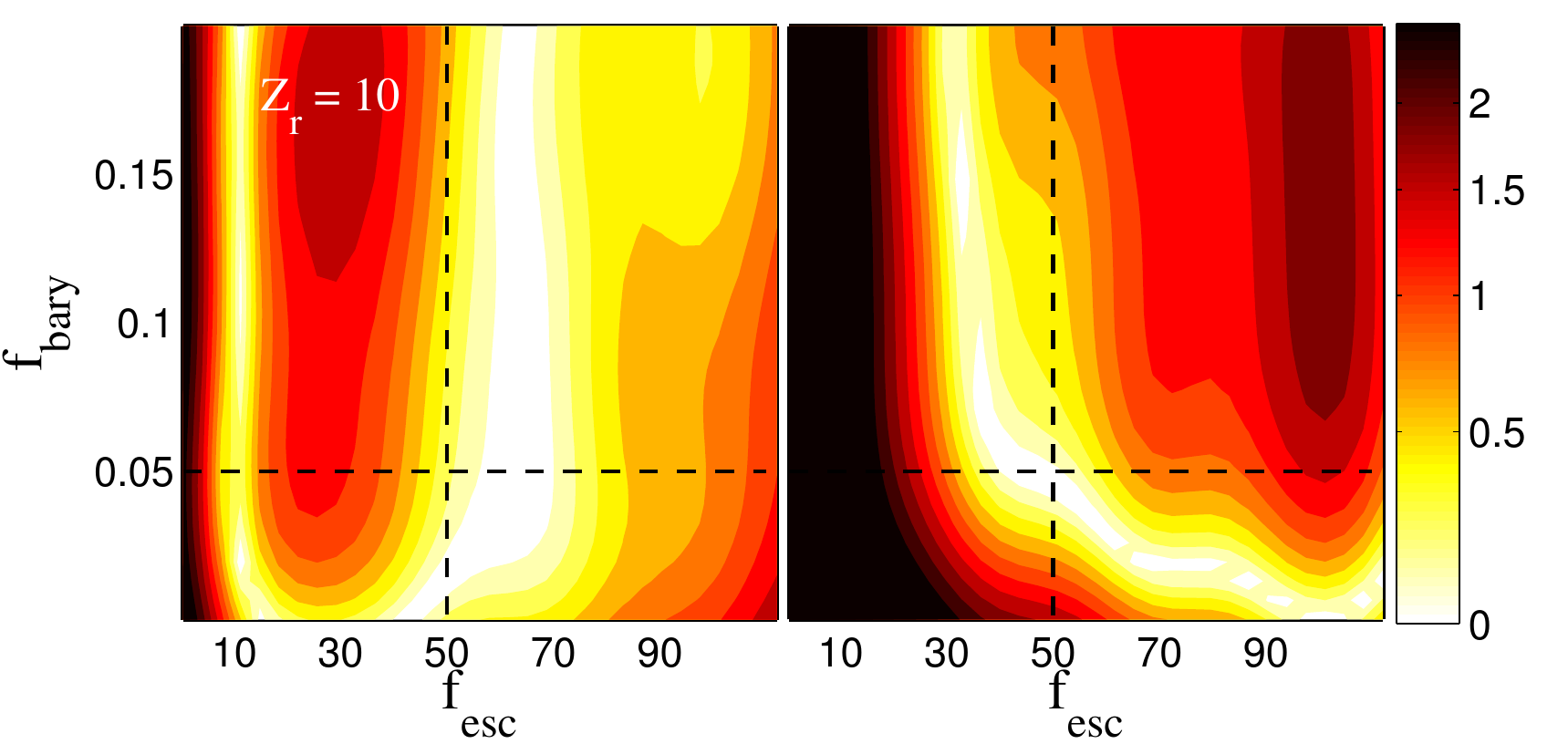}
\caption{As in \figref{fig:implaus} for the two model emulators constructed
  for the $L$-$Z$ relation.  The left and right panels show sections of
  the implausibility surfaces associated with the slope and the intercept,
  respectively.  Note that these model emulators can provide strong
  constrains to the model parameter $f_{\rm esc}$.}
\label{fig:implaus2}
\end{figure}

In \figref{fig:implaus} we show three different sections of each of the
implausibility surfaces obtained from the five model emulators constructed
for the LF's outputs.  The 3-dimensional implausibility surfaces are sliced
with three orthogonal planes as defined by the components of
$\mathitbf{x}_{\rm obs}$.  The top row panels show the $f_{\rm esc} = 50$
section of the $I(\mathitbf{x})$ surfaces. The black dashed lines indicate
the values of the remaining two components of $\mathitbf{x}_{\rm obs}$.
Given an input parameter vector $\mathitbf{x_{t}}$, the larger the value of
the $I(\mathitbf{x_{t}})$, the less likely is that a good fit to the
observable data can be obtained.  From the left-most panel (\emph{i.e.},
$M_{\rm v} = -16.5$) it becomes clear that the parameter controlling the
amount of available gas to form stars, $f_{\rm bary}$, is strongly
constrained by the number of satellite galaxies at the bright end of the
satellite galaxy luminosity function.  Furthermore, within the range of
values considered here, the number of satellites at this $M_{\rm v}$ is
independent of the redshift of the epoch re-ionization, $z_{\rm r}$.  The
most plausible parameter values are near the true value of $f_{\rm bary} =
0.05$.  As we move towards the faint end of the luminosity function the
model parameter $z_{\rm r}$ becomes progressively more constrained and the
total number of satellite galaxies becomes less dependent on $f_{\rm
  bary}$.  For $M_{\rm v} = -3.5$ (top right-most panel).  The
corresponding model emulator strongly constrains the input parameter space
around values of $z_{\rm r} \approx 10$, but it gives equally good fits for
nearly all possible values of $f_{\rm bary}$.  The second row of panels
show sections of the $I(\mathitbf{x})$ surfaces at $f_{\rm bary} = 0.05$.
The satellite galaxy luminosity function appears to be completely
independent of the value adopted for the escape factor of metals, $f_{\rm
  esc}$. This reflects the lack of a metallicity dependence on the baryon
budget and star formation rate. At the bright end of the luminosity
function, any combination of $z_{\rm r}$ and $f_{\rm esc}$ would yield an
equally good fit to the mock observable data.  However at the faint end
values of $z_{\rm r} \approx 10$ are required to fit the mock data.  A
similar result can be obtained for the third row of panels showing the
remaining sections, \emph{i.e.}, $z_{\rm r} = 10$.  Again, a good fit to
the "observable" data can be obtained with values of $f_{\rm bary} \approx
0.05$ for any possible value of $f_{\rm esc}$.

It is possible to put constraints on the parameter $f_{\rm esc}$ by
exploring cross-sections of the implausibility surfaces constructed for the
satellite galaxy luminosity-metallicity relation's slope and intercept
model emulators.  The middle and bottom panels of \figref{fig:implaus2}
show the sections defined by $f_{\rm bary} = 0.05$ and $z_{\rm r} = 10$ ,
respectively.  Comparison with \figref{fig:implaus} reveals implausibility
surfaces with a more complex topology.  Although both emulators present
regions of low implausibility for a wide range of $f_{\rm esc}$ values,
these regions are strongly associated with $z_{\rm r}$ and $f_{\rm bary}$.
These two parameters are also strongly constrained by the slope and
intercept of the $L$-$Z$ relation, as shown in the top row panels.

\begin{figure}
\includegraphics[width=85mm,clip]{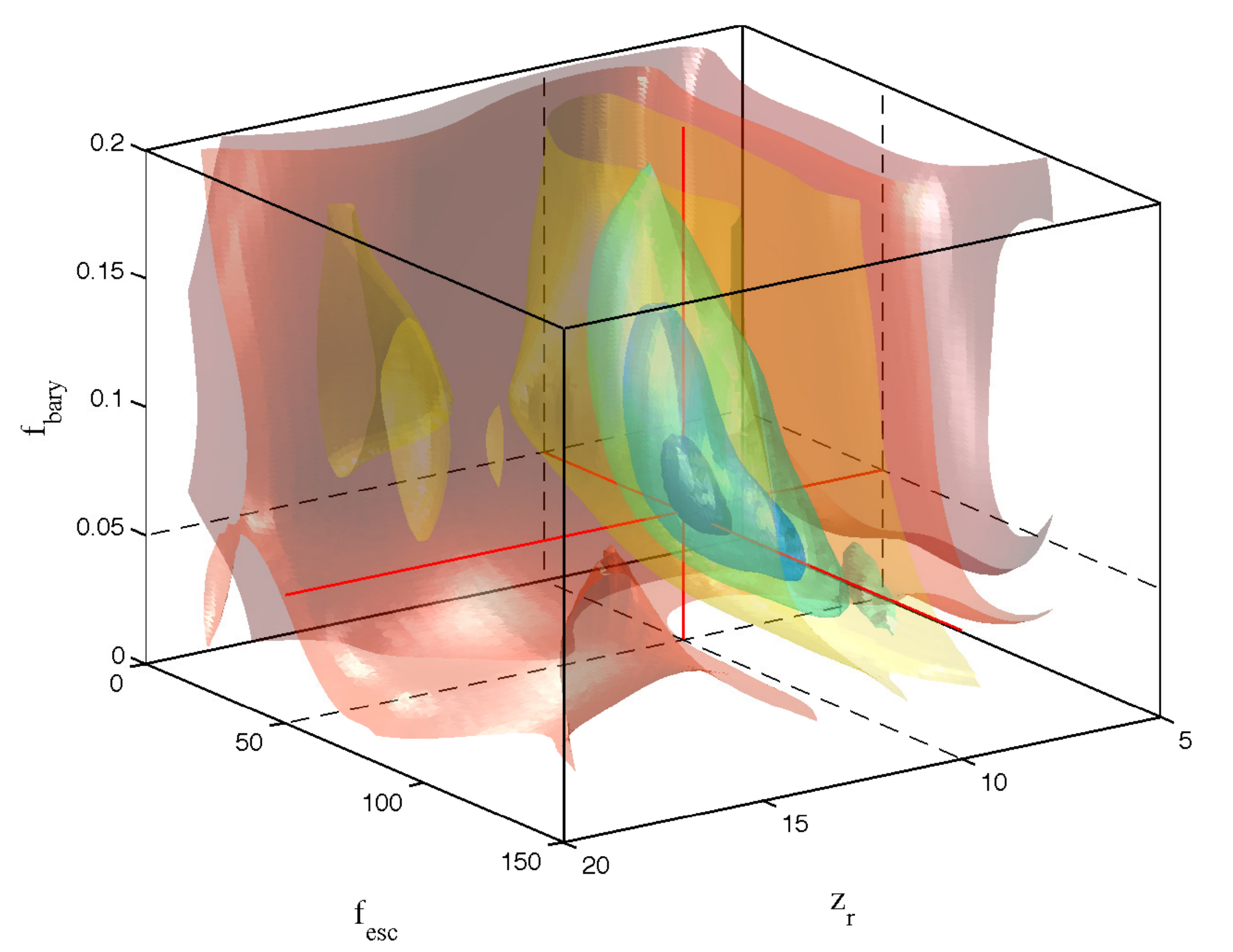}
\caption{Iso-implausibility surfaces extracted from the joint
  implausibility measure $J(\mathitbf{x})$.  Redder colors indicate larger
  values of $J(\mathitbf{x})$.  As the value of $J(\mathitbf{x})$ decreases
  the volume enclosed by each iso-surface becomes smaller, converging
  towards the values associated with $\mathitbf{x}_{\rm obs}$, as shown by
  the red solid lines.  The region of lowest implausibility (and thus
  highest plausibility) is shown by the opaque blue volume at the
  intersection of the red lines.}
\label{fig:implaus3d}
\end{figure}

\begin{figure*}
\includegraphics[width=180mm,clip]{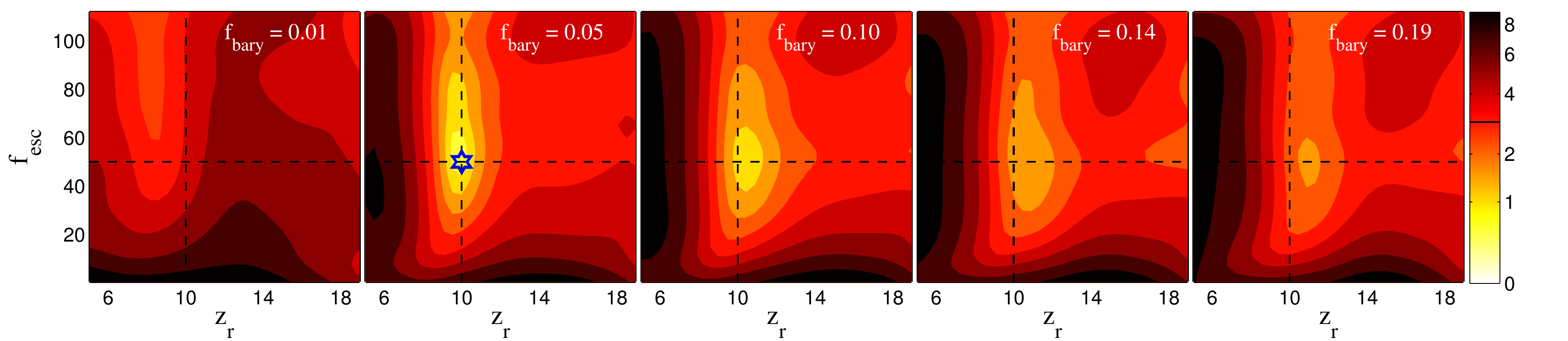}
\\
\includegraphics[width=180mm,clip]{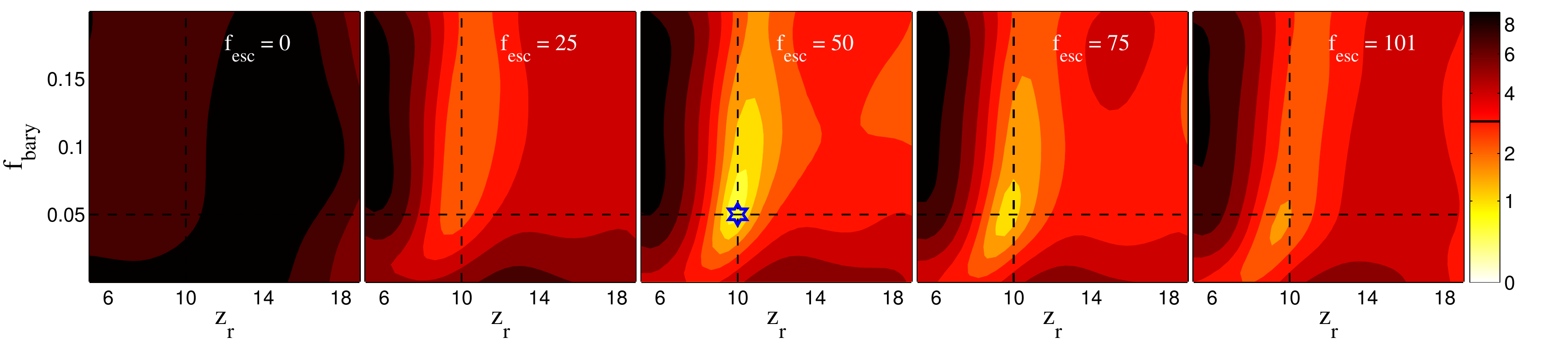}
\\
\includegraphics[width=180mm,clip]{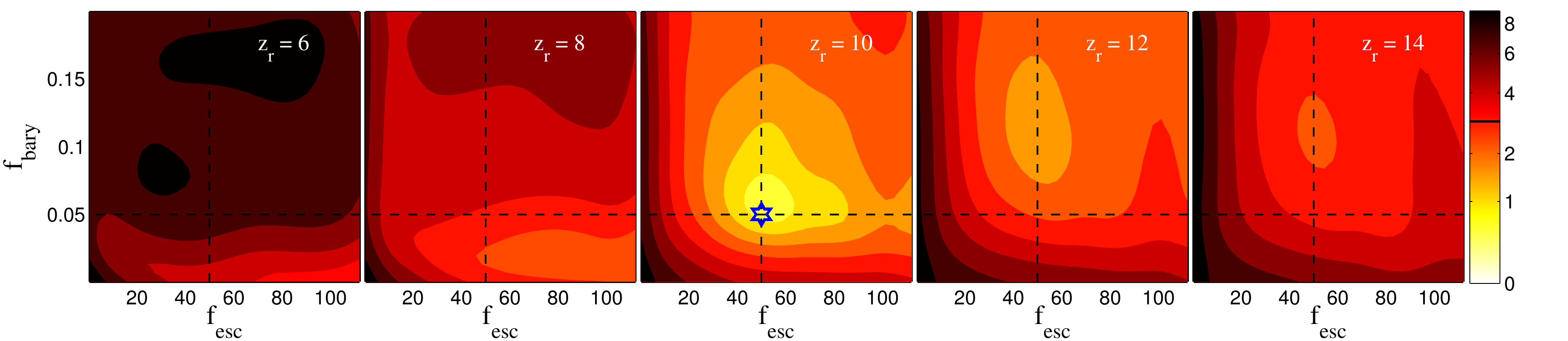}
\caption{Different  sections  of  the  Joint  implausibility  surface,
  $J(\mathitbf{x})$, obtained by combining information provided by the
  seven  model  emulators   shown  in  Figures  \ref{fig:implaus}  and
  \ref{fig:implaus2}.  The  different colors show  different values of
  $J(\mathitbf{x})$  in   logarithmic  scale.   Model   emulators  are
  compared  to  the  mock   observable  data  obtained  after  running
  ChemTreeN  with   the  input  parameter   vector  $\mathitbf{x}_{\rm
    obs}$. Both mock observables and training data set are obtained by
  coupling  ChemTreeN with  the  $N$-body simulation  \MW1.  The  top,
  middle  and bottom row  panels show  different sections  of constant
  $f_{\rm  esc}$, $f_{\rm  bary}$ and  $z_{\rm r}$,  respectively.  On
  each row, the  black dashed lines indicate the values  of two of the
  components of $\mathitbf{x}_{\rm obs}$.  If the three components are
  simultaneously   present    in   a   section,    the   location   of
  $\mathitbf{x}_{\rm  obs}$  is  indicated  with  a  blue  star.   The
  horizontal black solid  line on the color bars  indicate the imposed
  threshold:  a value  above  this  threshold shows  that  it is  very
  implausible  to obtain  a good  fit to  the observed  data  with the
  corresponding   values   of  the   model   parameters.   Note   that
  $J(\mathitbf{x})$  can strongly constrain  the full  parameter space
  under  study.  Note as  well that  the values  of the  components of
  $\mathitbf{x}_{\rm obs}$  are located in the  most plausible regions
  of the space. }
\label{fig:implaus3}
\end{figure*}

Individually, none of  the previously explored implausibility surfaces
can constrain the full parameter space.  This is not the case with the
joint  implausibility measure  $J(\mathitbf{x})$,  which combines  the
information  obtained   from  the  seven  model   emulators  into  one
quantity\footnote{Recall  that,  when   dealing  with  more  than  one
  observable  simultaneously, model emulators  are constructed  in the
  principal    component     space    of    the     training    set.}.
Figure~\ref{fig:implaus3d} shows different iso-implausibility surfaces
of  the resulting  $J(\mathitbf{x})$.   Notice that  as  the value  of
$J(\mathitbf{x})$  decreases the volume  enclosed by  each iso-surface
becomes  smaller,  converging   towards  the  values  associated  with
$\mathitbf{x}_{\rm obs}$, as  shown by the red solid  lines.  This can
be  more clearly  appreciated in  \figref{fig:implaus3}.  Each  row of
panels shows  different sections $J(\mathitbf{x})$ as  we traverse one
of the three  possible dimensions.  The black solid  line on the color
bars show the cutoff applied  to the joint implausibility.  A value of
$J(\mathitbf{x})  > 3$  (in  $t=7$ dimensions)  indicates  that it  is
implausible  to  obtain a  good  fit to  the  observed  data with  the
corresponding values  of the model  parameters.  Thus, regions  of the
parameter  space lying above  this threshold  can be  disregarded.  We
find  that $J(\mathitbf{x})$  strongly constrains  the  full parameter
space  under study.   Furthermore,  the values  of  the components  of
$\vx_{\rm  obs}$ are  located in  the  most plausible  regions of  the
space, as indicated by the star symbols in the corresponding panels.

As described by B10, it is possible to further constrain the input
parameters by performing an iterative approach.  The idea consists in
isolating the region below the chosen implausibility threshold and
generating a new training set using design points located within these
regions.  The kind of ``controlled'' experiments performed here, in
addition to the low dimensionality of our previous example, renders this
approach unnecessary.  Thus, we refer the reader to B10 for thorough
description of this procedure.

\begin{figure*}[ht]
\includegraphics[width=85mm,clip]{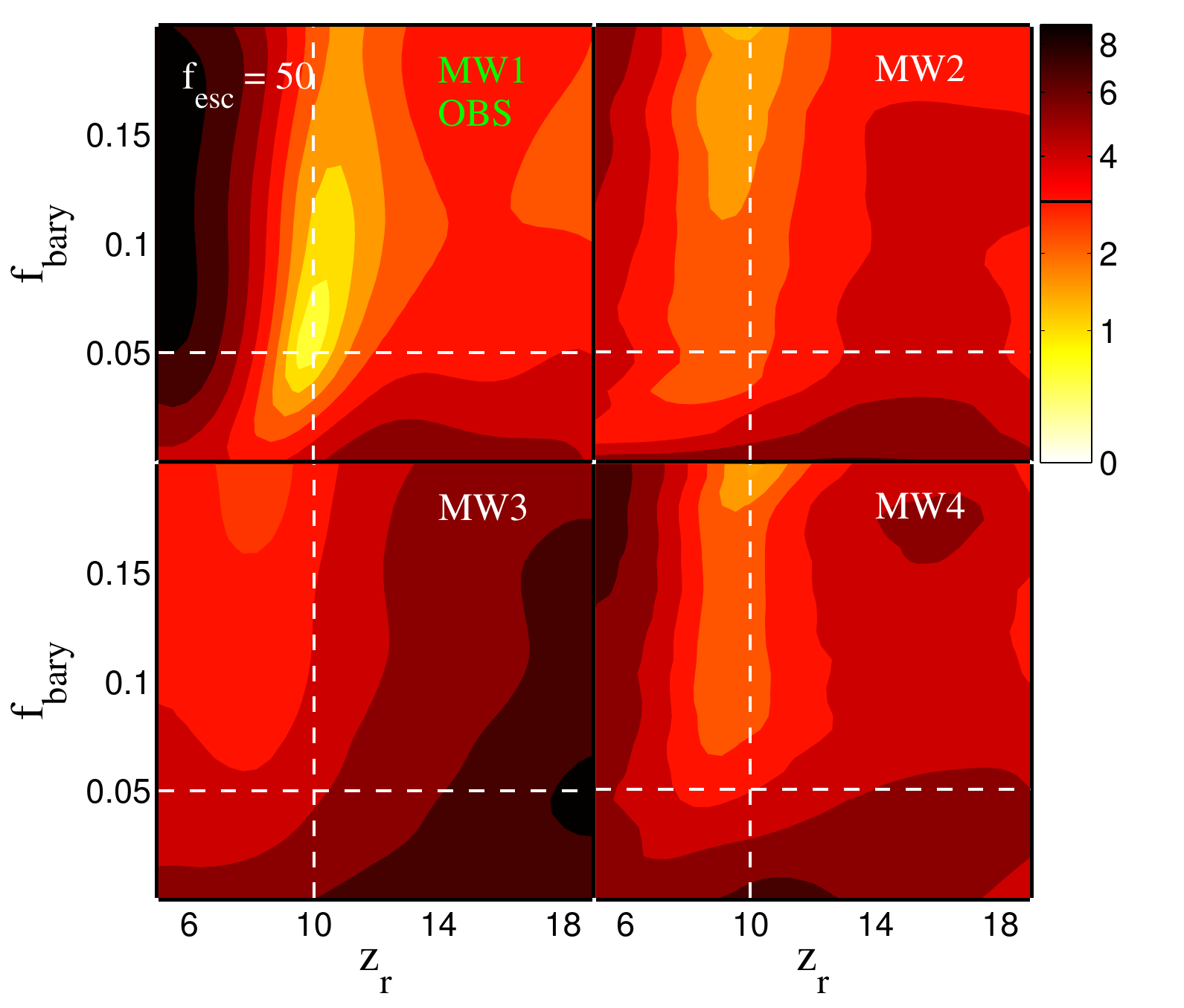}
\includegraphics[width=85mm,clip]{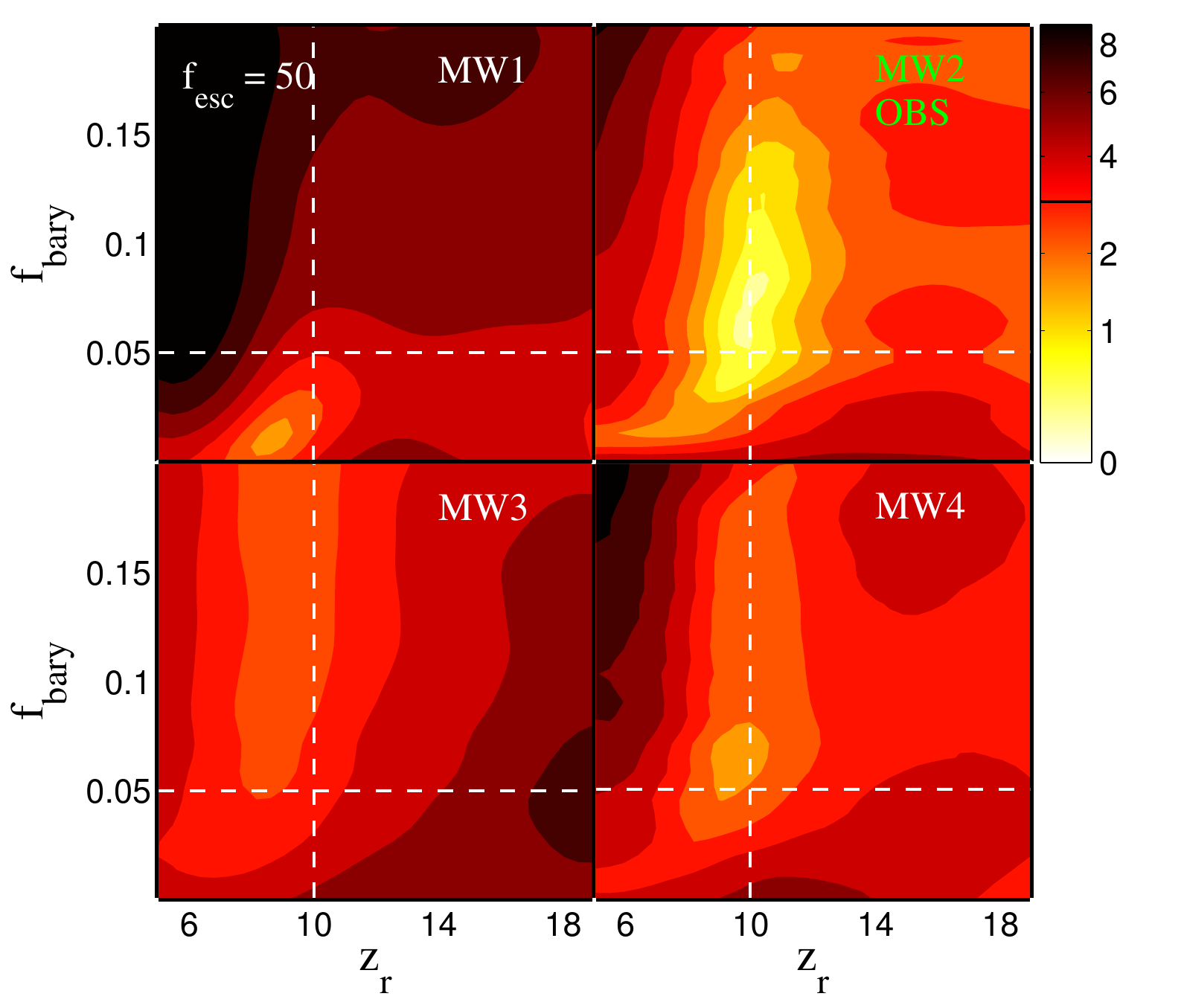}
\\
\includegraphics[width=85mm,clip]{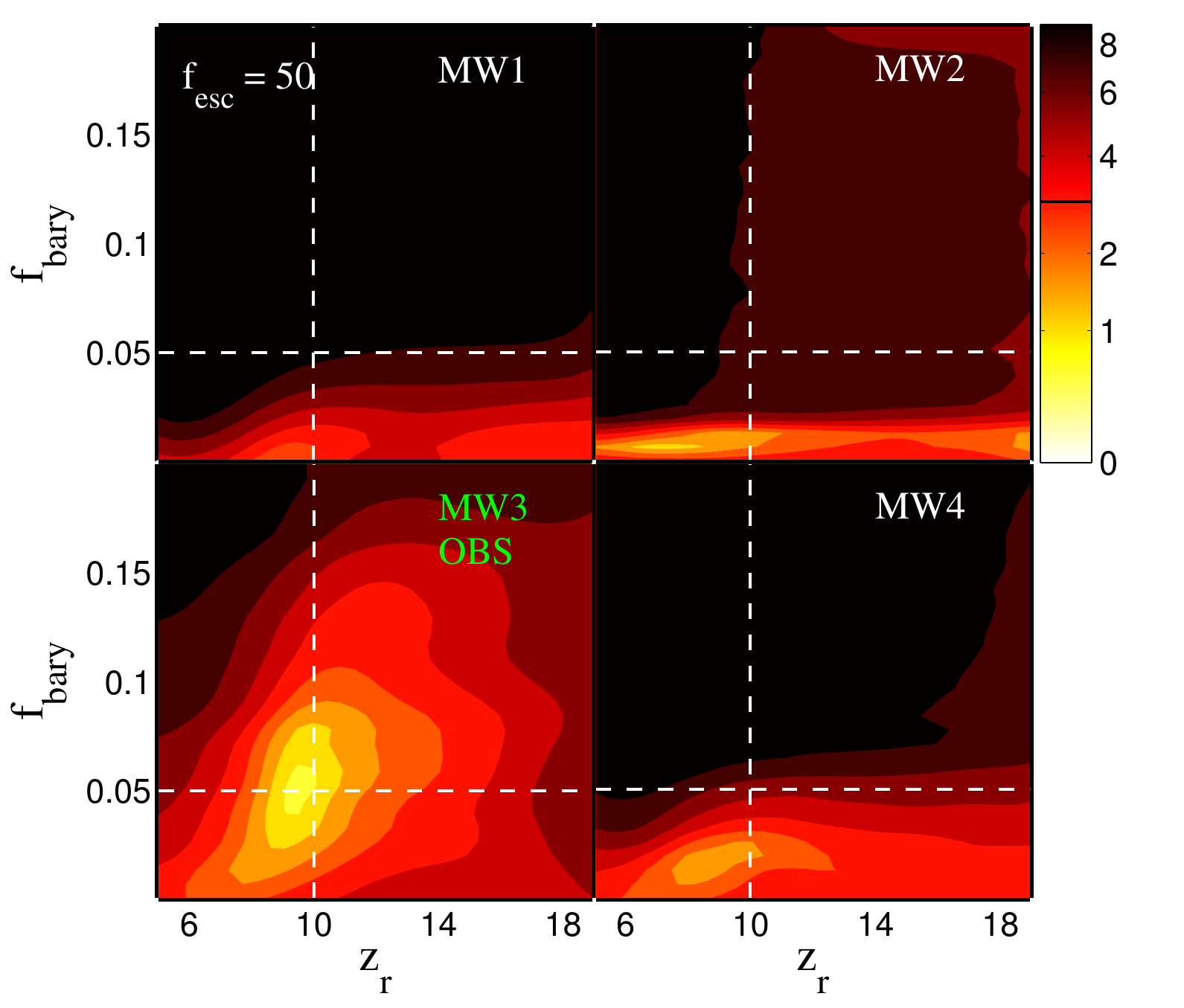}
\includegraphics[width=85mm,clip]{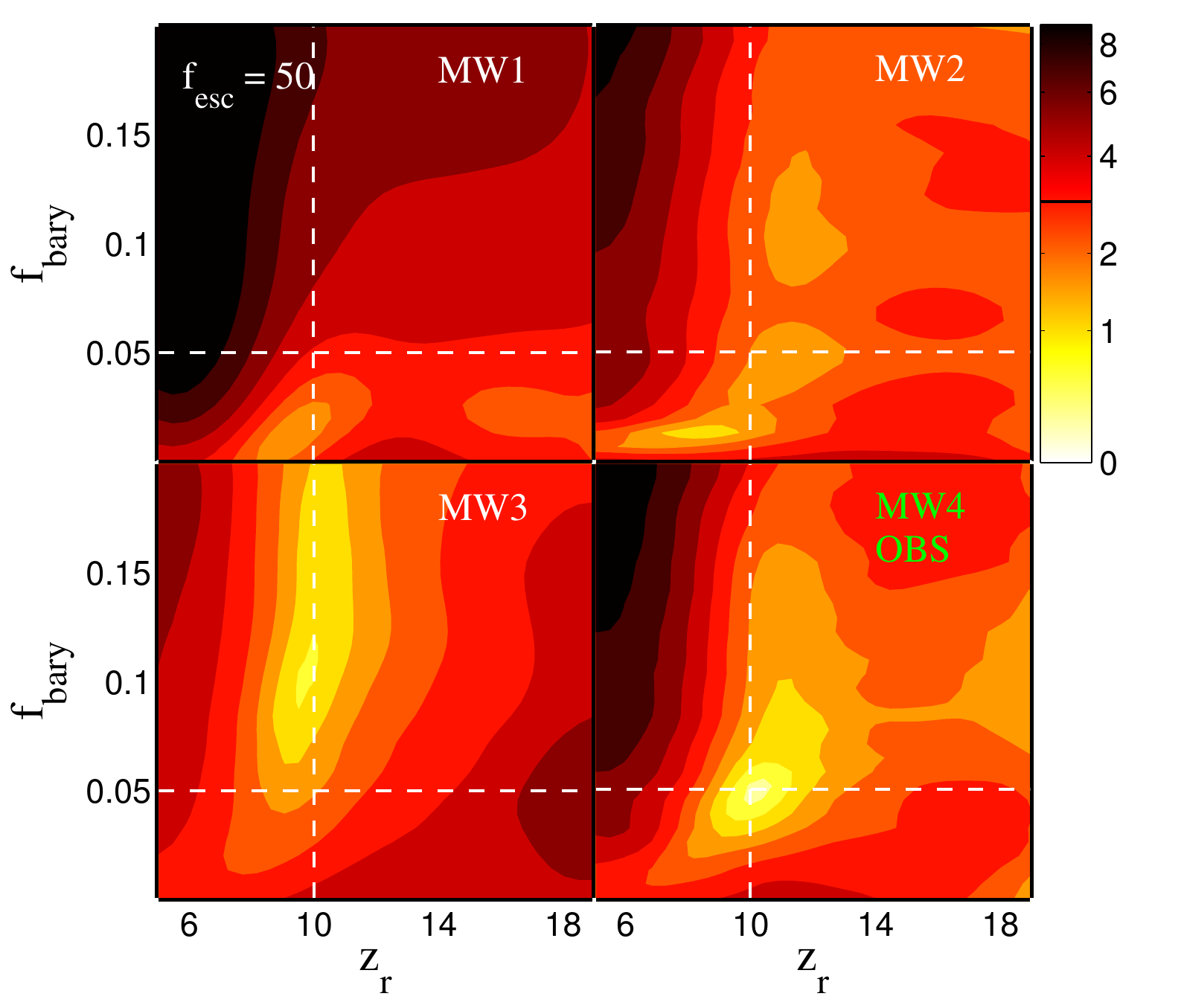}
\caption{Sections of Joint Implausibility surfaces' at constant $f_{\rm
    esc}$, obtained after comparing different models and mock observables.
  The different colors show different values of $J(\mathitbf{x})$ in
  logarithmic scale.  Each block of panel shows the results of comparing a
  given \MW{i}-observables to the four sets of \MW{k}-emulators (see text),
  where $k,~i = 1,~2,~3$ and $4$.  \MW{i}-observables are obtained by
  running ChemTreeN on the merger tree extracted from simulation \MW{i},
  using the input parameter vector $\mathitbf{x}_{\rm obs}$.  The labels on
  the top left corner of each panel indicates the \MW{k}-emulators being
  consider.  In green we indicate the \MW{i}-observables associated with
  each block. The section were selected such that the three components of
  $\mathitbf{x}_{\rm obs}$ are present.  On each panel, the white dashed
  lines indicate the values of two of the components.  The horizontal black
  line on the color bars indicate the imposed threshold: a value
  above this threshold shows that it is very implausible to obtain a good
  fit to the observed data with the corresponding values of the model
  parameters.  Note that, in many cases, similarly good fits to a given set
  of observables can be obtained with different parameter's values simply
  by modifying the host's merger history.  These values may significantly
  differ from those used to generate the mock observables.}
\label{fig:implaus4}
\end{figure*}

\subsection{Impact of merger histories}

In the previous subsection we showed that it is possible to recover the set
of input parameter chosen to create a mock Milky Way-like observational
data set using a suite of model emulators.  In this ``controlled''
experiment, both training and mock observable data were obtained by
coupling ChemTreeN to a merger tree obtained from a single
simulation. \emph{Therefore, we have implicitly assumed that the exact
  merger history of our Milky Way-like galaxy is a known variable of the
  problem}.  In reality, this merger history is poorly known, and could be
regarded as an extra input parameter of the model.  It is thus important to
study how different merger histories can compromise our ability to
meaningfully constrain the input parameter space.  To this end, we perform
the following set of controlled experiments.  Using the merger trees
extracted from the four available $N$-body simulations we generate four
different training sets (each training set containing $n=200$ design
points) and construct, for each set, the suite of model emulators discussed
previously.  Hereafter, we will refer to these emulators' sets as
``\MW{i}-emulators'', with $i =$ 1, 2, 3 and 4.  The input parameter vector
$\mathitbf{x}_{\rm obs} = (z_{\rm r},~f_{\rm esc},~f_{\rm bary}) =
(10,50,0.05)$ is used to obtain a mock observational data set from each
merger tree.  We will refer to these mock observables as
``\MW{i}-observables''.  We then ask the following question:
\begin{quote}\em
  Is it possible to recover the input parameter vector, $\mathitbf{x}_{\rm
    obs}$, if we use training data obtained from a merger tree different
  than that used to generate the mock observables?
\end{quote}
In \figref{fig:implaus4} we show the outcome of this experiment.  Each
block  of  four panels  shows  sections  of  the joint  implausibility
surface  obtained after  comparing given  \MW{i}-observables  with the
four  \MW{i}-emulators.    The  merger  tree  used   to  generate  the
\MW{i}-observables in each block is indicated with the green label and
marked  ``OBS''.  For example,  the top  left corner  shows comparison
results using \MW1-observables.  As in \figref{fig:implaus3}, when the
model emulators are  trained on the same merger  tree used to generate
the  mock  observables,  we   can  successfully  constrain  the  input
parameter  space  and  recover  the components  of  $\mathitbf{x}_{\rm
  obs}$.     However,   when    model    emulators   constructed    on
\emph{different}  merger  trees  are  considered, the  most  plausible
regions are located  around values of $f_{\rm bary}$  much larger than
those used  to obtain  the mock observables.   This is  not surprising
since,  as  shown  in  \figref{fig:diff_models},  \MW1  is  the  Milky
Way-like halo  that contains the  largest number of satellites  at all
$M_{\rm  v}$.   To achieve  a  good  fit  to \MW1-observables  in  the
remaining simulations requires  a larger amount of gas  to form stars.
Note as well  that the joint implausibility surface  obtained with the
\MW3-emulators never  falls significantly below  the chosen threshold,
so  \MW1-observables cannot  be  reproduced using  the merger  history
extracted from halo \MW3.  Another interesting example is shown on the
lower  right  panels   of  \figref{fig:implaus4},  where  we  consider
\MW4-observables.  Very good fits to these observables can be obtained
for either larger  (\MW3-emulators) or smaller (\MW2-emulators) values
of $f_{\rm bary}$ than that  used to generate the mock observables.  A
similar  situation is observed  for the  input parameter  $z_{\rm r}$.
Note that  we have only considered  the $f_{\rm esc} =  50$ section of
each    joint    implausibility     surface.     As    described    in
Section~\ref{sec:general_results},      the      satellite      galaxy
luminosity-metallicity relation is not sensitive to the merger history
of  the host  halo.  The  parameter  $f_{\rm esc}$  is therefore  well
constrained in all cases.   This forces the $J(\mathitbf{x})$ surfaces
to  have   the  most  plausible   regions  in  the  vicinity   of  the
aforementioned section of $J(\mathitbf{x})$.
 
The previous analysis clearly shows how a particular merger history can
influence the model parameter selection: \emph{similarly good fits to a
  given set of observables can be obtained with different model parameter
  values simply by modifying the host's merger history}.  In our
experiments these values may differ from those used to generate the mock
observables.  When comparing with real observational data, a given set of
best fitting parameter's values may be significantly off from the values
that could best parametrize the desired underlying physical processes.
This in turn may have important implications on other observable quantities
that we would like to study and which have not been used for model
parameter selection.

\subsection{Model comparison}

The previous analysis illustrated the importance that a host's merger
history has on model parameter estimation. For a given set of mock
observables, locations of low-joint implausibility regions may
significantly vary simply by changing the host merger history. However, it
remains to be explored whether differences in the resulting joint
implausibility surfaces obtained with, e.g., \MW4-observables (bottom right
panels of Figure~\ref{fig:implaus4}) could allow us to isolate a ``most
likely'' merger history for this mock data.  Is it possible to constrain
the host merger history by statistically comparing these surfaces?

This question can be addressed within a Bayesian framework for model
quality assessment.  The posterior probability for a model $M_{i}$, given
an observable data set $Y_{f}$, is given by the Bayes' theorem
\citep{gregory05},
\begin{equation}
P(M_i\mid Y_{f}) = \dfrac{P(Y_{f}\mid M_i)~P(M_i)}{P(Y_{f})}
\end{equation} 
where the denominator $P(Y_{f}) = \sum_{j=1}^{4} P(Y_{f}\mid M_j) P(
M_{j})$ is the marginal probability density of $Y_F$, and
$P(Y_{f}\mid M_i)$ is the probability density of $Y_F$ under model $M_i$,
marginalized over the model parameters, $\vx$, representing the probability
density for the observable $Y_{f}$ under the assumption of model $M_i$.  In
this analysis, a model $M_i$ represents the 
coupling of ChemTreeN with a given cosmological simulation.  In model
selection problems it is often more useful to consider the ratio of the
probability of two models rather than their probabilities directly.  The
quantity
\begin{equation}
\label{eqn:bayes}
  O_{ij} = \dfrac{P(M_i\mid Y_{f})}{P(M_j\mid Y_{f})} 
        = \dfrac{P(Y_{f}\mid M_i)}{P(Y_{f}\mid M_j)} \dfrac{P(M_i)}{~P(M_j)}
\end{equation} 
is known as the odds ratio in favor of model $M_i$ over model $M_j$.  The
first factor on the right side of \eqref{eqn:bayes} is known as the Bayes
factor,
\begin{equation}
\label{eqn:bayes_fac}
  B_{ij} =\dfrac{P(Y_{f}\mid M_i)}{P(Y_{f}\mid M_j)} = \dfrac{\int
    P(\vx\mid M_i) ~P(Y_{f}\mid \vx,M_i) ~d\vx} {\int
    P(\vx\mid M_j) ~P(Y_{f}\mid \vx,M_j) ~d\vx }.
\end{equation} 
If we assign equal prior probabilities for all models, then odds ratios and
Bayes factors coincide:
\begin{equation}
O_{ij} =B_{ij}\notag
\end{equation} 

In our problem we approximate $B_{ij}$ by calculating the likelihoods,
$\mathcal{L} = P(Y_{f}\mid \vx,M_i)$, under a multivariate normal
approximation from the corresponding joint implausibility, \emph{i.e.},
$\mathcal{L} \propto \int e^{-J_{i}(\vx)^2 / 2} ~ d\vx$, with uniform
priors $P(\vx\mid M_i)$.  Here, $J_{i}(\vx)$ is obtained using
\MW{i}-emulators.

While the use of Bayes factors to compare and select among models is well
established, a variety of scales have been proposed to facilitate their
quantitative interpretation \citep{fk2012}.  We will employ Jeffreys' scale
\citep{jeffreys98}, presented in \tabref{table:jeffrey}.  According to this
scale, values of $\log_{10} B_{ij}$ between 0 and 0.5 indicate that models
$M_{i}$ and $M_{j}$ are equally supported by the observable data set
$Y_{f}$.  Conversely, a value larger than 2 gives a decisive evidence for
model $M_{i}$ against model $M_{j}$, given $Y_{f}$.

\begin{figure}[htb]
\includegraphics[width=90mm,clip]{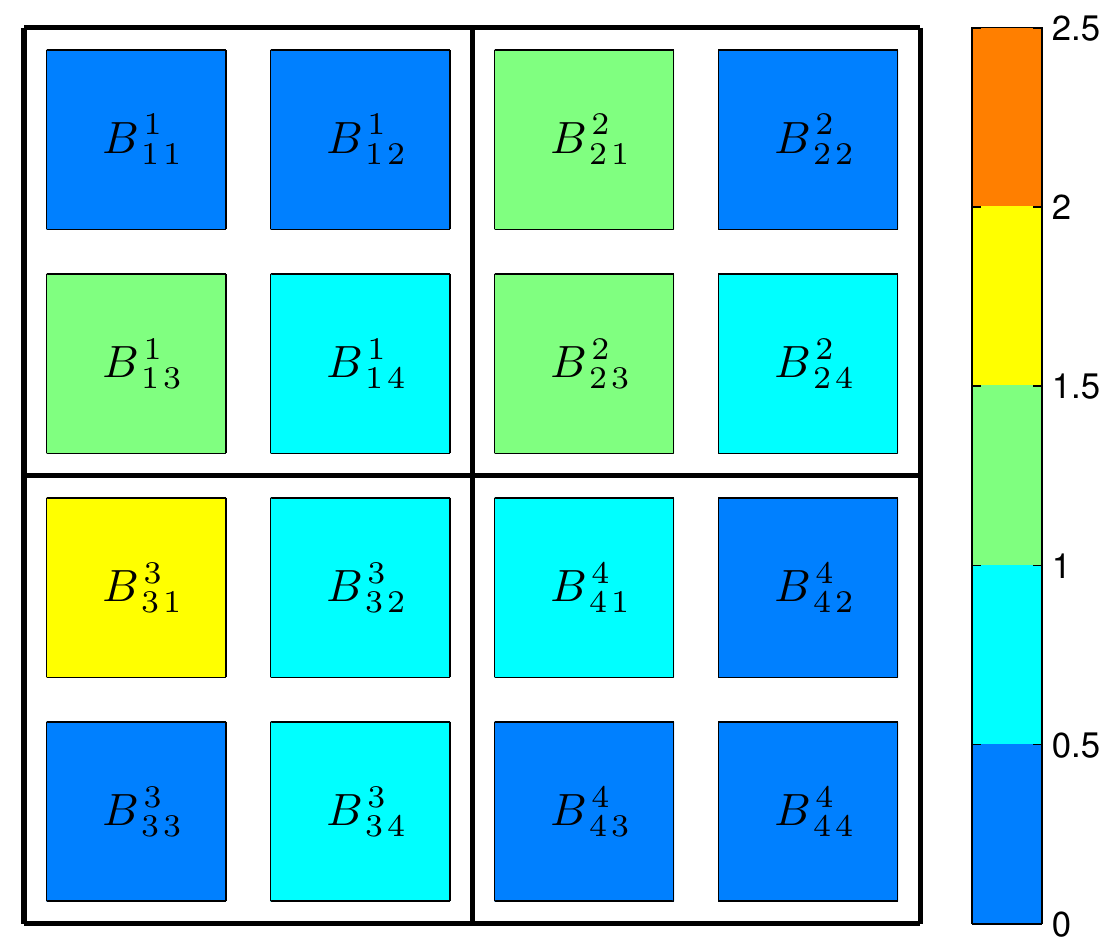}
\caption{Values of the $\log_{10}$ Bayes factors obtained by comparing two
  different models, after fixing a given mock observable data set. The
  values are presented in logarithmic scale.  Each element $B_{ij}^{k}$
  shows the result of comparing models $M_i$ and $M_j$ using
  \MW{k}-observables.  The matrix is divided in four block, each of them
  associated with a different \MW{k}-observable.  For comparison, the
  panels are distributed as in Figure~\ref{fig:implaus4}.  The values of
  the different Bayes factor are color coded according to the Jeffery's
  scale, introduced in \tabref{table:jeffrey}.  Each element $\log_{10}
  B_{ii}^i = 0$, since model $M_{i}$ is being compared to itself.  More
  than one model can be equally-well supported by a given mock observable
  data, as shown by the multiple white and light-grey elements of each
  matrix's blocks.  Note that the ``best fitting'' parameters extracted from
  each model may differ significantly across data sets (see
  Figure~\ref{fig:implaus4}).}
\label{fig:jeffrey}
\end{figure}

In this work we are dealing not only with four different models, but also
with four different mock observable data sets.  From now on, we will refer
as $B_{ij}^{k}$ to the Bayes factor obtained comparing models $M_i$ and
$M_j$ using \MW{k}-observables.  The results of this analysis are shown in
\figref{fig:jeffrey} in the form of a matrix.  The elements of this matrix
show, coded by color, the values of the different Bayes factors.  Note that
the matrix is divided into four blocks, each one of them associated with a
different \MW{k}-observable.  For comparison, the panels are distributed as
in Figure~\ref{fig:implaus4}.  For example, the first block on the top left
corner shows the probability ratios of model $M_{1}$ and $M_{j}$, with
$j=1,2,3$ and $4$, using \MW1-observables.  The first element of this block
is, not surprisingly, $B_{11}^1 = 0$ as we are comparing model $M_{1}$ with
itself.  It is interesting however to observe that, on the basis of
\MW1-observables, model $M_{1}$ and $M_{2}$ are equally plausible, as
indicated by $B_{12}^1$.  Comparison with \figref{fig:implaus4} (top left
panels) shows that the combination of model $M_{2}$ and \MW1-observables
yield plausible values of $f_{\rm bary}$ that are much larger than that
used to generate the mock observables (\emph{i.e.}, $f_{\rm bary} = 0.05$).
A similar situation arises when considering \MW4-observables.  In this
case, models $M_{2}$, $M_{3}$ and $M_{4}$ are all equally supported by the
mock observable data.  However, the corresponding joint implausibility
surface's sections, shown on the bottom left panels' block of
\figref{fig:implaus4}, suggest that the most plausible values of the input
parameter for models $M_{2}$ and $M_{3}$ differ significantly from the
``true'' values used to generate the \MW4-observables.

This analysis has shown us that multiple models with different merger tree
histories can be equally supported by a given observational data set.  The
``best fitting'' parameters extracted from these models are, however,
significantly different from each other.  On the basis of this analysis it
is therefore not possible to assess what set of parameters would be best
suited to model the formation of the stellar halo of the Milky Way.

\begin{deluxetable}{@{}cc}
  \tabletypesize{\scriptsize} \tablecaption{Jeffreys' scale for grading the
    significance associated with different ranges of the logarithmic Bayes
    factor.\label{table:jeffrey}} \tablewidth{180pt}
  \tablehead{\colhead{$\log_{10} B_{ij}$} & \colhead{Significance}}
  \startdata
  $0-0.5$ & Barely worth mentioning\\
  $0.5 - 1$ & Substantial  \\
  $1 - 1.5$ & Strong \\
  $1.5 - 2$ & Very strong\\
  $> 2$ & Decisive
  \enddata
\end{deluxetable}

\section{Discussion and Conclusions}
\label{sec:conclusions}

In this work we combined modern statistical methods with cosmological
simulations of the formation of the satellite galaxy population and stellar
halo of Milky Way-like galaxies to explore how one might meaningfully
constrain a galaxy's formation history using observables related to these
components.  The semi-analytic galaxy formation model ChemTreeN, coupled to
merger trees extracted from four cosmological $N$-body simulations, is used
to track the evolution of the baryonic properties of Milky Way-like
galaxies and their satellite populations. Gaussian Process model emulators
are used to emulate the outputs of ChemTreeN at any location of its
$p$-dimensional input parameter space.  This allow us to explore the full
input parameter space orders of magnitude faster than could be done with
the ChemTreeN code itself.  To compare observational data to the model
using these emulators we introduce the joint implausibility measure (see
also B10).  We show that, using this quantity as a measurement of
correctness, we can successfully recover the input parameter vector used to
generate mock observational data sets (namely, the satellite galaxy
luminosity function of our chosen Milky Way-like dark matter halos and the
slope and normalization of the satellite galaxy luminosity-metallicity
relation).

From our analysis we draw the three following main conclusions:
\begin{enumerate}

\item Using our Gaussian Process model emulator, we have explored the
  dependence of a variety of observational quantities on model parameters.
  In some cases, this highlights straightforward relationships between
  observational quantities and model parameters: for example, the number of
  low-luminosity dwarf galaxies depends very sensitively on the redshift of
  re-ionization, but is insensitive to other model parameters.  On the
  other hand, the number of high-luminosity dwarfs is insensitive to the
  redshift of re-ionization, but sensitive to the reservoir of gas
  available for star formation.  For simplicity, throughout this work we
  have focused our analysis on a three-dimensional input parameter space.
  However, as shown in Appendix~\ref{sec:appen-5} and B10, the method can
  be extended in a straightforward manner to much a higher-dimensional
  parameter space. Thus, techniques like the one presented here are well
  suited not only to explore the input parameters of complex models but
  also to expose and visualize the non-linear coupling between the
  parametrization of the different physical processes being considered.

\item The detailed merger history of a galaxy can very sensitively affect
  the outcome of the procedure for finding the ``best fitting'' parameters
  to a given mock observational data set.  Similarly good fits can be
  obtained with different parameter values simply by modifying the host
  galaxy's merger history.  More importantly, in our experiments these
  values may strongly differ from those used to generate the mock
  observables.  This is true even when the input $N$-body simulations have
  been carefully screened to all have Milky Way-like masses and times of
  last major mergers.  When comparing with real observational data, the
  resulting best fitting parameter values may differ significantly from the
  values that could best parametrize the desired underlying physical
  processes.  This in turn may have important implications on other
  observable quantities that we would like to study and which have not been
  used for model parameter selection.

\item Using a Bayesian framework for model quality assessment, we have
  explored whether it is possible to identify (or at least constrain) the
  merger history of the galaxy from which observables have been obtained.
  This exercise showed that multiple models with different merger histories
  can be equally supported by a given observational data set.  The ``best
  fitting'' parameters extracted from these models do, however, differ
  significantly from each other.  On the basis of this analysis it is
  therefore not possible to assess what set of parameters would be best
  suited to model the formation of the Milky Way's satellite galaxy
  population and its stellar halo.
\end{enumerate}

The results presented in this work highlight the dangers of selecting model
parameters based on simulations and observations of individual objects such
as the Milky Way.  A way to partially circumvent this problem is to fix the
values of a subset of parameters such that it is possible to reproduce
properties of the galaxy population in the local Universe as well as at
higher redshift.  This approach has been successfully adopted by several
authors to study the formation of the Milky Way Stellar halo and
characterize its satellite population \citep[see, e.g.][]{lh,
  2010MNRAS.401.2036L, cooper, 2011MNRAS.417.1260F}.  It is however not
surprising that important physical process involved in the evolution of,
e.g., the faintest Milky Way dwarf galaxies \citep[see
e.g.,][]{2005AJ....129.2692W, 2009MNRAS.397.1748B, 2007ApJ...654..897B,
  2007ApJ...662L..83W} are poorly constrained by the properties of the much
larger-scale galaxy population.  Statistical methods such as the one
presented here are very well suited to visualize the interplay between
different physical mechanism and thus can be used to further constrain the
parameters that are more relevant to the evolution of dwarf galaxies.

We have seen that different models of the Milky Way stellar halo, each with
a different formation history, can be equally supported by a given
observational data set even though the values extracted for the best
fitting parameter may significantly differ.  This seems to compromise the
possibility of constraining the formation history of the Milky Way through
this kind of analysis. Note, however, that our results are based on merger
trees extracted from only four cosmological simulations.  Therefore, it
remains to be explored whether statistical constraints on the MW's merger
history could be obtained by analyzing a large set of high resolution
cosmological simulations densely sampling full range of possible formation
histories.  We defer this analysis to a later paper.

\acknowledgments FAG, BWO, CCS and RLW are supported through the NSF Office
of Cyberinfrastructure by grant PHY-0941373.  FAG and BWO are supported in
part by the Michigan State University Institute for Cyber-Enabled Research
(iCER).  BWO was supported in part by the Department of Energy through the
Los Alamos National Laboratory Institute for Geophysics and Planetary
Physics and by NSF grant PHY 08-22648: Physics Frontiers Center/Joint
Institute for Nuclear Astrophysics (JINA). RLW is also supported by in part
by NSF grant DMS--0757549 and by NASA grant NNX09AK60G.

\bibliographystyle{apj}
\bibliography{apj-jour,halo_emulator}

\begin{thebibliography}{81}
\expandafter\ifx\csname natexlab\endcsname\relax\def\natexlab#1{#1}\fi

\bibitem[{{Agertz} {et~al.}(2011){Agertz}, {Teyssier}, \&
  {Moore}}]{2011MNRAS.410.1391A}
{Agertz}, O., {Teyssier}, R., \& {Moore}, B. 2011, \mnras, 410, 1391

\bibitem[{{Agertz} {et~al.}(2007){Agertz}, {Moore}, {Stadel}, {Potter},
  {Miniati}, {Read}, {Mayer}, {Gawryszczak}, {Kravtsov}, {Nordlund}, {Pearce},
  {Quilis}, {Rudd}, {Springel}, {Stone}, {Tasker}, {Teyssier}, {Wadsley}, \&
  {Walder}}]{2007MNRAS.380..963A}
{Agertz}, O., {et~al.} 2007, \mnras, 380, 963

\bibitem[{{Bayarri} {et~al.}(2002){Bayarri}, {Berger}, {Higdon}, {Kennedy},
  {Kottas}, {Paulo}, {Sacks}, {Cafeo}, {Cavendish}, {Lin}, \&
  {Tu}}]{Baya:Berg:Paul:etal:2007}
{Bayarri}, M.~J., {et~al.} 2002, A framework for validation of computer models,
  Tech. rep.

\bibitem[{{Belokurov} {et~al.}(2007){Belokurov}, {Zucker}, {Evans}, {Kleyna},
  {Koposov}, {Hodgkin}, {Irwin}, {Gilmore}, {Wilkinson}, {Fellhauer},
  {Bramich}, {Hewett}, {Vidrih}, {De Jong}, {Smith}, {Rix}, {Bell}, {Wyse},
  {Newberg}, {Mayeur}, {Yanny}, {Rockosi}, {Gnedin}, {Schneider}, {Beers},
  {Barentine}, {Brewington}, {Brinkmann}, {Harvanek}, {Kleinman}, {Krzesinski},
  {Long}, {Nitta}, \& {Snedden}}]{2007ApJ...654..897B}
{Belokurov}, V., {et~al.} 2007, \apj, 654, 897

\bibitem[{{Belokurov} {et~al.}(2009){Belokurov}, {Walker}, {Evans}, {Gilmore},
  {Irwin}, {Mateo}, {Mayer}, {Olszewski}, {Bechtold}, \&
  {Pickering}}]{2009MNRAS.397.1748B}
---. 2009, \mnras, 397, 1748

\bibitem[{{Benson}(2012)}]{2012NewA...17..175B}
{Benson}, A.~J. 2012, NewA, 17, 175

\bibitem[{{Benson} \& {Bower}(2010)}]{2010MNRAS.405.1573B}
{Benson}, A.~J., \& {Bower}, R. 2010, \mnras, 405, 1573

\bibitem[{{Benson} {et~al.}(2003){Benson}, {Bower}, {Frenk}, {Lacey}, {Baugh},
  \& {Cole}}]{2003ApJ...599...38B}
{Benson}, A.~J., {Bower}, R.~G., {Frenk}, C.~S., {Lacey}, C.~G., {Baugh},
  C.~M., \& {Cole}, S. 2003, \apj, 599, 38

\bibitem[{{Bond} {et~al.}(2010){Bond}, {Ivezi{\'c}}, {Sesar}, {Juri{\'c}},
  {Munn}, {Kowalski}, {Loebman}, {Ro{\v s}kar}, {Beers}, {Dalcanton},
  {Rockosi}, {Yanny}, {Newberg}, {Allende Prieto}, {Wilhelm}, {Lee},
  {Sivarani}, {Majewski}, {Norris}, {Bailer-Jones}, {Re Fiorentin}, {Schlegel},
  {Uomoto}, {Lupton}, {Knapp}, {Gunn}, {Covey}, {Allyn Smith}, {Miknaitis},
  {Doi}, {Tanaka}, {Fukugita}, {Kent}, {Finkbeiner}, {Quinn}, {Hawley},
  {Anderson}, {Kiuchi}, {Chen}, {Bushong}, {Sohi}, {Haggard}, {Kimball},
  {McGurk}, {Barentine}, {Brewington}, {Harvanek}, {Kleinman}, {Krzesinski},
  {Long}, {Nitta}, {Snedden}, {Lee}, {Pier}, {Harris}, {Brinkmann}, \&
  {Schneider}}]{2010ApJ...716....1B}
{Bond}, N.~A., {et~al.} 2010, \apj, 716, 1

\bibitem[{{Bower} {et~al.}(2006){Bower}, {Benson}, {Malbon}, {Helly}, {Frenk},
  {Baugh}, {Cole}, \& {Lacey}}]{2006MNRAS.370..645B}
{Bower}, R.~G., {Benson}, A.~J., {Malbon}, R., {Helly}, J.~C., {Frenk}, C.~S.,
  {Baugh}, C.~M., {Cole}, S., \& {Lacey}, C.~G. 2006, \mnras, 370, 645

\bibitem[{{Bower} {et~al.}(2010){Bower}, {Vernon}, {Goldstein}, {Benson},
  {Lacey}, {Baugh}, {Cole}, \& {Frenk}}]{2010MNRAS.407.2017B}
{Bower}, R.~G., {Vernon}, I., {Goldstein}, M., {Benson}, A.~J., {Lacey}, C.~G.,
  {Baugh}, C.~M., {Cole}, S., \& {Frenk}, C.~S. 2010, \mnras, 407, 2017

\bibitem[{{Brinchmann} {et~al.}(2004){Brinchmann}, {Charlot}, {White},
  {Tremonti}, {Kauffmann}, {Heckman}, \& {Brinkmann}}]{2004MNRAS.351.1151B}
{Brinchmann}, J., {Charlot}, S., {White}, S.~D.~M., {Tremonti}, C.,
  {Kauffmann}, G., {Heckman}, T., \& {Brinkmann}, J. 2004, \mnras, 351, 1151

\bibitem[{{Bullock} \& {Johnston}(2005)}]{bj05}
{Bullock}, J.~S., \& {Johnston}, K.~V. 2005, \apj, 635, 931

\bibitem[{{Carollo} {et~al.}(2007){Carollo}, {Beers}, {Lee}, {Chiba}, {Norris},
  {Wilhelm}, {Sivarani}, {Marsteller}, {Munn}, {Bailer-Jones}, {Fiorentin}, \&
  {York}}]{2007Natur.450.1020C}
{Carollo}, D., {et~al.} 2007, \nat, 450, 1020

\bibitem[{{Carollo} {et~al.}(2010){Carollo}, {Beers}, {Chiba}, {Norris},
  {Freeman}, {Lee}, {Ivezi{\'c}}, {Rockosi}, \& {Yanny}}]{c10}
---. 2010, \apj, 712, 692

\bibitem[{{Chil{\`e}s} \& {Delfiner}(1999)}]{Chil:Delf:1999}
{Chil{\`e}s}, J.-P., \& {Delfiner}, P. 1999, {Geostatistics: Modeling Spatial
  Uncertainty} (John Wiley \& Sons, New York, NY)

\bibitem[{{Choi} \& {Nagamine}(2009)}]{choi}
{Choi}, J.-H., \& {Nagamine}, K. 2009, \mnras, 393, 1595

\bibitem[{{Cooper} {et~al.}(2010){Cooper}, {Cole}, {Frenk}, {White}, {Helly},
  {Benson}, {De Lucia}, {Helmi}, {Jenkins}, {Navarro}, {Springel}, \&
  {Wang}}]{cooper}
{Cooper}, A.~P., {et~al.} 2010, \mnras, 406, 744

\bibitem[{{Cressie}(1993)}]{Cres:1993}
{Cressie}, N. A.~C. 1993, {Statistics for Spatial Data} ({John Wiley \& Sons,
  New York, NY})

\bibitem[{{Dame} {et~al.}(2001){Dame}, {Hartmann}, \&
  {Thaddeus}}]{2001ApJ...547..792D}
{Dame}, T.~M., {Hartmann}, D., \& {Thaddeus}, P. 2001, \apj, 547, 792

\bibitem[{{De Lucia} \& {Helmi}(2008)}]{lh}
{De Lucia}, G., \& {Helmi}, A. 2008, \mnras, 391, 14

\bibitem[{{Dekel} \& {Woo}(2003)}]{2003MNRAS.344.1131D}
{Dekel}, A., \& {Woo}, J. 2003, \mnras, 344, 1131

\bibitem[{{Diemand} {et~al.}(2006){Diemand}, {Kuhlen}, \& {Madau}}]{fof}
{Diemand}, J., {Kuhlen}, M., \& {Madau}, P. 2006, \apj, 649, 1

\bibitem[{{Fadely} \& {Keeton}(2012)}]{fk2012}
{Fadely}, R., \& {Keeton}, C.~R. 2012, \mnras, 419, 936

\bibitem[{{Font} {et~al.}(2011{\natexlab{a}}){Font}, {McCarthy}, {Crain},
  {Theuns}, {Schaye}, {Wiersma}, \& {Dalla Vecchia}}]{2011MNRAS.416.2802F}
{Font}, A.~S., {McCarthy}, I.~G., {Crain}, R.~A., {Theuns}, T., {Schaye}, J.,
  {Wiersma}, R.~P.~C., \& {Dalla Vecchia}, C. 2011{\natexlab{a}}, \mnras, 416,
  2802

\bibitem[{{Font} {et~al.}(2011{\natexlab{b}}){Font}, {Benson}, {Bower},
  {Frenk}, {Cooper}, {De Lucia}, {Helly}, {Helmi}, {Li}, {McCarthy}, {Navarro},
  {Springel}, {Starkenburg}, {Wang}, \& {White}}]{2011MNRAS.417.1260F}
{Font}, A.~S., {et~al.} 2011{\natexlab{b}}, \mnras, 417, 1260

\bibitem[{{Ghez} {et~al.}(2008){Ghez}, {Salim}, {Weinberg}, {Lu}, {Do}, {Dunn},
  {Matthews}, {Morris}, {Yelda}, {Becklin}, {Kremenek}, {Milosavljevic}, \&
  {Naiman}}]{2008ApJ...689.1044G}
{Ghez}, A.~M., {et~al.} 2008, \apj, 689, 1044

\bibitem[{{Girardi} {et~al.}(2002){Girardi}, {Bertelli}, {Bressan}, {Chiosi},
  {Groenewegen}, {Marigo}, {Salasnich}, \& {Weiss}}]{girardi2002}
{Girardi}, L., {Bertelli}, G., {Bressan}, A., {Chiosi}, C., {Groenewegen},
  M.~A.~T., {Marigo}, P., {Salasnich}, B., \& {Weiss}, A. 2002, \aap, 391, 195

\bibitem[{{Girardi} {et~al.}(2004){Girardi}, {Grebel}, {Odenkirchen}, \&
  {Chiosi}}]{girardi2004}
{Girardi}, L., {Grebel}, E.~K., {Odenkirchen}, M., \& {Chiosi}, C. 2004, \aap,
  422, 205

\bibitem[{{Gnedin}(2000)}]{gne}
{Gnedin}, N.~Y. 2000, \apj, 542, 535

\bibitem[{{Greggio} \& {Renzini}(1983)}]{gregren}
{Greggio}, L., \& {Renzini}, A. 1983, \aap, 118, 217

\bibitem[{{Gregory}(2005)}]{gregory05}
{Gregory}, P. 2005, {Bayesian Logical Data Analysis for the Physical Sciences}
  (Cambridge University Press, Cambridge, UK)

\bibitem[{{Henriques} {et~al.}(2009){Henriques}, {Thomas}, {Oliver}, \&
  {Roseboom}}]{2009MNRAS.396..535H}
{Henriques}, B.~M.~B., {Thomas}, P.~A., {Oliver}, S., \& {Roseboom}, I. 2009,
  \mnras, 396, 535

\bibitem[{{Higdon} {et~al.}(2008){Higdon}, {Gattiker}, {Williams}, \&
  {Rightley}}]{Higd:Gatt:etal:2008}
{Higdon}, D., {Gattiker}, J., {Williams}, B., \& {Rightley}, M. 2008, Journal
  of the American Statistical Association, 103, 570

\bibitem[{{Ivezi{\'c}} {et~al.}(2008){Ivezi{\'c}}, {Sesar}, {Juri{\'c}},
  {Bond}, {Dalcanton}, {Rockosi}, {Yanny}, {Newberg}, {Beers}, {Allende
  Prieto}, {Wilhelm}, {Lee}, {Sivarani}, {Norris}, {Bailer-Jones}, {Re
  Fiorentin}, {Schlegel}, {Uomoto}, {Lupton}, {Knapp}, {Gunn}, {Covey},
  {Smith}, {Miknaitis}, {Doi}, {Tanaka}, {Fukugita}, {Kent}, {Finkbeiner},
  {Munn}, {Pier}, {Quinn}, {Hawley}, {Anderson}, {Kiuchi}, {Chen}, {Bushong},
  {Sohi}, {Haggard}, {Kimball}, {Barentine}, {Brewington}, {Harvanek},
  {Kleinman}, {Krzesinski}, {Long}, {Nitta}, {Snedden}, {Lee}, {Harris},
  {Brinkmann}, {Schneider}, \& {York}}]{2008ApJ...684..287I}
{Ivezi{\'c}}, {\v Z}., {et~al.} 2008, \apj, 684, 287

\bibitem[{{Jeffreys}(1998)}]{jeffreys98}
{Jeffreys}, H. 1998, {Theory of Probability} (Oxford University Press)

\bibitem[{{Jollife}(2002)}]{jollife02}
{Jollife}, I.~T. 2002, {Principal Component Analysis} (Springer Series in
  Statistics, Springer, New York)

\bibitem[{{Juri{\'c}} {et~al.}(2008){Juri{\'c}}, {Ivezi{\'c}}, {Brooks},
  {Lupton}, {Schlegel}, {Finkbeiner}, {Padmanabhan}, {Bond}, {Sesar},
  {Rockosi}, {Knapp}, {Gunn}, {Sumi}, {Schneider}, {Barentine}, {Brewington},
  {Brinkmann}, {Fukugita}, {Harvanek}, {Kleinman}, {Krzesinski}, {Long},
  {Neilsen}, {Nitta}, {Snedden}, \& {York}}]{2008ApJ...673..864J}
{Juri{\'c}}, M., {et~al.} 2008, \apj, 673, 864

\bibitem[{{Kalirai} {et~al.}(2006){Kalirai}, {Gilbert}, {Guhathakurta},
  {Majewski}, {Ostheimer}, {Rich}, {Cooper}, {Reitzel}, \& {Patterson}}]{kali}
{Kalirai}, J.~S., {et~al.} 2006, \apj, 648, 389

\bibitem[{{Keller} {et~al.}(2007){Keller}, {Schmidt}, {Bessell}, {Conroy},
  {Francis}, {Granlund}, {Kowald}, {Oates}, {Martin-Jones}, {Preston},
  {Tisserand}, {Vaccarella}, \& {Waterson}}]{2007PASA...24....1K}
{Keller}, S.~C., {et~al.} 2007, PASA, 24, 1

\bibitem[{{Kennedy} \& {O'Hagan}(2000)}]{Kenn:OHag:2000}
{Kennedy}, M.~C., \& {O'Hagan}, A. 2000, Biometrika, 1

\bibitem[{Kennedy \& O'Hagan(2001)}]{Kenn:OHag:2001}
Kennedy, M.~C., \& O'Hagan, A. 2001, 63, 425

\bibitem[{{Kroupa}(2001)}]{kroupa2001}
{Kroupa}, P. 2001, \mnras, 322, 231

\bibitem[{{Li} {et~al.}(2010){Li}, {De Lucia}, \&
  {Helmi}}]{2010MNRAS.401.2036L}
{Li}, Y.-S., {De Lucia}, G., \& {Helmi}, A. 2010, \mnras, 401, 2036

\bibitem[{{LSST Science Collaborations} {et~al.}(2009){LSST Science
  Collaborations}, {Abell}, {Allison}, {Anderson}, {Andrew}, {Angel}, {Armus},
  {Arnett}, {Asztalos}, {Axelrod}, \& et~al.}]{2009arXiv0912.0201L}
{LSST Science Collaborations} {et~al.} 2009, ArXiv e-prints

\bibitem[{{Lu} {et~al.}(2012){Lu}, {Mo}, {Katz}, \&
  {Weinberg}}]{2012MNRAS.421.1779L}
{Lu}, Y., {Mo}, H.~J., {Katz}, N., \& {Weinberg}, M.~D. 2012, \mnras, 421, 1779

\bibitem[{{Lu} {et~al.}(2011){Lu}, {Mo}, {Weinberg}, \&
  {Katz}}]{2011MNRAS.416.1949L}
{Lu}, Y., {Mo}, H.~J., {Weinberg}, M.~D., \& {Katz}, N. 2011, \mnras, 416, 1949

\bibitem[{{Ma} {et~al.}(2012){Ma}, {Morrison}, {Harding}, {Xue}, {Rix},
  {Rockosi}, {Johnson}, {Lee}, \& {Cudworth}}]{2012AAS...21925214M}
{Ma}, Z., {et~al.} 2012, in American Astronomical Society Meeting Abstracts,
  Vol. 219, American Astronomical Society Meeting Abstracts, 252.14

\bibitem[{{Newberg} {et~al.}(2009){Newberg}, {China}, {in LAMOST}, \&
  {(PLUS)}}]{2009AAS...21341614N}
{Newberg}, H.~J., {China}, L.~p.~o., {in LAMOST}, P., \& {(PLUS)}, U. 2009, in
  Bulletin of the American Astronomical Society, Vol.~41, American Astronomical
  Society Meeting Abstracts 213, 416.14

\bibitem[{{Nomoto} {et~al.}(1997){Nomoto}, {Iwamoto}, {Nakasato}, {Thielemann},
  {Brachwitz}, {Tsujimoto}, {Kubo}, \& {Kishimoto}}]{nomo1997}
{Nomoto}, K., {Iwamoto}, K., {Nakasato}, N., {Thielemann}, F.-K., {Brachwitz},
  F., {Tsujimoto}, T., {Kubo}, Y., \& {Kishimoto}, N. 1997, Nuclear Physics A,
  621, 467

\bibitem[{{Norman} {et~al.}(2007){Norman}, {Bryan}, {Harkness}, {Bordner},
  {Reynolds}, {O'Shea}, \& {Wagner}}]{2007arXiv0705.1556N}
{Norman}, M.~L., {Bryan}, G.~L., {Harkness}, R., {Bordner}, J., {Reynolds}, D.,
  {O'Shea}, B., \& {Wagner}, R. 2007, ArXiv e-prints

\bibitem[{{Oakley} \& {O'Hagan}(2002)}]{Oakl:Ohag:2002}
{Oakley}, J.~E., \& {O'Hagan}, A. 2002, Biometrika, 89, 769

\bibitem[{{Oakley} \& {O'Hagan}(2004)}]{Oakl:Ohag:2004}
---. 2004, Journal of the Royal Statistical Society: Series B (Statistical
  Methodology), 66

\bibitem[{O'Hagan(2006)}]{OHag:2006}
O'Hagan, A. 2006, Reliability Engineering \& System Safety, 91, 1290 , the
  Fourth International Conference on Sensitivity Analysis of Model Output (SAMO
  2004) - SAMO 2004

\bibitem[{{O'Shea} {et~al.}(2004){O'Shea}, {Bryan}, {Bordner}, {Norman},
  {Abel}, {Harkness}, \& {Kritsuk}}]{2004astro.ph..3044O}
{O'Shea}, B.~W., {Bryan}, G., {Bordner}, J., {Norman}, M.~L., {Abel}, T.,
  {Harkness}, R., \& {Kritsuk}, A. 2004, ArXiv Astrophysics e-prints

\bibitem[{{O'Shea} {et~al.}(2005){O'Shea}, {Nagamine}, {Springel}, {Hernquist},
  \& {Norman}}]{2005ApJS..160....1O}
{O'Shea}, B.~W., {Nagamine}, K., {Springel}, V., {Hernquist}, L., \& {Norman},
  M.~L. 2005, \apjs, 160, 1

\bibitem[{{Papovich} {et~al.}(2005){Papovich}, {Dickinson}, {Giavalisco},
  {Conselice}, \& {Ferguson}}]{2005ApJ...631..101P}
{Papovich}, C., {Dickinson}, M., {Giavalisco}, M., {Conselice}, C.~J., \&
  {Ferguson}, H.~C. 2005, \apj, 631, 101

\bibitem[{{Perryman} {et~al.}(2001){Perryman}, {de Boer}, {Gilmore}, {H{\o}g},
  {Lattanzi}, {Lindegren}, {Luri}, {Mignard}, {Pace}, \& {de
  Zeeuw}}]{2001A&A...369..339P}
{Perryman}, M.~A.~C., {et~al.} 2001, \aap, 369, 339

\bibitem[{Rasmussen \& Williams(2005)}]{Rasmussen05}
Rasmussen, C.~E., \& Williams, C. K.~I. 2005, {Gaussian Processes for Machine
  Learning (Adaptive Computation and Machine Learning)} (The MIT Press)

\bibitem[{{Richardson} {et~al.}(2009){Richardson}, {Ferguson}, {Mackey},
  {Irwin}, {Chapman}, {Huxor}, {Ibata}, {Lewis}, \& {Tanvir}}]{rich}
{Richardson}, J.~C., {et~al.} 2009, \mnras, 396, 1842

\bibitem[{{Rix} {et~al.}(2004){Rix}, {Barden}, {Beckwith}, {Bell}, {Borch},
  {Caldwell}, {H{\"a}ussler}, {Jahnke}, {Jogee}, {McIntosh}, {Meisenheimer},
  {Peng}, {Sanchez}, {Somerville}, {Wisotzki}, \& {Wolf}}]{2004ApJS..152..163R}
{Rix}, H.-W., {et~al.} 2004, \apjs, 152, 163

\bibitem[{{Sacks} {et~al.}(1989){Sacks}, {Welch}, {Mitchell}, \&
  {Wynn}}]{Sack:Welc:Mitc:Wynn:1989}
{Sacks}, J., {Welch}, W.~J., {Mitchell}, T.~J., \& {Wynn}, H.~P. 1989, Stat.\
  Sci., 4, 409

\bibitem[{{Santner} {et~al.}(2003){Santner}, {Williams}, \&
  {Notz}}]{Sant:Will:Notz:2003}
{Santner}, T.~J., {Williams}, B.~J., \& {Notz}, W. 2003, {The Design and
  Analysis of Computer Experiments}

\bibitem[{{Shapley}(2011)}]{2011ARA&A..49..525S}
{Shapley}, A.~E. 2011, \araa, 49, 525

\bibitem[{{Sijacki} {et~al.}(2011){Sijacki}, {Vogelsberger}, {Keres},
  {Springel}, \& {Hernquist}}]{2011arXiv1109.3468S}
{Sijacki}, D., {Vogelsberger}, M., {Keres}, D., {Springel}, V., \& {Hernquist},
  L. 2011, ArXiv e-prints

\bibitem[{{Spergel} {et~al.}(2007){Spergel}, {Bean}, {Dor{\'e}}, {Nolta},
  {Bennett}, {Dunkley}, {Hinshaw}, {Jarosik}, {Komatsu}, {Page}, {Peiris},
  {Verde}, {Halpern}, {Hill}, {Kogut}, {Limon}, {Meyer}, {Odegard}, {Tucker},
  {Weiland}, {Wollack}, \& {Wright}}]{wmap}
{Spergel}, D.~N., {et~al.} 2007, \apjs, 170, 377

\bibitem[{{Springel}(2005{\natexlab{a}})}]{2005MNRAS.364.1105S}
{Springel}, V. 2005{\natexlab{a}}, \mnras, 364, 1105

\bibitem[{{Springel}(2005{\natexlab{b}})}]{springel2005}
---. 2005{\natexlab{b}}, \mnras, 364, 1105

\bibitem[{{Springel}(2010)}]{2010MNRAS.401..791S}
---. 2010, \mnras, 401, 791

\bibitem[{{Tammann} {et~al.}(1994){Tammann}, {Loeffler}, \&
  {Schroeder}}]{1994ApJS...92..487T}
{Tammann}, G.~A., {Loeffler}, W., \& {Schroeder}, A. 1994, \apjs, 92, 487

\bibitem[{{Tasker} {et~al.}(2008){Tasker}, {Brunino}, {Mitchell}, {Michielsen},
  {Hopton}, {Pearce}, {Bryan}, \& {Theuns}}]{2008MNRAS.390.1267T}
{Tasker}, E.~J., {Brunino}, R., {Mitchell}, N.~L., {Michielsen}, D., {Hopton},
  S., {Pearce}, F.~R., {Bryan}, G.~L., \& {Theuns}, T. 2008, \mnras, 390, 1267

\bibitem[{{Teyssier}(2002)}]{2002A&A...385..337T}
{Teyssier}, R. 2002, \aap, 385, 337

\bibitem[{{Tominaga}(2009)}]{2009ApJ...690..526T}
{Tominaga}, N. 2009, \apj, 690, 526

\bibitem[{{Tumlinson}(2006)}]{tumlinson1}
{Tumlinson}, J. 2006, \apj, 641, 1

\bibitem[{{Tumlinson}(2010)}]{tumlinson2}
---. 2010, \apj, 708, 1398

\bibitem[{{Wadsley} {et~al.}(2004){Wadsley}, {Stadel}, \&
  {Quinn}}]{2004NewA....9..137W}
{Wadsley}, J.~W., {Stadel}, J., \& {Quinn}, T. 2004, New Astronomy, 9, 137

\bibitem[{{Walsh} {et~al.}(2007){Walsh}, {Jerjen}, \&
  {Willman}}]{2007ApJ...662L..83W}
{Walsh}, S.~M., {Jerjen}, H., \& {Willman}, B. 2007, \apjl, 662, L83

\bibitem[{{Willman} {et~al.}(2005){Willman}, {Blanton}, {West}, {Dalcanton},
  {Hogg}, {Schneider}, {Wherry}, {Yanny}, \& {Brinkmann}}]{2005AJ....129.2692W}
{Willman}, B., {et~al.} 2005, \aj, 129, 2692

\bibitem[{{Yanny} {et~al.}(2003){Yanny}, {Newberg}, {Grebel}, {Kent},
  {Odenkirchen}, {Rockosi}, {Schlegel}, {Subbarao}, {Brinkmann}, {Fukugita},
  {Ivezic}, {Lamb}, {Schneider}, \& {York}}]{2003ApJ...588..824Y}
{Yanny}, B., {et~al.} 2003, \apj, 588, 824

\bibitem[{{Zolotov} {et~al.}(2009){Zolotov}, {Willman}, {Brooks}, {Governato},
  {Brook}, {Hogg}, {Quinn}, \& {Stinson}}]{2009ApJ...702.1058Z}
{Zolotov}, A., {Willman}, B., {Brooks}, A.~M., {Governato}, F., {Brook}, C.~B.,
  {Hogg}, D.~W., {Quinn}, T., \& {Stinson}, G. 2009, \apj, 702, 1058

\bibitem[{{Zolotov} {et~al.}(2010){Zolotov}, {Willman}, {Brooks}, {Governato},
  {Hogg}, {Shen}, \& {Wadsley}}]{2010ApJ...721..738Z}
{Zolotov}, A., {Willman}, B., {Brooks}, A.~M., {Governato}, F., {Hogg}, D.~W.,
  {Shen}, S., \& {Wadsley}, J. 2010, \apj, 721, 738

\end{thebibliography}

\appendix\section{Multivarite Emulator formalism}
\label{sec:appen-pca-emu}

Here we outline the details of constructing an emulator for models with
multidimensional output $\mathitbf{y} = \{y_1, \ldots, y_t\}$. In theory
each component could be considered independent of the others and emulated
separately, as we have shown above by examining the implausibilities
$I(\vx)$ this does not usually give satisfactory results.  Principal
Components Analysis (PCA) is used to generate an orthogonal decomposition
of the output vector $\mathitbf{y}$. The resulting basis is orthogonal and
approximately independent. We retain the $r$ eigen-functions forming a
subset of the full decomposition responsible for $95\%$ of the observed
variation.  The set of training observations $\mathitbf{Y} =
\{\mathitbf{y}_1, \ldots, \mathitbf{y}_n\}$ is then projected into this new
orthogonal basis and independent emulators are generated for the weights of
the $r$ basis functions. Given our set of $n$ observations $\mathitbf{Y} =
\{\mathitbf{y}_1, \ldots, \mathitbf{y}_n\}$ we suppose that $\mathitbf{y}_i
\sim \mathrm{MVN}(\mathitbf{{\mu}},\mathbf{\Sigma})$ where
 \begin{align}
  \mu\approx   \hat{\mu}_j &= \frac{1}{n}\sum_{i=1}^{n}(y_i)_j,\\
\mathbf{\Sigma}\approx   \mathbf{\hat{\Sigma}} &= \frac{1}{n}\sum_{i=1}^{n}
   \left[(\mathitbf{y}_i - \hat{\mu}_i)^{T} 
         (\mathitbf{y}_i - \hat{\mu}_i)\right],
 \end{align}
 are the sample mean vector (of length $t$) and sample covariance matrix
 ($t\times t$) respectively.  An eigen-decomposition of the sample
 covariance matrix $\hat{\Sigma}$
\begin{equation}
  \label{eqn:appen-pca-1}
  \hat{\Sigma} = \mathbf{U} \mathbf{\Lambda} \mathbf{U}^{t},
\end{equation}
determines the PCA decomposition.  The columns of the matrix $\mathbf{U}$
are the eigenvectors of $\hat{\Sigma}$, while $\mathbf{\Lambda}$ is
diagonal with the corresponding eigenvalues in decreasing order.  The
eigenvectors define the new orthogonal basis and the eigenvalues determine
the weights of each basis function.

The sum of the eigenvalues or the trace of $\mathbf {\Lambda}$ is also the
trace of $\hat\Sigma$, the sum of the sample variances of $\mathitbf{Y}$, so
each eigenvalue represents the variance contribution of its associated
eigenfunction to the observed total.  The first eigenvector gives the
direction in which the data $\mathitbf{Y}$ are most variable, and the
remaining eigenvectors correspond to orthogonal directions with succesively
smaller amounts of variation.  Typically a small number $r\ll t$ of the
largest eigenvalues corresponding to the most variable combinations of the
observables $\mathitbf{y}$ are sufficient to describe nearly all of the
variation of the full data set. 
PCA can be used for dimension reduction by keeping only the $r$ most
influential components, and regarding the remaining $t-r$ dimensions as
noise and discarding them.
\begin{equation}
  \mathbf{\hat{\Sigma}} = \mathbf{U} \mathbf{\Lambda} 
  \mathbf{U}^{t} \approx \mathbf{U}_{r} \mathbf{\Lambda}_{r} \mathbf{U}_{r}^T.
\end{equation}
The variability fraction $V(r)=\sum_{i=1}^{r} \lambda_i / \mathrm{Tr}
(\mathbf{\Lambda})$ accounted for by the first $r$ components can be used
to help select of $r$; in this analysis we chose the smallest $r$ for which
$V(r) \ge 0.95$.  Typically with $t=7$ we could attain this with $r=5$.

After carrying out the eigen-decomposition and selecting an appropriate
$r$, we project our training set $Y'$ into the associated PCA basis as
\begin{equation}
  \mathbf{Z} = \frac{1}{\sqrt{\Lambda_r}}\mathbf{U}_r^{T}
                          (\mathitbf{Y} - \mathitbf{\hat{\mu}}),
\end{equation}
an $r\times n$ matrix of the projections of the training data projected
into the $r$ independent and orthogonal components of the PCA
decomposition.  Note that $\mathbf{Z}$ is standardized to have mean zero
and unit covariance.  Now we can construct $r$ independent emulators, each
one trained on a component of $Z$.

Each orthogonal emulator is constructed as outlined in \secref{sec:mod_emu}
giving a posterior mean $m_z(\vx)$ and variance $\Sigma_z(\vx)$
\eqref{eqn-emu-mean-var}. The mean and variance are functions of the
location in the design space $\vx$ it is important to remember that they
are functions which give output in the P.C space. To obtain predictions for
the model outputs $\mathitbf{Y}$ we need to undo the PCA decomposition by
reprojecting back into the natural observable space. Naturally by selecting
$r < t$ we have lost some of the original information however with
judicious choice of $r$ this is usually not a serious issue.

The projected mean, a vector of length $t$, $\mathitbf{m(\vx)}$ in the
$\mathitbf{Y}$ space is given as
\begin{equation}
  \mathitbf{m}(\vx) \approx \mathitbf{\hat{\mu}} + \mathbf{U}_r
  \Lambda_r^{1/2} \mathitbf{m}_z(\vx), 
\end{equation}
where $\mathitbf{m}_z(\vx)$ is the length $r$ vector of emulator means in
the P.C space. We label the emulated covariance of the $l$'th and $j$'th
model observables at the location $\vx_i$ as $K_{lj}(\vx_i)$, which is
\begin{equation}
\label{eqa7}
  K_{lj}(\vx_i) = \mbox{Cov}[y_l(\vx_i),y_j(\vx_i)] \approx
  \sum_{\alpha,\beta,\gamma=1}^r
  U_{l\alpha}\Lambda^{1/2}_{\alpha\beta}U_{j\gamma}\Lambda^{1/2}_{\gamma\beta}
  \mbox{Var}[Z_{\beta}(\vx_i)], 
\end{equation}
where $\mbox{Var}[Z_{\beta}(\vx_i)]$ is the emulated variance of the
$\beta$'th P.C dimension at the location $\vx_i$ in the parameter
space. This gives a useful estimate of the covariance between our
$\mathitbf{y}$ observables at as yet untried input locations and the $t
\times t$ matrix $\mathbf{K}$ is crucial for the joint implausibility
$J(\vx)$. 

\section{Five-dimensional parameter space exploration}
\label{sec:appen-5}

\begin{figure*}[ht]
\includegraphics[width=180mm,clip]{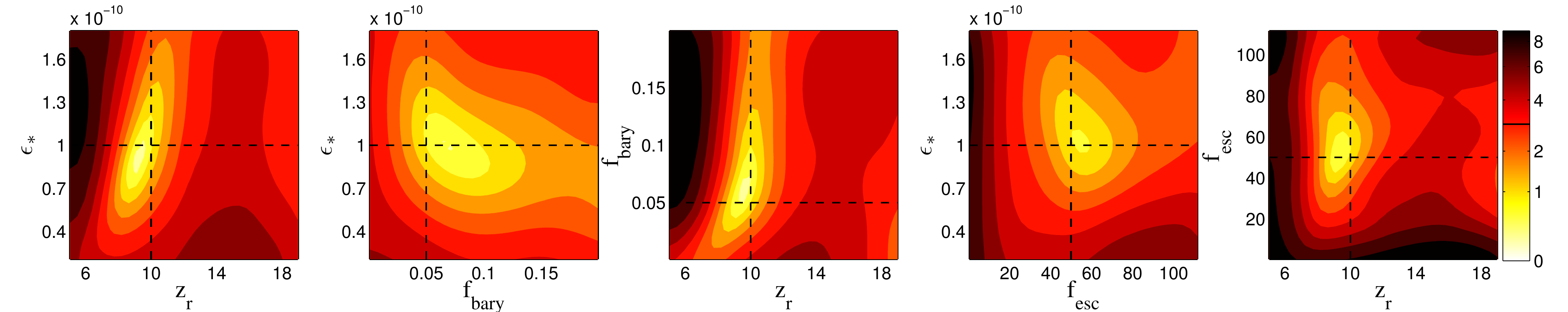}\\
\includegraphics[width=180mm,clip]{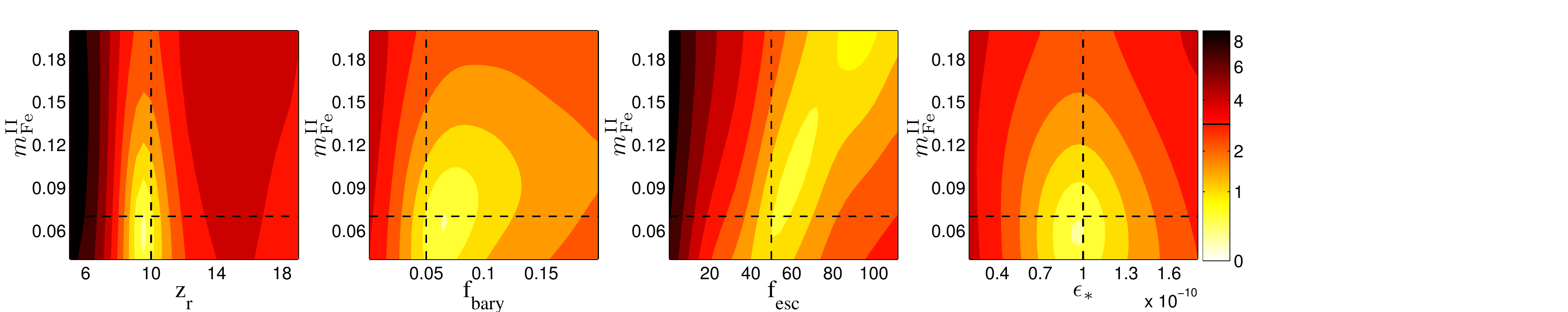}
\caption{Different sections of the Joint implausibility surface,
  $J(\mathitbf{x})$, obtained after training model emulators in a five
  dimensional input parameter space.  The mock observable data is generated
  by feeding ChemTreeN with the input parameter vector $\mathitbf{x}_{\rm
    obs} = (z_{\rm r},~f_{\rm esc},~f_{\rm bary},~\epsilon_{*},~m_{\rm
    Fe}^{\rm II}) = (10,~50,~0.05,~10^{-10},~0.07)$.  Each 2D section is
  obtained after fixing three out of the five parameters to the values
  associated with the component of $\mathitbf{x}_{\rm obs}$.  The black
  dashed lines indicate the values of the remaining two components.  The
  different colors show different values of $J(\mathitbf{x})$ in
  logarithmic scale.  The horizontal black solid line on the color bars
  indicate the imposed threshold: a value above this threshold
  shows that it is very implausible to obtain a good fit to the observed
  data with the corresponding values of the model parameters.  Note that
  $J(\mathitbf{x})$ can strongly constrain the five dimensional input
  parameter space under study.  Note as well that the values of the
  components of $\mathitbf{x}_{\rm obs}$ are always located in the most
  plausible regions of the space.}
\label{fig:appen1}
\end{figure*}

As  discussed  in \secref{sec:param_explor},  model  emulators can  be
easily generalized to deal with  input parameter spaces of much larger
dimensionality  than  previously  considered.   In  this  appendix  we
demonstrated  this  by  training  a  suite of  model  emulators  in  a
five-dimensional  input   parameter  space.   The   dimensionality  is
increased by including in  the analysis the star formation efficiency,
$\epsilon_{*}$, and the Type II  SNe yield, $m_{\rm Fe}^{\rm II}$. The
training  data set  is constructed  by running  ChemTreeN on  a design
$\mathcal{D} = \{\mathitbf{x_1}, \ldots, \mathitbf{x_n}\}$, containing
a  number $n=240$  design points.   Note the  larger number  of design
points with  respect to previous  examples, intended to  better sample
the larger volume of this input parameter space.  A mock observational
data  set  is generated  by  considering  the  input parameter  vector
$\mathitbf{x}_{\rm    obs}   =    (z_{\rm    r},~f_{\rm   esc},~f_{\rm
  bary},~\epsilon_{*},~m_{\rm          Fe}^{\rm         II})         =
(10,~50,~0.05,~10^{-10},~0.07)$.  Both,  mock observables and training
data  set are  obtained by  coupling  ChemTreeN with  the merger  tree
extracted  from the  dark matter-only  $N$-body simulation  \MW1.  The
results  are  shown  on   \figref{fig:appen1}.   Each  panel  shows  a
different  section  of  the  resulting Joint  implausibility  surface.
These  2D  sections are  obtained  after  fixing  the remaining  three
parameters  to the  values associated  with  $\mathitbf{x}_{\rm obs}$.
The black solid line on the  color bars show the threshold applied to the
joint implausibility.  Values above this threshold indicate that it is
very implausible to  obtain a good fit to  the mock observational data
using the  corresponding values of the model  parameters.  As observed
in  the  three-dimensional  example  shown  in  \figref{fig:implaus3},
$J(\mathitbf{x}_{\rm obs})$ can  strongly constrain the full parameter
space under study.  Even  in this higher-dimensional space, the values
of the  components of $\mathitbf{x}_{\rm  obs}$ are always  located in
the most  plausible regions  of the space,  as indicated by  the black
dashed lines.

\label{lastpage}
\end{document}